\documentclass[12pt,a4paper]{article}

\usepackage{amssymb}
\usepackage{subcaption}
\usepackage{mhchem}
\usepackage{graphicx}
\usepackage{booktabs}
\usepackage[T1]{fontenc}
\usepackage{tabularx}
\usepackage{siunitx}
\usepackage{eurosym}
\usepackage{xcolor}
\usepackage{color, colortbl}
\usepackage{longtable}
\usepackage{wrapfig}
\usepackage{booktabs}
\usepackage{amssymb}
\usepackage{placeins}
\usepackage{float}

\usepackage[colorlinks=true,linkcolor=blue,citecolor=blue,urlcolor=blue]{hyperref}
\usepackage[left=2.5cm,top=2.5cm,bottom=2.5cm,right=2.5cm]{geometry}

\input{author_formatting}
\DeclareAffiliation{uerj}{Faculty of Engineering, Rio de Janeiro State University, Brazil}
\DeclareAffiliation{umn}{School of Physics and Astronomy, University of Minnesota, USA}
\DeclareAffiliation{ess}{European Spallation Source ERIC, Sweden}
\DeclareAffiliation{columbia}{Department of Physics, Columbia University, USA}
\DeclareAffiliation{ornl}{Oak Ridge National Laboratory, USA}
\DeclareAffiliation{stockholm}{Department of Physics, Stockholm University, Sweden}
\DeclareAffiliation{cnr}{Institute for Plasma Science and Technology, National Research Council of Italy (CNR), Italy}
\DeclareAffiliation{bicocca}{Department of Physics ``Giuseppe Occhialini'', University of Milano-Bicocca, Italy}
\DeclareAffiliation{tum}{Physik-Department, Technical University of Munich, Germany}
\DeclareAffiliation{chalmers}{Department of Physics, Chalmers University of Technology, Sweden}
\DeclareAffiliation{lund}{Department of Physics, Lund University, Sweden}
\DeclareAffiliation{lundph}{Department of Philosophy, Lund University, Sweden}
\DeclareAffiliation{utk}{Department of Physics and Astronomy, University of Tennessee, USA}
\DeclareAffiliation{unicamp}{``Gleb Wataghin'' Institute of Physics, State University of Campinas, Brazil}
\DeclareAffiliation{consultant}{European Spallation Source consultant, Norway}
\DeclareAffiliation{ifjpan}{Henryk Niewodniczański Institute of Nuclear Physics, Polish Academy of Sciences, Poland}
\DeclareAffiliation{uff}{Department of Natural Sciences, Fluminense Federal University, Brazil}
\DeclareAffiliation{denver}{Department of Physics, University of Denver, USA}
\DeclareAffiliation{hip}{Helsinki Institute of Physics, University of Helsinki, Finland}
\DeclareAffiliation{hip2}{Department of Physics, University of Helsinki, Finland}
\DeclareAffiliation{uj}{Marian Smoluchowski Institute of Physics, Jagiellonian University, Poland}
\DeclareAffiliation{indiana}{Department of Physics, Indiana University Bloomington, USA}
\DeclareAffiliation{sussex}{Department of Physics and Astronomy, University of Sussex,  United Kingdom}
\DeclareAffiliation{uppsala}{Department of Physics and Astronomy, Uppsala University, Sweden}
\DeclareAffiliation{edinburgh}{College of Science and Engineering, University of Edinburgh, United Kingdom}
\DeclareAffiliation{ncsu}{Department of Physics and Astronomy, North Carolina State University, USA}
\DeclareAffiliation{cbpf}{Centro Brasileiro de Pesquisas Físicas (CBPF), Rio de Janeiro, Brazil}
\DeclareAffiliation{aarhus}{Department of Physics and Astronomy, Aarhus University, Denmark}
\DeclareAffiliation{uiuc}{The Grainger College of Engineering, University of Illinois Urbana-Champaign, USA}
\DeclareAffiliation{spain1}{Escuela de Ciencias, Ingeniería y Diseño, Universidad Europea de Valencia, Valencia, Spain}
\DeclareAffiliation{spain2}{Departamento de Física, Universidad de Castilla-La Mancha, Toledo, Spain}
\DeclareAffiliation{maryland}{Department of Physics, University of Maryland, College Park, MD 20742, USA}
\DeclareAffiliation{maryland}{Department of Physics, Indiana University Bloomington, USA}
\DeclareAffiliation{mlz}{Heinz Maier-Leibnitz Zentrum, Technical University of Munich, Germany}
\DeclareAffiliation{yan}{C. N. Yang Institute for Theoretical Physics and Department of Physics and Astronomy, Stony Brook University, New York, USA}
\DeclareAffiliation{nagoya}{Department of Physics, Nagoya University, Japan}
\DeclareAffiliation{alabama}{Department of Physics, University of Alabama, Tuscaloosa, Alabama 35487, USA}
\DeclareAffiliation{washington}{Department of Physics and McDonnell Center for the Space Sciences, Washington University, USA}
\DeclareAffiliation{Okl}{Department of Physics, Oklahoma State University, USA}
\DeclareAffiliation{Infn}{Laboratori Nazionali del Gran Sasso, National Institute for Nuclear Physics (INFN), Italy}
\DeclareAffiliation{cas}{Institute of High Energy Physics, Chinese Academy of Science, China}
\DeclareAffiliation{csns}{China Spallation Neutron Source, Dongguan, China}
\DeclareAffiliation{surrey}{School of Maths and Physics, University of Surrey, United Kingdom}
\DeclareAffiliation{ant}{Department of Physics, University of Antwerp, Belgium}
\DeclareAffiliation{CERN}{CERN, European Organization for Nuclear Research, Geneva, Switzerland}
\DeclareAffiliation{ferrara}{Sezione di Ferrara, Istituto Nazionale di Fisica Nucleare (INFN), Italy}
\DeclareAffiliation{liverpool}{Department of Physics, University of Liverpool, UK}
\DeclareAffiliation{naples}{Sezione di Napoli, Istituto Nazionale di Fisica Nucleare (INFN), Italy}
\DeclareAffiliation{montclair}{Physics and Astronomy, College of Science and Mathematics, Montclair State University, USA}
\DeclareAffiliation{mirrotron}{Mirrotron Ltd, Hungary}
\DeclareAffiliation{hunren}{HUN-REN Centre for Energy Research, Hungary}
\DeclareAffiliation{caltech}{Kellogg Radiation Laboratory, California Institute of Technology, USA}
\DeclareAffiliation{zacatecas}{Unidad Académica de Física, Universidad Autónoma de Zacatecas, Mexico}
\DeclareAffiliation{stfc}{Particle Physics Department, Rutherford Appleton Laboratory (STFC), Harwell, United Kingdom}
\SetCorrespondingAuthor{valentina.santoro@fysik.lu.se}

\AddAuthor{J.~Amaral}{uerj}
\AddAuthor{J.~Barrow}{umn}
\AddAuthor{M.~Bartis}{ess}
\AddAuthor{D.H.~Beck}{uiuc}
\AddAuthor{P.M.~Bentley}{ess}
\AddAuthor{J.~Bijnens}{lund}
\AddAuthor{G.~Brooijmans}{columbia}
\AddAuthor{L.J.~Broussard}{ornl}
\AddAuthor{A.~Burgman}{stockholm}
\AddAuthor{G.~Cibinetto}{ferrara}
\AddAuthor{G.~Croci}{cnr,bicocca}
\AddAuthor{D.D.~Di~Julio}{ess}
\AddAuthor{P.~Fierlinger}{tum}
\AddAuthor{B.W.~Filippone}{caltech}
\AddAuthor{C.~Gatto}{naples}
\AddAuthor{H.~Gisbert}{spain1}
\AddAuthor{H.~Gogoi}{ifjpan}
\AddAuthor{G.~Gorini}{bicocca}
\AddAuthor{A.~Gutiérrez-Rodríguez}{zacatecas}
\AddAuthor{A.~Koberinski}{lundph}
\AddAuthor{M.~Kaczmarek}{uj}
\AddAuthor{M.~Silarski}{uj}
\AddAuthor{M.~O'Riordan}{lund}

\AddAuthor{A.~Heinz}{chalmers}
\AddAuthor{J.~Henderson}{surrey}
\AddAuthor{M.~Holl}{lund}
\AddAuthor{H.T.~Johansson}{chalmers}
\AddAuthor{B.~Jönsson}{ess}
\AddAuthor{H.~Henriksson}{lund}
\AddAuthor{L.~Jönsson}{lund}
\AddAuthor{Y.~Kamyshkov}{utk}
\AddAuthor{M.~P.~Kuhl}{lund}
\AddAuthor{P.~Kinhult}{ess}
\AddAuthor{R.~Kolevatov}{consultant}
\AddAuthor{A.~Kozela}{ifjpan}
\AddAuthor{W.T.~Lee}{ess}
\AddAuthor{K.K.H.~Leung}{montclair}
\AddAuthor{T.-J.~Liang}{cas, csns}
\AddAuthor{G.~Luzin}{lund}
\AddAuthor{B.~Meirose}{lund,chalmers}
\AddAuthor{D.~Milstead}{stockholm}
\AddAuthor{R.~Mohapatra}{maryland}
\AddAuthor{G.~Muhrer}{ess}
\AddAuthor{A.~Muraro}{cnr}
\AddAuthor{A.~Nepomuceno}{uff}
\AddAuthor{N.~Okada}{alabama}
\AddAuthor{A.~Oskarsson}{lund}
\AddAuthor{T.~Pałasz}{uj}
\AddAuthor{R.~Pasechnik}{lund}
\AddAuthor{V.~Larsson}{lund}
\AddAuthor{L.B.~Persson}{lund}
\AddAuthor{O.~Putignano}{cnr}

\AddAuthor{J.~Ramsey}{ornl}
\AddAuthor{C.~Snow}{denver}
\AddAuthor{L.~Werselius}{lund}
\AddAuthor{B.~Rataj}{lund}
\AddAuthor{A.~Rodríguez-Sánchez}{spain2}
\AddAuthor{A.~Saunders}{ornl}
\AddAuthorCorr{V.~Santoro}{ess,lund}
\AddAuthor{F.~Scioscioli}{cnr,bicocca}
\AddAuthor{R.~Shrock}{yan}
\AddAuthor{D.~Silvermyr}{lund}
\AddAuthor{H.~Sina}{ess}
\AddAuthor{W.M.~Snow}{indiana}
\AddAuthor{Y.V.~Stadnik}{sussex}
\AddAuthor{L.~Tchelidze}{ess} 
\AddAuthor{N.~Tsapatsaris}{ess}
\AddAuthor{A.~Tureanu}{hip,hip2}
\AddAuthor{A.~Vilela~Pereira}{cbpf}
\AddAuthor{M.~Wolke}{uppsala}
\AddAuthor{J.~Womersley}{edinburgh}
\AddAuthor{S.~Xu}{lund}
\AddAuthor{A.R.~Young}{ncsu}
\AddAuthor{K.~Österberg}{hip,hip2}

\begin{document}

\begin{center}

{\Large {\bf FINESSE}: A Versatile Beamline for \textbf{F}undamental \textbf{I}nteractions with \textbf{N}eutrons at the \textbf{E}uropean \textbf{S}pallation \textbf{S}ourc\textbf{E}
\par}

\vspace{1em}

\PrintAuthors

\vspace{1em}

\PrintAffiliations

\vspace{0.5em}

{\corrauthormarker\ \corrauthortext}

\end{center}

\vspace{0.8em}

\vspace{1em}

\begin{abstract}
FINESSE (A Versatile Beamline for \textbf{F}undamental \textbf{I}nteractions with \textbf{N}eutrons at the \textbf{E}uropean \textbf{S}pallation \textbf{S}ourc\textbf{E}) is a proposed versatile beamline for fundamental interactions with neutrons at the European Spallation Source. Building on the HIBEAM design studies, FINESSE extends the original concept into a dual-guide beamline capable of operating either in a high-flux configuration or in a low-background configuration within a single ESS beam port. The high-flux branch is optimized for rare-process searches, including neutron--antineutron oscillations, neutron--sterile-neutron transitions, axion-like-particle searches, neutron charge measurements, and rare neutron decays. The low-background branch is designed for precision measurements requiring strong suppression of prompt and line-of-sight backgrounds, with a cryogenic neutron electric-dipole-moment experiment used as a benchmark case. A baseline exchangeable-guide concept has been developed in which only short guide sections inside the ESS bunker are exchanged, while the upstream extraction system and chopper positions remain fixed. A staged implementation, Short FINESSE, would use the upstream section of the high-flux beamline together with a compact magnetically controlled experimental region, enabling an early physics and R\&D programme before completion of the full facility. This paper summarizes the scientific case, beamline concept, guide simulations, chopper and timing system, magnetic-field control, detector developments, and staged implementation strategy. FINESSE would establish a long-term particle-physics infrastructure at ESS, providing a beamline for both discovery-driven searches and precision neutron measurements throughout the lifetime of the facility.
\end{abstract}

\noindent\textbf{Keywords:}
fundamental neutron physics; European Spallation Source; neutron beamline; neutron oscillations; neutron electric dipole moment

\vspace{0.8em}

\begingroup
\renewcommand{\thefootnote}{}
\footnotetext{\scriptsize
This manuscript has been co-authored by UT-Battelle, LLC, under contract
DE-AC05-00OR22725 with the US Department of Energy (DOE). The US government
retains and the publisher, by accepting the article for publication,
acknowledges that the US government retains a nonexclusive, paid-up,
irrevocable, worldwide license to publish or reproduce the published form of
this manuscript, or allow others to do so, for US government purposes. DOE
will provide public access to these results of federally sponsored research
in accordance with the DOE Public Access Plan
(\url{https://www.energy.gov/doe-public-access-plan}).}
\endgroup

\section{Introduction}

The European Spallation Source (ESS) will provide the highest-brightness pulsed cold-neutron beams currently available, opening new opportunities for experiments in fundamental particle physics based on intense and well-controlled neutron beams~\cite{Santoro:2023izd,Abele:2022iml}. The unique combination of high cold-neutron flux, long-pulse time structure, and the possibility of constructing long free-flight beamlines with controlled magnetic and background conditions makes ESS particularly well suited for a broad programme of particle physics with neutrons.

The proposed FINESSE instrument builds directly on the HIBEAM design studies~\cite{Santoro:2023izd} and represents the evolution of HIBEAM into a broader and more versatile beamline for fundamental interactions with neutrons. Supported by an international community, FINESSE exploits a dual-guide approach for high-flux and low-background operation. Since the high-flux configuration has already been described in detail in the HIBEAM design paper~\cite{Santoro:2023izd}, the focus here is on the extension to low-background operation and on the exchangeable-guide system that enables both configurations to be implemented within a single beamline infrastructure. The possibility of initiating the scientific programme with a shorter version of the high-flux beamline, referred to as Short FINESSE, is also considered.

The scientific case spans several key directions in physics beyond the Standard Model and precision tests of fundamental interactions. These include searches for neutron--antineutron oscillations and other neutron oscillation phenomena, which probe baryon-number violation and dark-sector scenarios; searches for axion-like particles and other weakly coupled new physics; tests of the electrical neutrality of the neutron; searches for the neutron electric dipole moment as a probe of new sources of CP violation; and precision measurements of neutron beta decay observables and hadronic weak interactions.

The flexibility of the FINESSE concept also allows complementary nuclear-physics applications. In particular, the scientific case developed within the SINE proposal (\textit{Scientific Investigations of Nuclei at ESS}), recently submitted to the ESS Instrument Roadmap call\footnote{\url{https://ess.eu/instrument-roadmap}}, has been incorporated into the broader FINESSE programme. This extends the scientific reach of FINESSE to neutron-induced reactions, high-resolution gamma and electron spectroscopy, nuclear-structure studies, and nuclear-data measurements.

Different classes of experiments impose markedly different requirements on the beam delivery system. Searches for extremely rare processes are primarily flux-limited and benefit from preserving the full moderator brightness, a large beam cross-section, and long magnetically controlled free-flight regions. By contrast, precision experiments and nuclear-spectroscopy measurements are often limited by environmental backgrounds, phase-space definition, polarization control, timing, detector response, and systematic effects, and therefore require low-background beam delivery with carefully controlled beam optics and shielding. FINESSE is designed to address both regimes within a single long-term ESS instrument through its complementary high-flux and low-background configurations.

The paper first presents the scientific motivation and physics programme, followed by the beamline concept and the high-flux and low-background configurations. The exchangeable-guide system, chopper and timing system, detector concepts, and staged implementation based on Short FINESSE are then described, followed by an outlook for the full FINESSE facility.

\section{Scientific motivation and physics programme}
Particle physics has entered a phase in which the intensity frontier is becoming increasingly important for searches for rare processes, tiny symmetry-violating effects, and high-precision deviations from Standard-Model (SM) predictions.  As the world's brightest pulsed neutron source, ESS offers a unique programme of such measurements and searches~\cite{Santoro:2023izd,Abele:2022iml}. FINESSE utilises the high cold-neutron brightness, pulsed time structure and possibility of a long, magnetically controlled free-flight region. This enables fundamental open questions to be addressed including the origin of the matter--antimatter asymmetry, the nature of dark matter, baryon-number and CP violation, and the existence of new physics beyond the SM~\cite{Canetti:2012zc,Babu:2013yca,Mohapatra:2022nnbarBaryo,Bertone:2004pz,Mohapatra:2024mirrorDM}. The 2026 European Strategy for Particle Physics emphasizes the importance of exploiting European Laboratories that can provide unique physics programmes~\cite{ESPP2026Recommendations,ESPP2026BriefingBook}. The NuPECC Long Range Plan 2024 identifies the need for fundamental neutron physics at ESS as an important future European capability with FINESSE matching the prescribed infrastructure properties~\cite{NuPECCLRP2024}.

The physics programme is based on two complementary operating modes. High-Flux FINESSE is optimized for searches for which sensitivity scales directly with neutron yield. This programme includes ultralight axion-like dark matter and other spin-dependent new-physics searches, free neutron--antineutron conversion, sterile-neutron transitions, neutron-charge determination, exotic neutron decays, and neutron-capture searches for dark-sector particles. Low-Background FINESSE is optimized for precision experiments in which environmental backgrounds and systematic control are central, including high-precision measurements of neutron decay, hadronic parity violation and neutron electric-dipole-moment. Short FINESSE is the staged initial implementation of High-Flux FINESSE, using the straight high-flux guide and providing a magnetically controlled experimental region of length around 10~m to allow early physics and R\&D prior to the development of the full instrument. 

Unless explicitly stated otherwise, all sensitivities and performances for high-flux, low-background FINESSE and Short FINESSE stated in this section assume an ESS linac power of $2\,$MW. 

An advantage of the ESS is the pulsed beam, which can be used for much more than prompt-background rejection. For High-Flux FINESSE, pulse timing is valuable for Ramsey neutron-spin measurements. In the axion-like-particle benchmark, time-of-flight information gives the neutron velocity on a pulse-by-pulse basis and enables velocity-dependent RF manipulation, allowing a broad velocity distribution to contribute coherently. In the neutron-oscillation programme, the pulse structure also allows velocity-dependent effects to be resolved rather than averaged over, while timing gates, sidebands and pulse-to-pulse consistency checks help characterize beam-related backgrounds. For Low-Background FINESSE, time-of-flight information can localize neutron pulses within decay-spectrometer fiducial volumes and provide wavelength tagging, for example in PERC-like neutron-decay measurements (Section~\ref{sec:beta_decay}). It can also help distinguish genuine spin-precession or parity-violating signals from important classes of velocity-dependent false effects in beam-EDM studies and hadronic-parity-violation measurements. 

The principal High-Flux FINESSE measurements and representative running scenarios are summarized in Table~\ref{tab:highflux_science}. The quoted sensitivities should be interpreted as representative physics targets rather than fixed performance guarantees, since the final reach depends on the detailed beamline configuration, accepted phase space, running time, and ESS operating power. The complementary precision-measurement programme foreseen for Low-Background FINESSE is summarized in Table~\ref{tab:lowbackground_science}. 

\begin{table}[H]
\centering
\caption{Representative physics programme for High-Flux FINESSE. Unless otherwise stated, the sensitivities refer to the full High-Flux configuration at an ESS linac power of 2~MW.}
\label{tab:highflux_science}

\begin{tabularx}{\textwidth}{p{2.3cm} p{2.8cm} X}
\toprule
\textbf{Search or} & \textbf{Representative} & \textbf{Projected FINESSE} \\
\textbf{channel} & \textbf{exposure} & \textbf{reach or role} \\
\midrule

Free \(n\rightarrow\bar n\) conversion
& \(\sim 3\)--\(5\) years
& Substantial improvement in discovery sensitivity relative to the historical ILL free-neutron search. Even with the compact WASA-based detector configuration, corresponding to an approximately 40~cm transverse acceptance, a five-year programme at 2~MW provides a sensitivity gain of several times that of ILL. With an 80~cm transverse acceptance, approximately one order of magnitude improvement can be approached in about three years at 2~MW. Further gains are possible through larger transverse acceptance, increased beam power, and coherent \(n/\bar n\) reflection. \\

\(n\rightarrow n'\) disappearance
& \(\sim 100\) days
& Projected pointwise 95\% confidence sensitivity to \(\tau_{nn'}\) of approximately 144~s in the \(0.012\)--\(12\)~peV mass-splitting range (see Section~\ref{sec:neutron_oscillations}). \\

\(n\rightarrow n'\rightarrow n\) regeneration
& \(\sim 100\) days
& Projected pointwise 95\% confidence sensitivity to \(\tau_{nn'}\) of approximately 92~s in the \(0.012\)--\(12\)~peV mass-splitting range, assuming a detected background rate of 0.1 counts/s (see Section~\ref{sec:neutron_oscillations}). \\

\(n\rightarrow n'\rightarrow\bar n\)
& \(\sim 100\) days
& Projected sensitivity to \(\sqrt{\tau_{nn'}\tau_{n'\bar n}}\) of approximately 187~s at 95\% confidence, under the idealized assumptions described in Section~\ref{sec:neutron_oscillations}. \\

Ultralight axion / axion-like dark matter
& \(\sim 150\) days
& Ramsey neutron-spin search covering approximately \(10^{-22}\)--\(10^{-16}\)~eV, with potential improvements of up to \(2\)--\(3\) orders of magnitude over current laboratory constraints in favourable regions of parameter space. \\

Neutron electric charge
& \(\sim 100\) days
& Direct sensitivity improvement of up to approximately a factor \(700\) relative to the present beam limit, depending on the final neutron-optics configuration and accepted phase space. \\

Exotic neutron decays
& Parasitic with \(n\rightarrow\bar n\)
& Searches for weakly interacting or long-lived particles in single-photon and partially invisible final states using the common annihilation-detector infrastructure. \\

Neutron-capture dark-sector searches
& To be determined
& Exploratory search for weakly coupled keV--MeV-scale particles using correlated satellite-line structures in neutron-capture \(\gamma\)-ray spectra. \\

\bottomrule
\end{tabularx}
\end{table}

\subsection{The High-Flux FINESSE scientific programme}

\subsubsection{Baryon-number violation and neutron oscillations}
\label{sec:neutron_oscillations}

High-Flux FINESSE supports several classes of neutral-neutron oscillation searches. The most prominent is direct free \(n\rightarrow\bar n\) conversion, a baryon-number-violating process with \(\Delta B=2\). Baryon-number violation is one of the Sakharov conditions required for the dynamical generation of the cosmic matter--antimatter asymmetry, making such searches directly relevant to the origin of observable matter~\cite{Canetti:2012zc,Babu:2013yca,Mohapatra:2022nnbarBaryo}. FINESSE can also search for oscillations involving sterile neutron states \(n'\), including disappearance, regeneration, and \(n\rightarrow n'\rightarrow\bar n\) channels. Such states arise in hidden-sector scenarios that provide well-motivated extensions of the Standard Model and can also be connected to dark matter~\cite{mirror1,mirror2,Mohapatra:2024mirrorDM}. All of these searches exploit the intense cold-neutron flux, long free-propagation region, and controlled magnetic environment provided by High-Flux FINESSE.

\paragraph{Free neutron--antineutron conversion}

A flagship High-Flux FINESSE measurement is a search for free neutron--antineutron conversion,
\[
n \rightarrow \bar n ,
\]
which can arise in a range of well-motivated extensions of the Standard Model~\cite{Babu:2013yca,Mohapatra:2022nnbarBaryo}. Observation of this process would constitute direct evidence for baryon-number violation with \(\Delta B=2\) and would have profound implications for models addressing the origin of the cosmic matter--antimatter asymmetry.

For a free-neutron beam experiment, the discovery figure of merit scales approximately as
\[
\mathrm{FOM}\sim N_n\langle t_n^2\rangle ,
\]
where \(N_n\) is the neutron yield and \(t_n\) is the free-propagation time. A long free-flight region is therefore advantageous because of the quadratic dependence on propagation time. However, the experimental reach is determined by the combined effects of neutron intensity, propagation time, transverse phase-space acceptance, annihilation-target size, detector performance, and accumulated exposure.

Neutrons propagate through a magnetically controlled region with a residual field below around \(10\)~nT. This limits the energy splitting between the neutron and antineutron states arising from their opposite magnetic moments and preserves the quasi-free condition required for oscillation searches~\cite{Davis:2016uyk}. Magnetic-field control at this level was already demonstrated in the historical ILL experiment through a combination of passive magnetic shielding and active field compensation~\cite{BaldoCeolin:1994nnbar}. If an antineutron is produced, it annihilates in a thin carbon target. The annihilation releases an energy of order \(2m_n\simeq1.88\)~GeV and produces an approximately isotropic multi-pion final state, reconstructed in the dedicated annihilation detector described in Section~\ref{anndetector}.

The historical ILL experiment remains the benchmark for free-neutron oscillation searches~\cite{BaldoCeolin:1994nnbar}. It employed a neutron-propagation region of approximately \(100\)~m, including an 81~m oscillation drift vessel, and a 110~cm-diameter carbon annihilation target. The effective quasi-free propagation time was
\[
\sqrt{\langle t^2\rangle}=0.109\pm0.002~\mathrm{s},
\]
with a neutron intensity at the annihilation target of approximately
\[
1.25\times10^{11}~\mathrm{n\,s^{-1}}.
\]
The overall antineutron detection efficiency was \(0.52\pm0.02\), and the effective accumulated running time was \(2.4\times10^7\)~s~\cite{BaldoCeolin:1994nnbar}.

Using the measured neutron intensity and effective propagation time, the
corresponding ILL beam figure of merit is
\[
N_n\langle t_n^2\rangle_{\mathrm{ILL}}
=
\left(1.25\times10^{11}~\mathrm{n\,s^{-1}}\right)
\left(0.109~\mathrm{s}\right)^2
\simeq 1.50\times10^9~\mathrm{n\,s}.
\]
This value is used as the reference beam figure of merit throughout the
following comparison. The dependence of the FINESSE beam figure of merit on propagation length
and transverse acceptance is illustrated in Fig.~\ref{fig:finesse_fom_size}. The calculation shows that both parameters provide important and complementary optimization handles. Increasing the free-flight distance enhances the contribution from \(t_n^2\), while increasing the annihilation-target size allows a larger fraction of the transported neutron phase space to contribute to the measurement.

\begin{center}
  \includegraphics[width=0.65\linewidth]{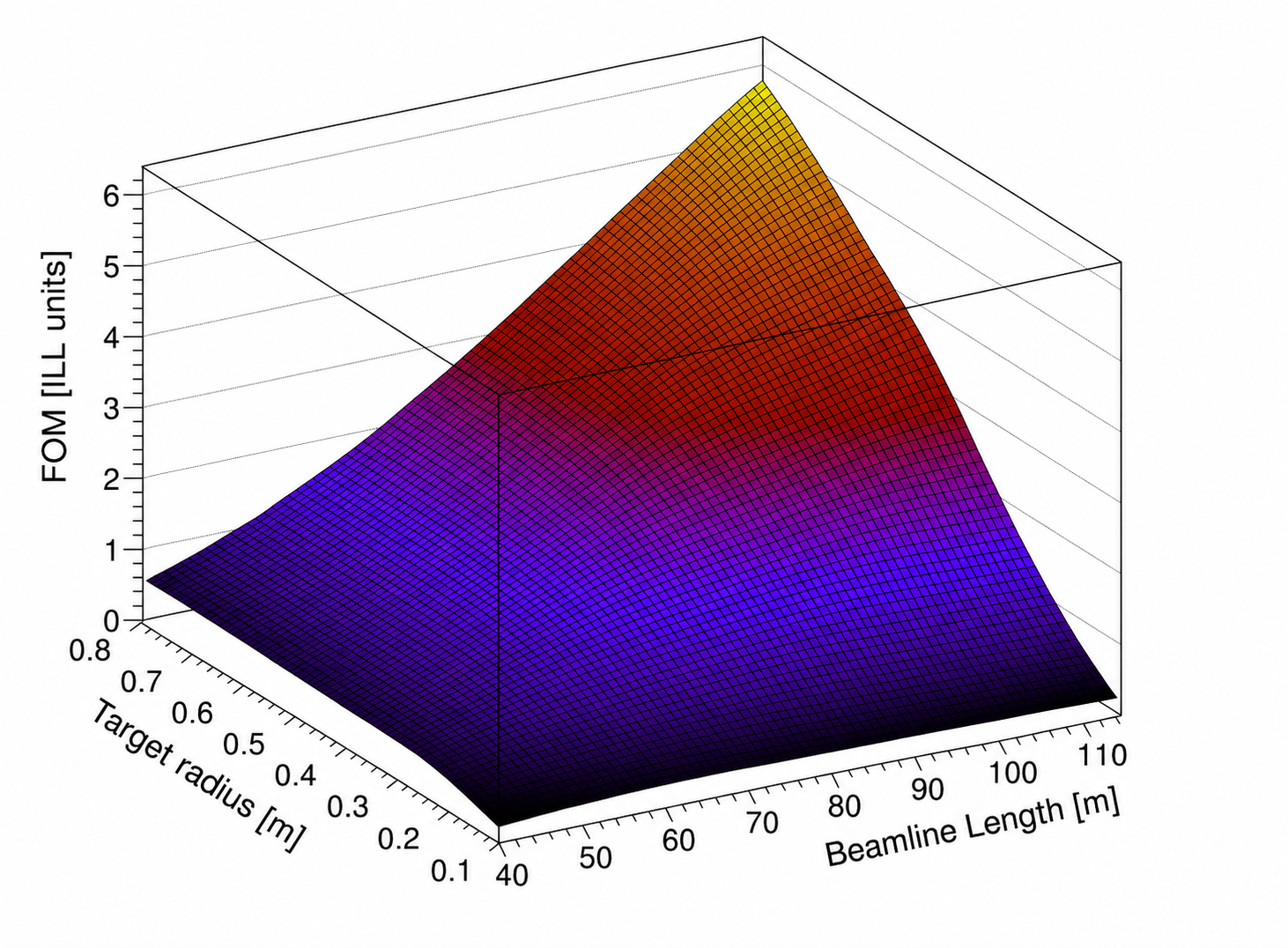}

  \refstepcounter{figure}\label{fig:finesse_fom_size}
  \small\textbf{Figure \thefigure.} 
  High-Flux FINESSE beam figure of merit for
\(n\rightarrow\bar n\), normalized to the historical ILL beam figure of merit,
\(N_n\langle t_n^2\rangle_{\mathrm{ILL}}
=1.50\times10^9~\mathrm{n\,s}\), as a function of propagation length and
annihilation-target size.
\end{center}

A representative High-Flux FINESSE propagation length of 75~m is used in
Table~\ref{tab:nnbar_scenarios} to illustrate the effect of transverse
acceptance. The 40~cm-diameter configuration corresponds approximately to
the acceptance presently available with the WASA-based~\cite{wasa} detector
option (see Section~\ref{anndetector}) and therefore does not exploit the
full transverse phase space delivered by the beamline. At 2~MW, this
configuration gives a beam figure of merit of \(1.34\times10^9\)~n\,s,
close to the historical ILL value. At 5~MW, the corresponding value rises to
\(3.35\times10^9\)~n\,s, already exceeding the ILL benchmark despite the
substantially smaller transverse acceptance. Increasing the target diameter
to 80~cm raises the beam figure of merit to \(3.05\times10^9\)~n\,s at
2~MW and \(7.61\times10^9\)~n\,s at 5~MW. A 1~m-diameter target, approaching
the transverse scale of the historical ILL experiment, further increases
the figure of merit to \(3.50\times10^9\)~n\,s at 2~MW and
\(8.74\times10^9\)~n\,s at 5~MW.

\begin{table}[H]
\centering
\caption{Representative High-Flux FINESSE configurations for free
\(n\rightarrow\bar n\) searches compared with the historical ILL experiment.
The beam figure of merit is \(N_n\langle t_n^2\rangle\). FINESSE values are shown
for ESS linac powers of 2 and 5~MW. The FINESSE configurations illustrate
the gain obtained by increasing the transverse acceptance at a fixed
75~m propagation length.}
\label{tab:nnbar_scenarios}

\begin{tabular}{lcccc}
\toprule
\textbf{Configuration} &
\textbf{Propagation} &
\textbf{Target} &
\multicolumn{2}{c}{\(\boldsymbol{N_n\langle t_n^2\rangle}\) \textbf{[n\,s]}} \\
\cmidrule(lr){4-5}
&
\textbf{length [m]} &
\textbf{diameter [cm]} &
\textbf{2 MW} &
\textbf{5 MW} \\
\midrule

FINESSE
& 75 & 40
& \(1.34\times10^{9}\)
& \(3.35\times10^{9}\) \\

FINESSE
& 75 & 80
& \(3.05\times10^{9}\)
& \(7.61\times10^{9}\) \\

FINESSE
& 75 & 100
& \(3.50\times10^{9}\)
& \(8.74\times10^{9}\) \\

\midrule

ILL
& \(\sim100\)
& 110
& \multicolumn{2}{c}{\(1.50\times10^{9}\)} \\

\bottomrule
\end{tabular}
\end{table}

The comparison demonstrates that propagation length alone does not determine
the reach of a free-neutron oscillation search. The 40~cm configuration is
primarily limited by the acceptance of the present WASA-based detector
option, whereas larger targets allow a substantially greater fraction of the
neutron phase space transported to E5 to contribute to the measurement.
For target sizes approaching that used in the historical ILL experiment,
the FINESSE beam figure of merit reaches more than twice the ILL value at
2~MW and nearly six times the ILL value at 5~MW. The large transverse
phase-space acceptance available at E5 therefore compensates substantially
for the shorter propagation distance relative to the historical ILL
geometry. 
Importantly, the beam figure of merit represents only one component of the overall experimental sensitivity. FINESSE also benefits from a substantially improved annihilation detector. A Time Projection Chamber (TPC) provides three-dimensional tracking and precision vertex reconstruction, while electromagnetic calorimetry enables photon and \(\pi^0\) reconstruction. Recent vertex-reconstruction studies obtain median transverse vertex errors of \(4\)--\(6\)~mm for the best-performing algorithms, compared with a transverse vertex resolution of order \(4\)--\(5\)~cm in the ILL experiment~\cite{Burgman:2026fer,BaldoCeolin:1994nnbar}. The expected FINESSE signal efficiency of approximately \(70\)--\(75\%\) also exceeds the approximately \(52\%\) overall antineutron-detection efficiency achieved at ILL~\cite{Santoro:2024lvc,BaldoCeolin:1994nnbar}. The WASA crystal calorimeter, available for FINESSE, provides nearly hermetic electromagnetic calorimetry and precision neutral-pion reconstruction, while an alternative larger lead-glass implementation~\cite{Santoro:2024lvc} would allow a larger annihilation target and correspondingly greater transverse acceptance, as discussed in Section~\ref{anndetector}.

The overall gain in discovery sensitivity relative to ILL can therefore be expressed schematically as
\[
G_{\mathrm{tot}}
=
\frac{N_n\langle t_n^2\rangle_{\mathrm{FINESSE}}}
     {N_n\langle t_n^2\rangle_{\mathrm{ILL}}}
\times
G_{\mathrm{det}}
\times
G_{\mathrm{runtime}},
\]

where \(G_{\mathrm{det}}\) accounts for the improved annihilation-detector efficiency and reconstruction performance, and \(G_{\mathrm{runtime}}\) accounts for the accumulated exposure relative to the historical ILL experiment.

For the purpose of comparing integrated sensitivities, one nominal year of
FINESSE operation is taken to correspond to \(2.4\times10^7\)~s of live time,
equal to the effective accumulated running time of the historical ILL
experiment.

Combining the beam figure of merit, improved detector performance, and
accumulated exposure gives FINESSE substantial discovery reach beyond the
historical free-neutron search. Even with the compact WASA-based detector
configuration, corresponding to an approximately 40~cm transverse acceptance,
a five-year programme at 2~MW provides an overall sensitivity gain of several
times that of the ILL experiment. Increasing the transverse acceptance to
80~cm raises the beam figure of merit to approximately twice the ILL value;
including the expected improvement in detector performance, a three-year
programme at 2~MW corresponds to an overall gain of approximately a factor
of nine, i.e. close to one order of magnitude relative to ILL. Operation at
5~MW provides additional sensitivity and considerably greater margin beyond
this benchmark, while the 1~m configuration illustrates the further reach
available when the transported neutron phase space can be more fully exploited.

These results demonstrate that a substantial advance in free
\(n\rightarrow\bar n\) sensitivity can be achieved within the E5 geometry.
Even the compact-detector configuration provides a clear improvement over the
historical ILL experiment, while larger transverse acceptance and increased
beam power offer substantial additional gains. The combination of high neutron
intensity, improved annihilation detection, and multi-year exposure therefore
provides several complementary routes to increased sensitivity.

A further possible enhancement is provided by coherent neutron and antineutron reflection from a dedicated guide geometry. In this approach, subcritical neutron and antineutron components can undergo coherent mirror reflections without immediately destroying the oscillation amplitude. Realistic studies indicate that this ``bounce'' method could provide an additional gain of approximately a factor of two in discovery potential~\cite{Gudkov:2021wvn,Nesvizhevsky:2018tft}. This provides an additional optimization path beyond the baseline sensitivity estimates presented above.

Short FINESSE will provide an important early configuration for free-\(n\rightarrow\bar n\) studies. In the staged programme, approximately \(100\) days of early operation can be devoted to \(n\rightarrow\bar n\) R\&D, including measurements of the E5 beam-related background environment and validation of the annihilation detector, magnetic shielding, and event reconstruction under beam conditions~\cite{FINESSEProposal2026}. As discussed in Section~\ref{sec:shortfinesse_optics}, neutron-optics simulations show that the uncollimated Short FINESSE configuration reaches a beam figure of merit of the same order as that of the historical ILL free-neutron search, supporting its use for \(n\rightarrow\bar n\) R\&D, commissioning, and validation of the experimental concept under realistic ESS beam conditions.

Within the broader experimental landscape, free-neutron searches and bound-neutron searches, such as those performed at Super-Kamiokande~\cite{Abe:2020ywm}, provide complementary information. The two approaches probe different realizations and operator structures of baryon-number-violating physics and should therefore be regarded as complementary rather than redundant~\cite{Barrow:2025rhm,Barbieri:1981yr,Berezhiani:2015afa}.

\paragraph{Sterile-neutron searches and dark-sector physics} \label{sterile}

FINESSE can also search for transitions between ordinary neutrons and sterile neutron states ($n'$), motivated by mirror-matter models in which ordinary particles have counterparts in a hidden sector \cite{mirror1, mirror2}. In this framework, mixing between the neutron and its mirror counterpart can give rise to $n \leftrightarrow n'$ oscillations \cite{zurab2, Kamyshkov:2022nmn}. FINESSE can probe these transitions through three complementary channels.  A disappearance search ($n \rightarrow n’$) looks for a deficit in the transmitted flux, normalised to an upstream beam monitor, as the applied magnetic field is varied. In a regeneration search ($n \rightarrow n’ \rightarrow n$), an absorber separates two propagation regions. The absorber removes ordinary neutrons while transmitting the sterile neutrons. Neutrons which are regenerated downstream can then be detected \cite{disappearance}. A third channel searches for $n\rightarrow n'\rightarrow\bar n$, a magnetically induced neutron-antineutron conversion via a sterile state (rather than the free neutron conversion considered earlier), which is identified through its annihilation signature.

Here we consider sensitivity estimates for a mass splitting ($\Delta m = m_{n’}-m_n$), assuming a vanishing mirror magnetic field ($B’=0$). An applied magnetic field shifts the spin states of the ordinary neutrons relative to the sterile states, allowing resonant conversion when $|\Delta m|=|\mu_n B|$, where $|\mu_n|=\SI{6.03e-14}{eV/\micro\tesla}$. Scanning the applied field therefore probes a range of mass splittings $\Delta m$. At a well-resolved resonance and in the weak-mixing limit, one spin component of an unpolarized beam dominates, giving
\begin{align}
P_{n\rightarrow n'}\simeq\frac{t^2}{2\tau_{nn'}^2}, \qquad P_{n\rightarrow n'\rightarrow n}\simeq\frac{t_1^2t_2^2}{2\tau_{nn'}^4},
\end{align}
where $t$ is the interaction time for disappearance and $(t_1, t_2)$ are the interaction times in the two regeneration regions. Here we define $\tau_{nn'}=\hbar/|\epsilon_{nn'}|$, where $\epsilon_{nn'}$ is the neutron–sterile-neutron mixing matrix element. For neutron regeneration, the applied fields in the two regions have equal magnitude and direction. The same spin component thus participates in both regeneration steps.

The estimates use weighted neutron trajectories obtained with
\textsc{McStas}~\cite{mcstas}, a Monte Carlo neutron ray-tracing package
widely used for the design and optimization of neutron instruments. The
simulations use the high-flux optics model (see Section~\ref{highfluxguide}) and are normalized to an ESS beam
power of 2~MW.
 A circular detection aperture of 40~cm diameter is assumed, positioned approximately 75~m from the target for FINESSE and 25~m for Short FINESSE. The nominal interaction length is taken from the bunker exit, approximately 15 m from the target, giving approximately 60 m for FINESSE and 10 m for Short FINESSE, with conversion probabilities evaluated in the free-propagation approximation. For regeneration, these lengths are split equally between the two conversion regions. The transverse detector positions are optimised separately for disappearance and regeneration using the flight-time-weighted neutron distributions. 

A separate exposure of 100 live days is assumed for each of the search modes. The scan comprises 1,000 field magnitudes from 2 mG to 2 G in steps of 2 mG, with equal exposure at both polarities. The corresponding resonant mass-splitting range is approximately (0.012–12) peV. For neutron regeneration, a constant detected background rate of 0.1 counts/s is assumed. The background mean is treated as known, and pointwise 95\% confidence sensitivities are obtained from the median background-only Poisson outcome. Unit neutron-detection efficiency is assumed within the accepted aperture. For disappearance, 90 days are allocated to the scan and 10 days to a shared signal-free reference measurement. The estimate includes counting fluctuations in the transmitted beam, monitor and reference, with wavelength-dependent monitor absorption normalised to 30\% mean interception. 

\begin{figure}[H]
    \centering
\includegraphics[width=0.85\textwidth]{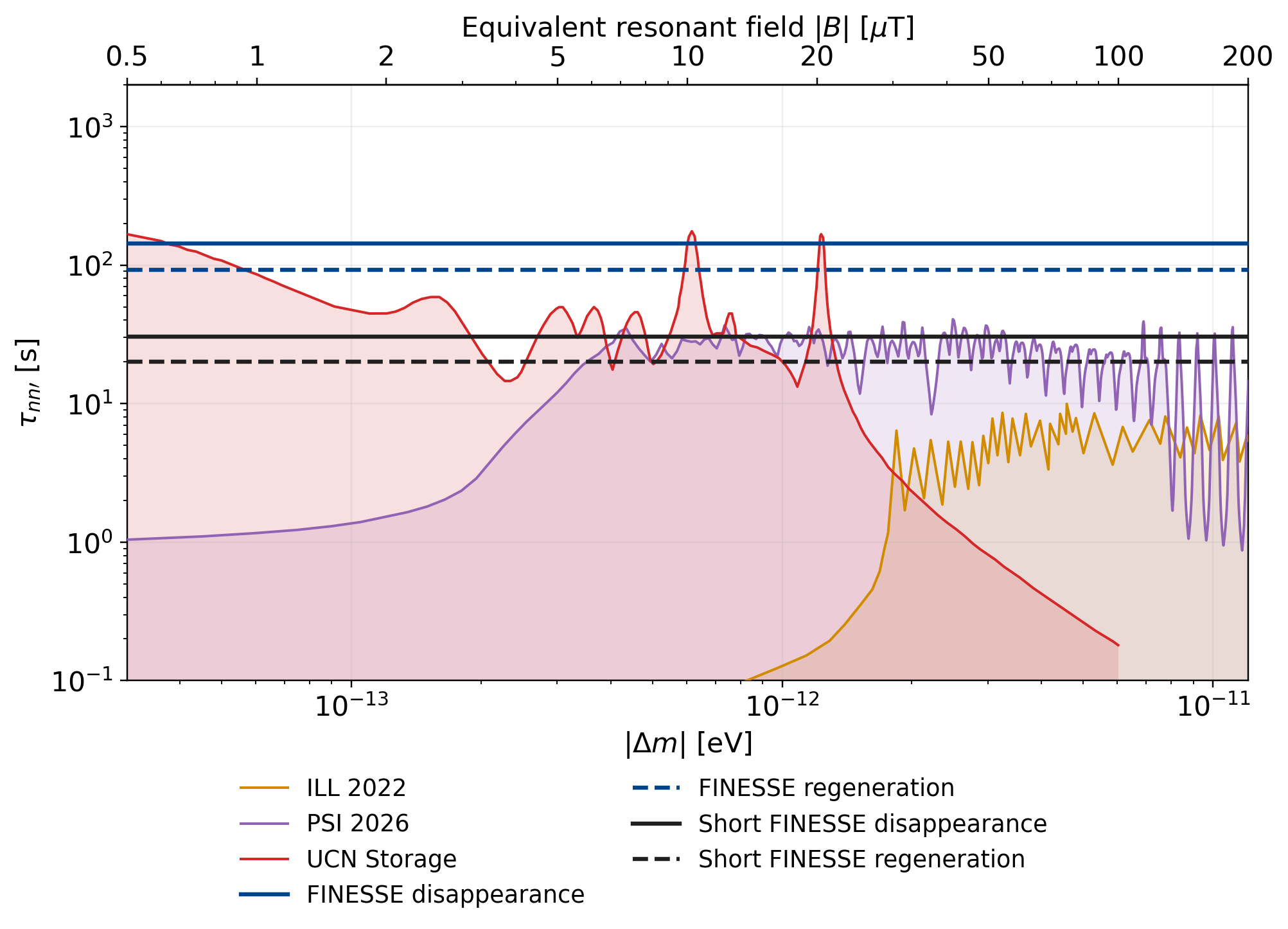}
\caption{Projected pointwise 95\% confidence sensitivities to neutron disappearance and regeneration in the mass-splitting scenario ($\Delta m\neq0$, $B'=0$). Both FINESSE configurations assume 2 MW beam power, a 40 cm diameter detection aperture, and interaction lengths of approximately 60 m and 10 m, respectively. A separate exposure of 100 live days is assumed for each search mode. The applied field is scanned from 2~mG to 2~G in 2~mG steps, with both polarities measured, giving 2,000 measurement settings. Regeneration thus receives 4,320~s per signed setting and assumes a known detected background rate of 0.1 counts/s. Disappearance allocates 90 days to scanning and 10 days to a shared signal-free reference. Horizontal lines show the ideal-field sensitivities evaluated at resonance. Shaded regions indicate existing exclusions. The UCN storage boundary is the upper envelope of the experimental results in Refs. \cite{Ban2007, Altarev2009b, Serebrov2009, Berezhiani2018, Abel2021}. The subsequent ILL \cite{Ban2023} and PSI measurements \cite{Ayres2026} are shown separately. The upper axis gives the equivalent resonant applied field.}
    \label{fig:sterile}
\end{figure}

Under these assumptions, the projected disappearance sensitivity to $\tau_{nn’}$ is approximately 144 s for FINESSE and 30 s for Short FINESSE. The corresponding neutron regeneration sensitivities are approximately 92 s and 20 s. Figure \ref{fig:sterile} compares these projections with existing constraints. 

For the two-stage $n\rightarrow n'\rightarrow\bar n$ benchmark, a single sterile intermediate state is assumed. The absorber separates two regions with applied fields of equal magnitude and opposite direction, allowing the same spin component to participate in both resonant transitions. At resonance, 
\begin{align}
P_{n\rightarrow n'\rightarrow\bar n}\simeq \frac{t_1^2t_2^2}{2\tau_{nn'}^2\tau_{n'\bar n}^2}.\end{align}
Here $\tau_{n'\bar n}=\hbar/|\epsilon_{n'\bar n}|$ denotes the characteristic mixing time for the second transition. Assuming complete background rejection and unit annihilation-detection efficiency within the accepted aperture, the same 100-day scan gives a projected sensitivity to $\sqrt{\tau_{nn'}\tau_{n'\bar n}}$ of approximately 187~s for  FINESSE and 41~s for Short FINESSE at a confidence level of 95\%.  

Note that the estimates in this section are statistics-based and do not include detailed studies of magnetic-field uniformity, detector performance or systematics. Further dedicated studies on these topics are planned. Furthermore, choosing to scan a more narrow magnetic field region for a given running time would lift the sensitivity higher still. It should also be added that the same apparatus could also support searches for a neutron transition magnetic moment by including configurations that tune magnetic and matter potentials \cite{nTMM}.

\subsubsection{Axion-like particles and other spin-dependent new-physics searches} \label{ALP}

High-Flux FINESSE enables a Ramsey neutron-beam search for ultralight axion-like dark matter through its derivative coupling to neutron spin~\cite{PhysRevLett.133.181001}. A coherently oscillating axion field can act as a pseudo-magnetic field. This generates an additional phase as polarized neutrons traverse a magnetically controlled region.

The ESS pulse structure provides a strong advantage. Time-dependent radiofrequency spin manipulation can compensate for the neutron velocity distribution. This allows a broad neutron spectrum to contribute coherently to the measurement. Velocity compensation alone can improve phase sensitivity by approximately a factor of four, while its combination with a $0.3\,$ms chopper window can yield an improvement as large as approximately $44$~\cite{RamseyProp}.

The projected full-length sensitivity extends over axion masses of approximately $10^{-22}$--$10^{-16}\,$eV. Recent laboratory searches have substantially improved existing constraints, particularly at the lowest masses and through new broadband spin-based interferometry~\cite{Zhang2025AxionReanalysis,GavilanMartin2025SpinInterferometer}. Nevertheless, the full-length FINESSE Ramsey experiment retains projected improvements of up to $2$--$3$ orders of magnitude over current laboratory constraints in parts of the intermediate-mass region~\cite{PhysRevLett.133.181001,Zhang2025AxionReanalysis,GavilanMartin2025SpinInterferometer}. This measurement does not require the full-length FINESSE facility. For Short FINESSE, a representative $150$-day exposure at the $2\,$MW operation benchmark retains more than an order-of-magnitude improvement over current laboratory constraints in favourable parts of the same parameter space. This is consistent with the statistical scaling of the published Ramsey sensitivity and with the staged sensitivity estimate in the proposal~\cite{PhysRevLett.133.181001,FINESSEProposal2026}. The full High-Flux configuration extends the coupling reach by approximately a further order of magnitude in the statistics-dominated regime.

Ultralight axion dark matter acts as a benchmark for a broader class of spin-dependent new-physics searches which employ precision neutron-spin interferometry. The same basic experimental infrastructure, a polarized neutron beam, a magnetically controlled interaction region, Ramsey RF coils, spin analysis and time-of-flight information, can be adapted to search for other interactions that generate anomalous neutron-spin precession. Examples include matter-sourced spin-dependent fifth forces, Lorentz/CPT-violating spin couplings, axion-induced oscillating neutron electric dipole moments, and selected ultralight or transient dark-sector fields with spin-dependent couplings~\cite{Piegsa2012,Cong2025,Altarev2009,Parnell2020,Chu2013,Abel2017,Schulthess2022,Wang2026,mulkey2025newupperboundsexotic,geerits2024measuringangularmomentumneutron,HADDOCK2018227}. Such searches generally require experiment-specific modifications to the interaction region or analysis strategy.

\subsubsection{Neutron electric charge}

The observation of a finite neutron electric charge would be of fundamental significance, addressing open questions such as electric charge quantisation. The FINESSE concept uses a collimated intense neutron beam, slit patterns and two achromatic mirrors to search for a non-zero neutron charge. A transverse electric field would produce a minute deflection if the neutron carried non-zero charge, resulting in a measurable change in transmitted intensity. Reversal of the field in the double-mirror geometry provides strong suppression of first-order systematic effects.

The present direct limit is
\[
q_n=(-0.4\pm1.1)\times10^{-21}e
\]
~\cite{Siemensen:2018charge}. Earlier full-beamline studies project improvements in direct sensitivity of up to around $700$~\cite{Santoro:2023izd} for a representative $100$-day exposure, with the achievable gain depending on the final neutron-optics configuration and accepted phase space. 

For Short FINESSE, the corresponding staged benchmark is approximately $100$ days at $2\,$MW, for which the proposal estimates an improvement of about a factor of $70$ relative to the present direct sensitivity~\cite{FINESSEProposal2026}.

\subsubsection{Neutron-capture searches for dark-sector particles}

A further High-Flux FINESSE opportunity is neutron-capture $\gamma$ spectroscopy as part of a search for weakly coupled particles emitted in nuclear de-excitation~\cite{Mmcapture,Massarczyk2022M1Axion}. The proposed ``satellite-line comb'' searches for multiple weak $\gamma$ lines shifted by a common energy offset below known capture transitions. Combining information from parent transitions and target nuclei allows nuclear-structure ambiguities and instrumental backgrounds to be constrained while sensitivity can be kept for keV--MeV-scale dark sector particles which couple to nucleons. This is exploratory and promising work. 

\subsubsection{Exotic neutron decays}

The annihilation-detector infrastructure can also be used to search for non-standard neutron decays into weakly interacting or long-lived particles~\cite{Meirose:2025fzq,Fornal:2023review}. Single-photon and partially invisible final states can be difficult to observe in large underground detectors but can be observed in a free-neutron experiment with precise vertex and kinematic reconstruction.

Searches can be performed parasitically during the $n\rightarrow\bar n$ programme. This broadens the discovery reach of High-Flux FINESSE without requiring a separate beamline or a comparably large additional experimental infrastructure.

\subsubsection{Early physics with Short FINESSE}

Short FINESSE represents the staged initial implementation of the High-Flux FINESSE configuration. It uses the upstream section of the high-flux guide together with an approximately 10~m magnetically controlled experimental region, allowing a substantial early physics programme before completion of the full beamline; a detailed description is given in Section~\ref{shortfinesse}. Short FINESSE does not itself provide the low-background beam required for the precision programme described below, but it has an important preparatory role for Low-Background FINESSE. It can be used to commission common beamline infrastructure and to provide direct measurements of the E5 neutron spectrum, pulse timing, and beam-related backgrounds, while also supporting detector and DAQ tests and, where appropriate, R\&D on spin manipulation, polarimetry, and magnetic-field control. Competitive neutron-decay, hadronic-parity-violation, and EDM measurements require the full Low-Background bender, chopper, and polarization infrastructure described in Section~\ref{lowbackgroundbeamline}.

The representative early physics programme for Short FINESSE is summarized in Table~\ref{tab:shortfinesse_science}. Unless otherwise stated, the sensitivities quoted for Short FINESSE assume an ESS linac power of 2~MW. The programme combines competitive early measurements with commissioning and R\&D activities that directly prepare the subsequent full FINESSE facility.

For free neutron--antineutron oscillations, an initial running period of approximately 100 days can be used for R\&D and commissioning of the annihilation detector, magnetic shielding, reconstruction, and the characterization of the E5 beam-background environment. Beyond this initial programme, the neutron-optics studies described in
Section~\ref{sec:shortfinesse_optics} indicate that the uncollimated Short
FINESSE configuration reaches a beam figure of merit of the same order as
that of the historical ILL free-neutron search, supporting its use for
\(n\rightarrow\bar n\) R\&D, commissioning, and validation of the
experimental concept under realistic ESS beam conditions.

Short FINESSE also supports a dedicated sterile-neutron programme using the approximately 10~m magnetically controlled region. For the representative 100-day magnetic-field scan described in Section~\ref{sterile}, the projected sensitivity to the neutron--sterile-neutron oscillation time is approximately \(30\)~s for disappearance and \(20\)~s for regeneration. The same configuration can also be used to search for sterile-neutron-mediated baryon-number violation through \(n\rightarrow n'\rightarrow\bar n\), with a projected sensitivity of approximately
\[
\sqrt{\tau_{nn'}\tau_{n'\bar n}}\simeq 41~\mathrm{s}.
\]

The Ramsey neutron-beam search for ultralight axion-like dark matter is among the most mature early measurements foreseen for Short FINESSE. For a representative exposure of approximately 150 days at 2~MW, the short configuration is expected to retain more than an order-of-magnitude improvement over current laboratory constraints in favourable regions of parameter space, as discussed in Section~\ref{ALP}. 

The same magnetically controlled infrastructure can also support other spin-dependent new-physics searches. The sensitivity of polarized neutron searches for spin dependent effects from $Z'$ vector boson exchange and pseudoscalar boson exchange with matter for light boson masses corresponding to force ranges from millimeters to fermis can be improved by a few orders of magnitude. 

A direct neutron electric-charge measurement can likewise be performed with the short high-flux configuration. For a representative 100-day exposure at 2~MW, statistics-dominated scaling from the staged benchmark corresponds to an improvement of approximately a factor of \(70\) relative to the present direct sensitivity.

Finally, Short FINESSE can support exploratory neutron-capture searches for weakly coupled dark-sector particles. The detailed sensitivity of these measurements depends on the detector configuration and target programme, but the short beamline already provides access to an intense cold-neutron beam and therefore offers a useful early opportunity to investigate this class of searches.

\begin{table}[H]
\centering
\caption{Representative early physics programme for Short FINESSE at an ESS
linac power of 2~MW. The quoted sensitivities correspond to the assumptions
discussed in the relevant physics sections.}
\label{tab:shortfinesse_science}

\begin{tabularx}{\textwidth}{p{2.3cm} p{2.8cm} X}
\toprule
\textbf{Search / channel} &
\textbf{Representative exposure} &
\textbf{Short FINESSE role / reach} \\
\midrule

Free \(n\rightarrow\bar n\)
& \(\sim 100\) days
& R\&D and commissioning of the annihilation detector, magnetic shielding,
event reconstruction, and characterization of the E5 beam-background
environment under realistic ESS beam conditions
(Section~\ref{sec:shortfinesse_optics}). \\

\(n\rightarrow n'\rightarrow n\) regeneration
& \(\sim 100\) days
& Projected sensitivity of approximately
\(\tau_{nn'}\simeq20\)~s for the representative magnetic-field scan described
in Section~\ref{sterile}. \\

\(n\rightarrow n'\rightarrow\bar n\)
& \(\sim 100\) days
& Projected sensitivity of approximately
\(\sqrt{\tau_{nn'}\tau_{n'\bar n}}\simeq41\)~s for the representative
magnetic-field scan described in Section~\ref{sterile}. \\

\(n\rightarrow n'\) disappearance
& \(\sim 100\) days
& Projected sensitivity of approximately
\(\tau_{nn'}\simeq30\)~s for the representative magnetic-field scan described
in Section~\ref{sterile}. \\

Ultralight axion / axion-like dark matter
& \(\sim 150\) days
& More than an order-of-magnitude improvement over current laboratory
constraints in favourable regions of parameter space
(Section~\ref{ALP}). \\

Neutron electric charge
& \(\sim 100\) days
& Approximately a factor \(70\) improvement relative to the present direct
sensitivity. \\

Neutron-capture dark-sector searches
& To be determined
& Exploratory early programme exploiting the intense Short FINESSE cold-neutron
beam; the achievable sensitivity depends on the detector configuration and
target programme. \\

\bottomrule
\end{tabularx}
\end{table}

\subsection{The Low-Background FINESSE scientific programme}

The main precision-measurement opportunities enabled by
Low-Background FINESSE are summarized in
Table~\ref{tab:lowbackground_science}.
In contrast to the largely flux-limited High-Flux programme,
the achievable sensitivity of these measurements depends
critically on background suppression, polarization,
phase-space control, and experiment-specific systematic
uncertainties.

\begin{figure}[t!]
\begin{center}
  \includegraphics[height=0.15\textheight]{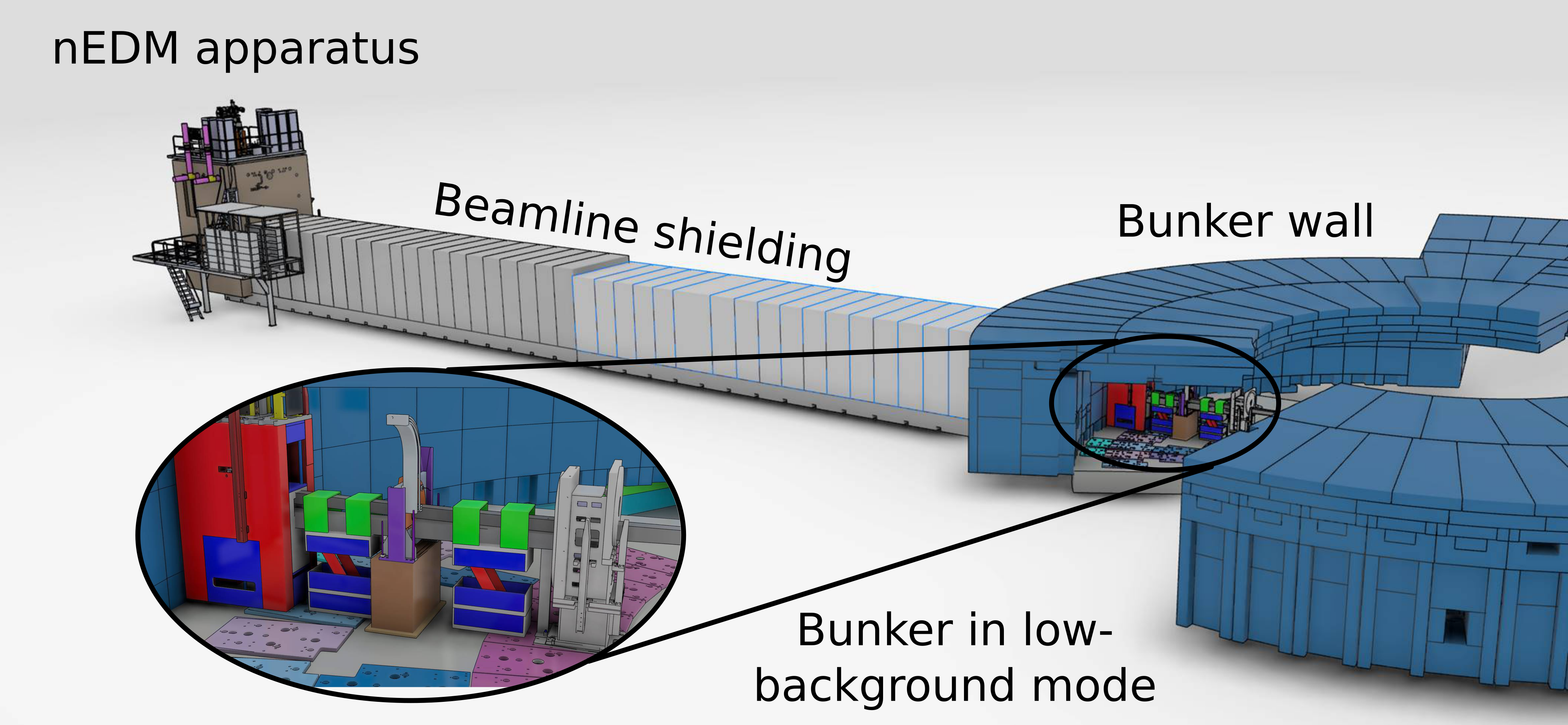}
  \caption{CAD model of the ESS facility illustrating the
  proposed Low-Background FINESSE beamline configuration
  for a cryogenic neutron electric dipole moment (nEDM)
  experiment. The concept builds on the experimental
  infrastructure and developments of the nEDM@SNS project.}
  \label{nEDMlayout}
\end{center}
\end{figure}

A conceptual layout of the Low-Background FINESSE beamline,
configured for a cryogenic neutron electric dipole moment
(nEDM) experiment, is shown in Fig.~\ref{nEDMlayout}.

\begin{table}[H]
\centering
\caption{Representative physics programme for Low-Background FINESSE. The quoted reach summarizes the physics role discussed in the text; ultimate sensitivities depend on the dedicated experimental apparatus and systematic control.}
\label{tab:lowbackground_science}

\footnotesize
\setlength{\tabcolsep}{4pt}
\renewcommand{\arraystretch}{1.08}

\begin{tabularx}{\textwidth}{
>{\raggedright\arraybackslash}p{2.0cm}
>{\raggedright\arraybackslash}p{3.2cm}
>{\raggedright\arraybackslash}X}
\toprule
\textbf{Search / channel} &
\textbf{Physics objective} &
\textbf{Projected FINESSE reach or role} \\
\midrule

Cryogenic nEDM
& Search for new sources of \(CP\) violation that could contribute to explaining the matter--antimatter asymmetry through baryogenesis.
& Flagship long-term application using in-situ UCN production in superfluid helium. A representative sensitivity of order \(10^{-28}\,e\,\mathrm{cm}\) after several years of running requires dedicated cryogenic, high-voltage, magnetic-shielding, and spin-analysis infrastructure. \\

Precision neutron \(\beta\) decay
& Precision tests of charged-current observables, including \(A\), \(a\), \(b\), transverse-polarization correlations, \(V_{ud}\), and non-standard scalar, tensor, right-handed, or \(CP\)-violating interactions.
& PERC-/PERKEO-like measurements benefit from high phase-space density and spatial pulse localization, while BRAND-like measurements provide complementary sensitivity through event-by-event reconstruction and transverse-electron polarimetry. \\

Hadronic parity violation
& Measurement of weak hadronic amplitudes through capture asymmetries and neutron spin rotation.
& Enables next-generation measurements with reduced prompt background, wavelength information, polarization control, and discrimination of velocity-dependent false effects. \\

Beam EDM
& Complementary EDM search based on Ramsey spin precession in a pulsed neutron beam.
& Provides a long interaction region and neutron-velocity information to distinguish EDM phases from velocity-dependent motional-field effects. Primarily an R\&D and longer-term precision opportunity. \\

CRES-based \(\beta\) spectroscopy
& Alternative determination of beta-spectrum observables, including sensitivity to the Fierz term \(b\).
& Longer-term R\&D opportunity with detector systematics qualitatively different from conventional magnetic spectrometers. \\

\bottomrule
\end{tabularx}
\end{table}

\subsubsection{Neutron electric dipole moment}\label{sec:nedm}

A permanent neutron electric dipole moment (nEDM) is a direct signal of time-reversal violation and, assuming CPT invariance, of $CP$ violation. EDM searches are therefore exceptionally sensitive probes of new sources of $CP$ violation, with direct relevance to the origin of the cosmic matter--antimatter asymmetry and to the strong-$CP$ problem~\cite{Chupp2019_EDMReview,Abel2020_PSI_nEDM}. A cryogenic nEDM experiment is a natural flagship long-term application of Low-Background FINESSE, analogous to the role of free $n\rightarrow\bar n$ in the High-Flux FINESSE programme~\cite{Ahmed_2019,Abele:2022iml}. A detailed E5-specific optimization of the Low-Background FINESSE guide for this cryogenic benchmark is presented in Section~\ref{sec:cryoedm_guide}.

The Low-Background FINESSE guide is designed around the stringent beam requirements of this class of experiment. This requires removal of direct line of sight to the source and strong suppression of backgrounds from fast neutrons and prompt radiation, while the instrument must accommodate the large cryogenic, high-voltage, magnetic-shielding and spin-analysis systems required for a world-class measurement. The projected sensitivity for the cryogenic concept is of order
\[
10^{-28}\,e\,\mathrm{cm},
\]
after several years of running~\cite{Ahmed_2019,Abele:2022iml}, corresponding to an improvement of approximately two orders of magnitude over the present experimental limit. Reaching this sensitivity would require a major dedicated experiment and a collaboration and infrastructure programme commensurate with its scale.

Beam-EDM measurements provide a complementary possible approach with qualitatively different systematics from ultracold-neutron EDM experiments. In the pulsed-beam concept, Ramsey spin precession is measured as a function of neutron time of flight: a true EDM phase and important motional-field effects, in particular the relativistic $\mathbf{v}\times\mathbf{E}$ contribution, have different velocity dependences and can therefore be separated experimentally~\cite{Piegsa2013_BeamEDM,Chanel2019_PulsedBeamEDM,Abele:2022iml}. Proof of principle has been demonstrated on a shorter apparatus. The long, pulsed Low-Background FINESSE flight path provides a possible environment for extending the method to substantially longer interaction lengths and higher neutron statistics, while exploiting time-of-flight information to distinguish velocity-dependent false effects.

\subsubsection{Precision neutron beta decay}\label{sec:beta_decay}

Precision neutron $\beta$ decay tests the charged-current weak interaction through an over-constrained set of experimentally distinct observables. PERC is a magnetically guided source of neutron-decay electrons and protons, designed to support precision measurements including the beta asymmetry $A$, the electron--antineutrino correlation $a$, the proton asymmetry $C$, the Fierz interference term $b$, spectral-shape observables and recoil-order effects such as weak magnetism~\cite{PERC,PERCProgress2012,MaerkischCKM2023}. The beta asymmetry gives a particularly precise determination of $\lambda=g_A/g_V$; combined with an independent neutron-lifetime measurement of commensurate precision, this determines $V_{ud}$. The targeted PERC precision can potentially allow a neutron-based $V_{ud}$ determination to be competitive with superallowed $0^+\!\rightarrow0^+$ nuclear $\beta$ decays, while the additional observables provide independent tests of scalar, tensor and right-handed interactions~\cite{MaerkischCKM2023,HardyTowner2020,Falkowski2021Beta}. BRAND-like measurements are complementary: they target transverse-electron-polarization correlations such as $N$ and the time-reversal-odd coefficient $R$, together with conventional decay observables, using electron tracking, proton coincidence and Mott polarimetry~\cite{BRAND,Bodek:2021BrandReview}.


Several of these measurements have important statistical and systematic components. Exploiting the available statistical gain therefore requires corresponding improvement or control of the dominant systematics. These can arise from detector calibration and non-linearity, electron backscattering and energy loss, proton detection efficiency, neutron polarimetry and spin-flip efficiency, magnetic-field non-uniformity, analysing powers, beam-profile and alignment effects, backgrounds, dead time and instrumental asymmetries~\cite{PERC,MARKISCH2009216,MaerkischPSI2022,Roick2019,Saul2020_b}. The pulsed FINESSE beam helps with several of these through spatial pulse localization, wavelength tagging, time-of-flight-dependent consistency checks and timing sidebands, but it does not by itself remove apparatus-limited systematics~\cite{MARKISCH2009216,Soldner_2019,Abele:2022iml}. These measurements form part of a broader and complementary programme together with global $\beta$-decay fits, Nab/pNab and high-$p_T$ collider constraints~\cite{Falkowski2021Beta,Cirigliano2024CLEW,Hayen2024Review,ATLAS2025ccDY,Baessler2024Nab,Broussard2025NabOps}. Different observables probe different combinations of charged-current EFT coefficients and have different experimental and theoretical systematics. The value of Low-Background FINESSE is therefore not simply additional statistics, but also access to complementary observables and to measurements for which pulse localization, wavelength information and time-of-flight-dependent checks can improve systematic control. CRES-based $\beta$ spectroscopy provides a promising longer-term approach with qualitatively different detector systematics, and is presently an R\&D opportunity~\cite{Byron2023CRES,Hayen2024Review}.

\subsubsection{Hadronic weak interaction and parity violation}\label{sec:hadronic_pv}

Hadronic parity violation provides a further strong case for Low-Background FINESSE, with several concrete observables rather than a single class of ``asymmetry''. Examples include the parity-violating photon angular asymmetry $A_\gamma$ in polarized-neutron capture on hydrogen, measured by NPDGamma; the proton--triton asymmetry in $\vec n+{}^3\mathrm{He}\rightarrow p+t$; the corresponding parity-violating observable in neutron--deuteron capture (NDTGamma); and the parity-violating neutron spin-rotation angle per unit length in hydrogen or helium~\cite{Gericke:2011qoo,McCrea:2020n3He,Snow2012PVRotation}. Together these observables constrain complementary amplitudes of the hadronic weak interaction.

The advantage of ESS is particularly clear where its pulsed structure provides neutron-velocity information. Time of flight permits the signal and instrumental asymmetries to be studied as functions of wavelength, provides timing sidebands for background characterization, and can separate genuine parity-violating effects from velocity-dependent false rotations or asymmetries. This is especially valuable for NDTGamma and hydrogen spin rotation. Not every measurement in this class is ESS-exclusive---helium spin rotation, for example, is also being pursued elsewhere---so the strongest FINESSE case is where the combination of intensity, pulse structure and Low-Background FINESSE transport gives demonstrably improved systematic control or access to a new observable.

Event-rate gains of order ten have been projected for several hadronic-parity-violation measurements relative to existing facilities~\cite{Soldner_2019,Abele:2022iml}, potentially translating into substantial improvements in statistical precision. For measurements not already statistically limited, the benefit of the increased rate must be matched by corresponding control of spin transport, polarization, detector gain stability, target-related effects, magnetic-field gradients and beam-correlated backgrounds.

\subsection{Nuclear-physics applications}
\label{sec:nuclear_physics_applications}

FINESSE also proposes to support a complementary nuclear-physics programme developed in the SINE (\textit{Scientific Investigations of Nuclei at ESS}) proposal~\cite{ONeill:SINE2026}. The concept is to exploit the combination of high neutron intensity, selectable low-background operation, and flexible detector integration for neutron-induced reaction studies. An important feature of this programme is the possibility of performing simultaneous high-resolution gamma-ray and electron spectroscopy. Representative measurements include HPGe gamma-ray spectroscopy, potentially with Compton suppression or LaBr$_3$ timing, combined with conversion-electron detection, as well as charged-particle or fission-fragment detection following neutron capture, fission and related reactions. The combined gamma--electron capability is particularly important for studies of nuclear shape and shape coexistence, since many relevant electric-monopole (E0) transitions cannot decay through single-gamma emission and, for transition energies below the pair-production threshold, are observed through conversion electrons. Such measurements can address nuclear level schemes and level densities, electromagnetic transition strengths, shape coexistence, capture cascades and reaction channels, with particular value for small or scarce isotope samples.

For measurements with thermal and cold neutrons, the chopper system can be used to define the incident wavelength range and to provide the pulse control required by specific experiments. Combined gamma--electron spectroscopy can provide complementary information on nuclear level structure and electromagnetic transition properties that is inaccessible through gamma spectroscopy alone. This is relevant to neutron-capture cross sections, nuclear-structure and shape studies, astrophysical $s$-process branching points, nuclear-model and evaluated-data benchmarks, and selected applications in fission, fusion, waste management, safeguards and isotope production. At this stage these should be regarded as complementary user-driven opportunities rather than a separate beamline programme: they exploit the same FINESSE capabilities of high neutron intensity, low-background transport, pulse control and flexible detector integration.

\section{Beamline concept and design requirements}
\begin{figure}[t!]
  \begin{center}
    \includegraphics[width=0.7\textwidth]{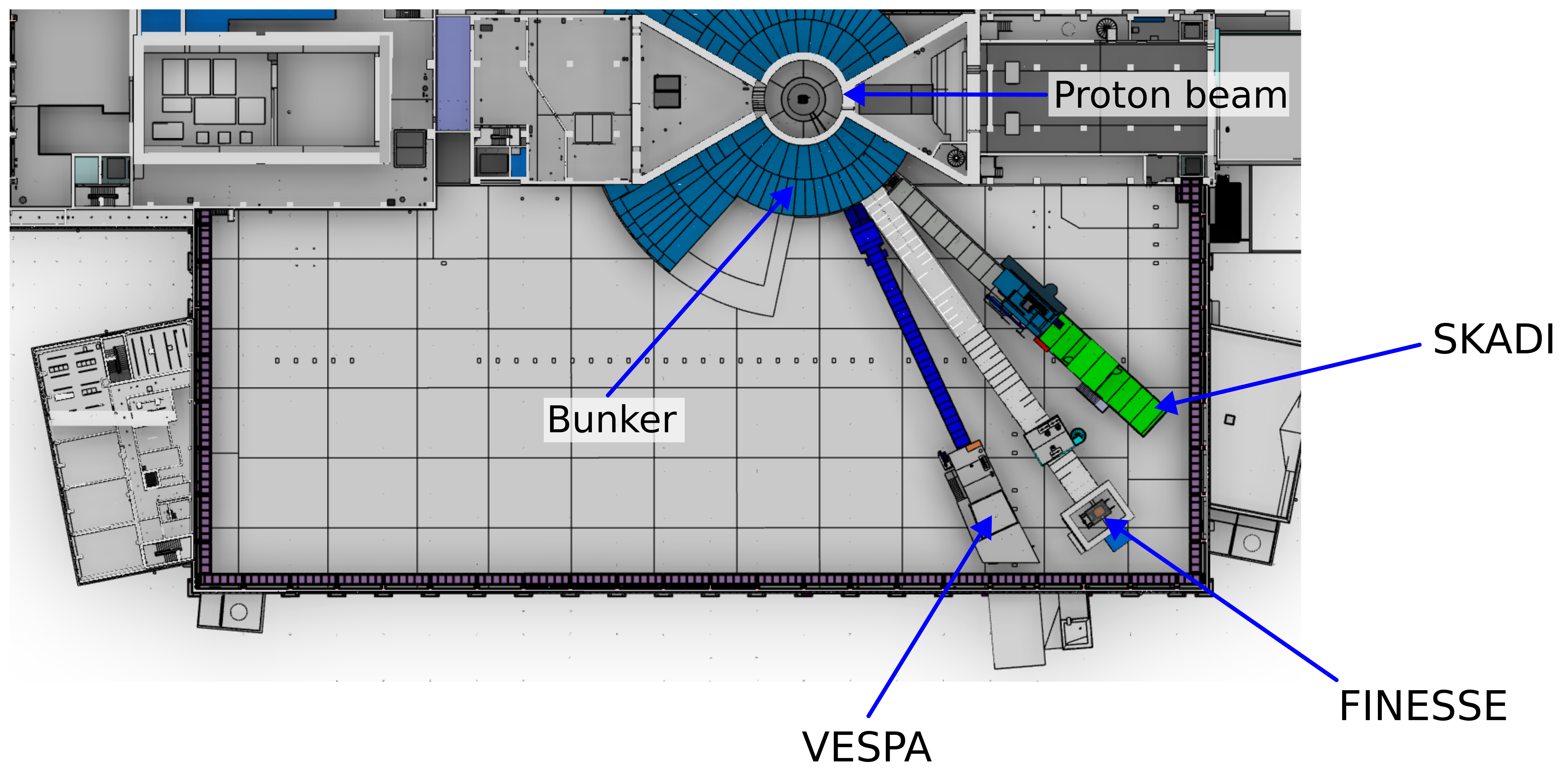}
    \caption{Overview of the beamline in the ESS experimental hall.}
    \label{beamline}
  \end{center}
\end{figure}

The beamline is proposed for construction at the ESS East Sector beam port E5, where a dedicated neutron extraction system, referred to as the monolith insert, has already been installed. A detailed description of this system can be found in Ref.~\cite{Santoro:2023izd}. An overview of the FINESSE instrument layout in the ESS experimental hall is shown in Fig.~\ref{beamline}.

The E5 location provides the space required for the FINESSE beamline and preserves the possibility of a future extension of the experimental area, if needed for large-scale infrastructure such as a cryogenic nEDM experiment. 

The design is driven by the need to support two distinct operating conditions: the \textit{high-flux mode} and the \textit{low-background mode}.

The central design principle is therefore a dual-guide beamline in which a short guide section inside the ESS bunker is exchanged to select one of two complete downstream optical configurations. This approach allows the same beam port and common infrastructure to support both operating conditions: the high-flux and low-background configurations. A detailed description of the exchangeable-guide concept is given in Section~\ref{exchguide}.

\section{High-flux beamline configuration}
\label{highfluxguide}

The high-flux guide configuration, including optics simulations, shielding studies, magnetic-field simulations, and detector performance studies, is described in detail in Ref.~\cite{Santoro:2023izd}. Here we summarize the main beamline characteristics relevant to the dual-guide concept.

\begin{figure}[t!]
\begin{center}
  \includegraphics[height=0.25\textheight]{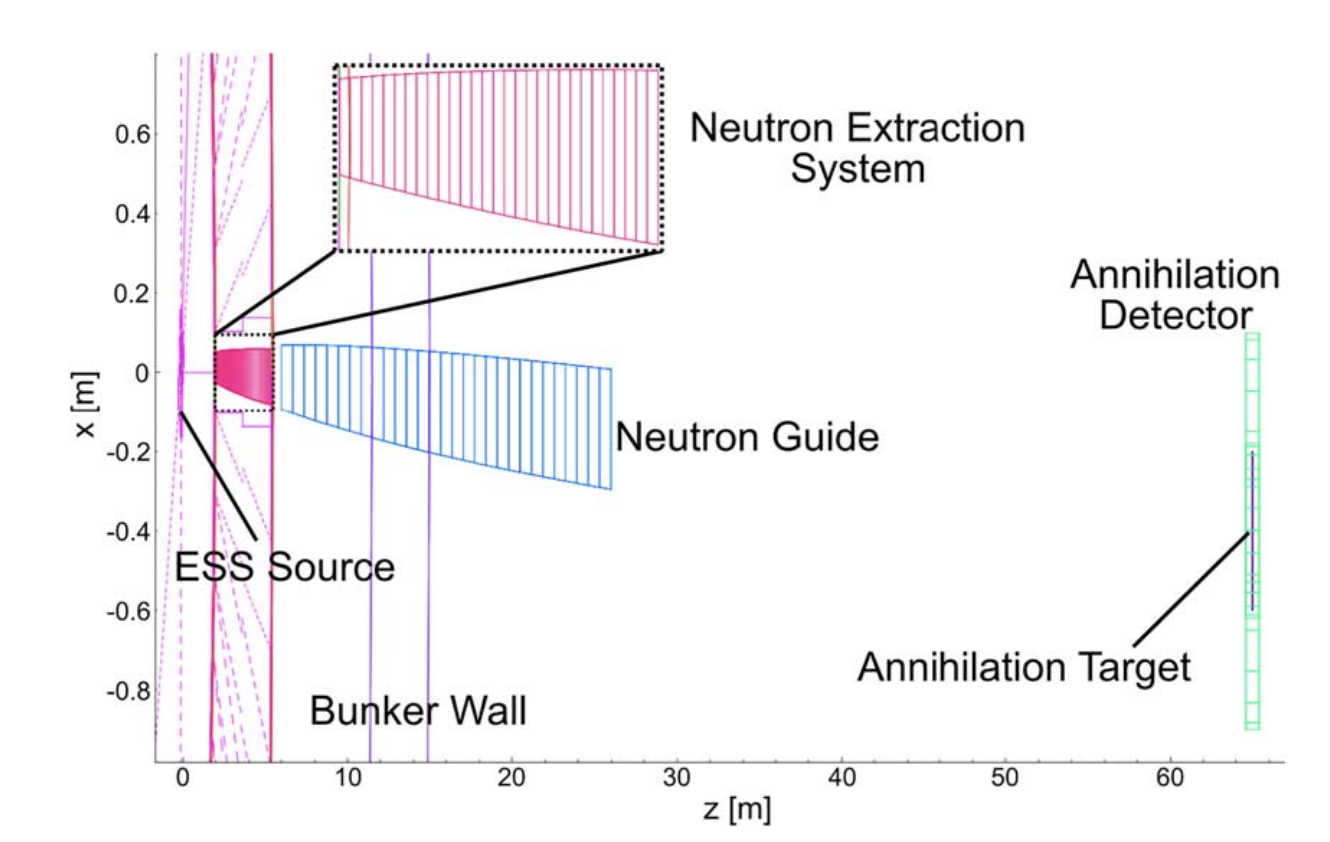}%
\caption{Overview of the guide geometry for the high-flux configuration.}
   \label{highfluxbeamline}
  \end{center}
\end{figure}

With this configuration, a neutron flux of $1.0 \times 10^{12}$~n\,s$^{-1}$ can be delivered at 75~m in the experimental cave, with an average wavelength of 3.8~\AA. The guide geometry is shown in Fig.~\ref{highfluxbeamline}. The current design consists of a 20~m long neutron guide whose reflector surfaces follow the quadratic form
\[
a_{xx} x^2 + a_{yy} y^2 + a_{xy}xy + a_x x + a_y y - 1 = 0.
\]
The geometry was optimized using \textsc{McStas}, together with the \texttt{guide\_bot}~\cite{Bertelsen2017GuideBot} framework. The optimization resulted in opening dimensions of $16.4 \times 14.4~\mathrm{cm^2}$ at the guide entrance and $30.5 \times 23.0~\mathrm{cm^2}$ at the guide exit.

Importantly, the high-flux configuration has been studied within a complete E5-specific simulation chain. The neutron extraction system, beamline optics, bunker region, shielding, experimental cave, and beam stop were modeled consistently within an integrated framework combining \textsc{McStas} and CombLayer/PHITS calculations~\cite{Santoro:2023izd,Holl2022_BeamlineSimulation}. This demonstrates that the concept has been developed as a realistic instrument layout, including radiological shielding performance and ESS floor-space constraints.

\section{Low-background beamline configuration}
\label{lowbackgroundbeamline}

\begin{figure}[t!]
\begin{center}
  \includegraphics[height=0.25\textheight]{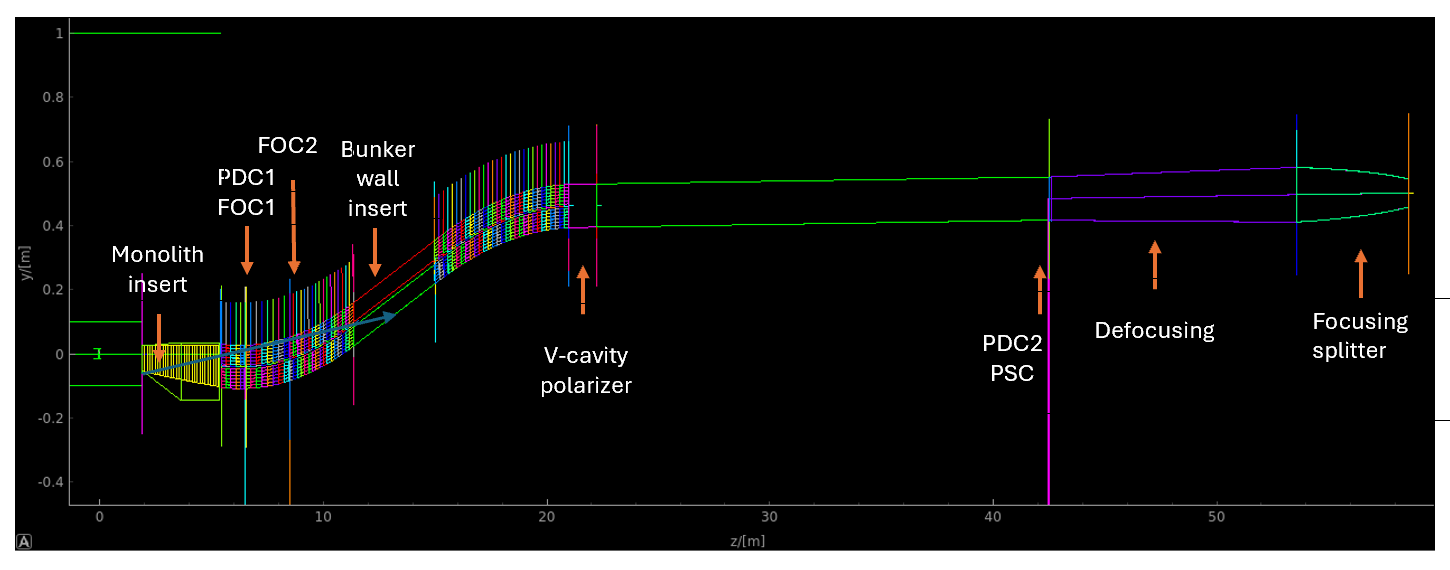}%
\caption{Overview of the guide geometry for the low-background configuration.}
   \label{lowfluxbeamline}
  \end{center}
\end{figure}

The Low-Background FINESSE configuration is designed to provide a clean and well-defined cold-neutron beam while satisfying the geometric and shielding constraints of the E5 beam port. The design is based on a vertically deflected double-bender system with an absorbing background scraper placed between two vertically stacked guide channels. This arrangement removes the direct line of sight to the moderator before the bunker wall while preserving a large fraction of the useful cold-neutron phase space. The overall beamline layout is shown in Fig.~\ref{lowfluxbeamline}.

\begin{figure}[t!]
\begin{center}
  \includegraphics[height=0.15\textheight]{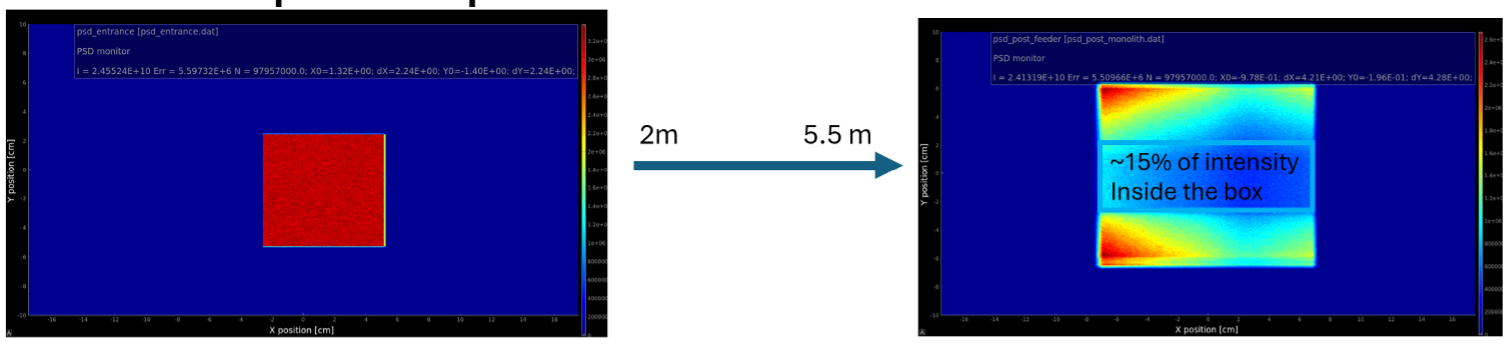}%
\caption{Beam-profile evolution between approximately 2 and 5.5~m from the moderator for the E5 beamport.}
   \label{beamprofile}
  \end{center}
\end{figure}

\begin{figure}[t!]
\begin{center}
  \includegraphics[height=0.17\textheight]{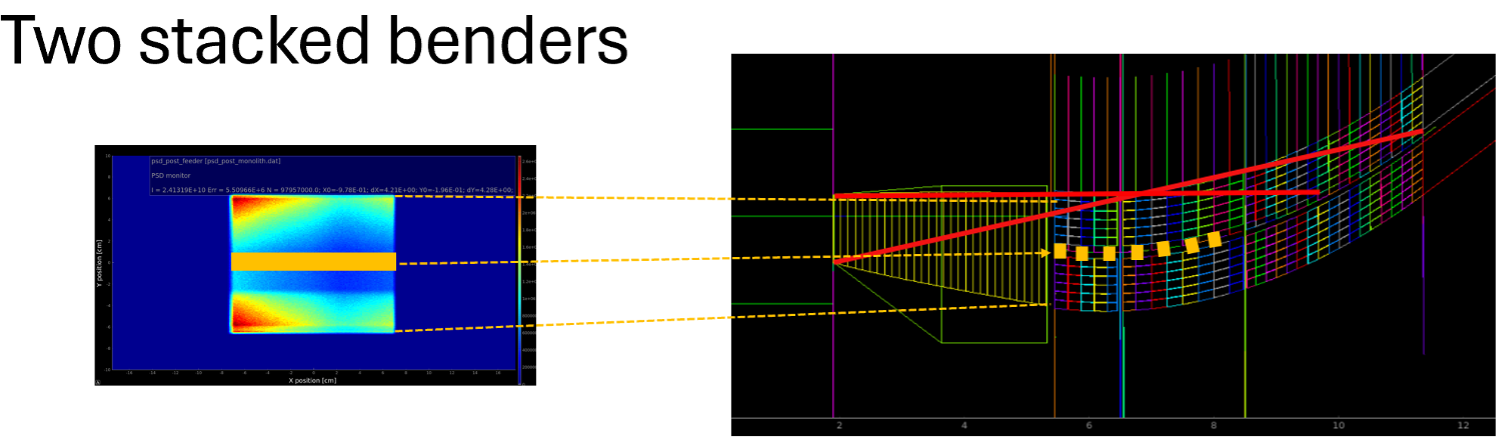}%
\caption{Background-scraper concept based on two vertically stacked benders.}
   \label{stackedbenders}
  \end{center}
\end{figure}

A key feature of the design is that it exploits the non-uniform transverse neutron distribution emerging from the E5 monolith extraction optics. The central part of the beam profile is naturally depleted, as shown in Fig.~\ref{beamprofile}, allowing an absorbing plate to be inserted between the upper and lower guide channels with only a moderate loss of useful neutron intensity, as illustrated in Fig.~\ref{stackedbenders}. The first bender deflects the beam upwards and is followed by a straight bunker-wall feedthrough. A second bender downstream of the bunker returns the beam towards the horizontal direction, providing an additional suppression of direct line-of-sight background.

The present baseline configuration uses a 1.5~cm central background scraper, with a first bender approximately 5.9~m long, a 3.5~m bunker-wall feedthrough, and a second bender approximately 6~m long. The first bender provides an upward deflection of approximately \(3.48^\circ\). The scraper thickness represents an important optimization parameter, and configurations with a 4.5~cm absorber have also been investigated.

Because the neutron intensity is already reduced in the central region of the extracted phase space, increasing the scraper thickness produces a considerably smaller loss of useful flux than would be expected from the corresponding reduction of the geometric guide opening. In the simulations, blocking approximately one third of the vertical opening reduces the total transmitted intensity by only about 15\%, while substantially increasing the amount of absorbing material intersecting trajectories associated with the original moderator line of sight.

The downstream bender can additionally be configured so that the upper and lower guide channels progressively merge towards its exit. The downstream guide height can thereby be reduced from approximately 13.5~cm to about 9~cm. Although this reduces the total transmitted intensity, it can increase the neutron flux density, which is advantageous for experiments with a limited transverse acceptance or decay volume. In such cases the relevant figure of merit can depend more strongly on neutron flux per unit area than on the total integrated beam intensity.

A second family of guide geometries with stronger curvature has also been studied in order to accommodate simultaneously the mechanical interfaces required for the high-flux and low-background guide options in the E5 bunker region. These configurations use a first-bender deflection of approximately \(4.48^\circ\) and provide an additional design option at the cost of some neutron transmission.

Preliminary discussions with neutron-guide manufacturers indicate that the stacked-bender concept is technically feasible. The principal engineering challenge is associated with the downstream merging region, where the separation between the upper and lower channels becomes small. Since the direct line of sight has already been removed upstream, the central separator does not need to extend continuously to the merging point. This provides additional freedom in the mechanical implementation, for example through the use of thin neutron-optical elements and, close to the merge, a common wall shared by neighbouring channels. No fundamental manufacturing obstacle has so far been identified, although the detailed geometry and tolerances will require further engineering optimization.

The Low-Background FINESSE guide should therefore be regarded as a well-defined baseline optical concept with a controlled set of optimization parameters rather than as a single immutable geometry. Variations in background-scraper thickness, bender curvature, guide cross section, and degree of channel merging allow the beam properties to be adapted to different experimental figures of merit. The baseline geometry provides the reference configuration for engineering development, while the alternative solutions quantify the available trade-offs between total transmitted flux, flux density, and background suppression.

For experiments requiring polarized neutrons, the second bender can be followed by a vertical V-cavity polarizer. The rejected spin state can either be removed progressively through losses at downstream guide surfaces or suppressed more rapidly using an absorbing structure associated with the polarizer. The same low-background transport system can therefore provide either an unpolarized or polarized cold-neutron beam depending on the experimental requirements.

The cryogenic neutron electric-dipole-moment experiment provides a particularly demanding benchmark for the Low-Background FINESSE concept. It requires efficient transport of neutrons near 8.9~\AA, suppression of prompt and line-of-sight backgrounds, neutron polarization, wavelength selection, and efficient delivery of the beam to the superfluid-helium UCN converter. The optimization described below demonstrates that these requirements can be met while retaining considerable flexibility for the broader FINESSE precision-neutron programme.

\subsection{Cryogenic nEDM guide optimization}
\label{sec:cryoedm_guide}

A dedicated \textsc{McStas} study was performed to optimize the Low-Background FINESSE configuration for a possible cryogenic nEDM experiment at E5, based on the nEDM@SNS concept~\cite{Ahmed_2019}.
 The experiment uses a polarized cold-neutron beam incident on two liquid-helium cells, in which cold neutrons are converted into ultracold neutrons.

The UCN production probability is strongly wavelength dependent and reaches its maximum close to 8.9~\AA, corresponding to a neutron velocity of approximately 444~m/s~\cite{KOROBKINA2002462}.

Beyond the common E5 monolith extraction optics, the nEDM configuration consists of the first vertically stacked bender, the bunker-wall feedthrough, the second bender, a V-cavity polarizer, a straight transport section, and a downstream defocusing and focusing splitter. The splitter distributes the neutron beam between two liquid-helium cells, each approximately 40~cm long with a transverse cross section of \(7.5\times10\)~cm$^2$.

The downstream splitter optics were optimized using the \texttt{guide\_bot} framework. The focusing surfaces were represented by asymmetric tapered sections described by second-order curves, and the geometric parameters were sampled using a particle-swarm optimization procedure. Defocusing lengths between approximately 11 and 18~m and focusing lengths between approximately 5 and 8~m were investigated. The best performance was obtained with a defocusing section approximately 15~m long followed by focusing sections approximately 6~m long. The same basic focusing solution remains effective for both the larger-aperture baseline guide and the reduced-height configuration in which the two bender channels merge towards a 9~cm-high downstream guide.

For the cryogenic nEDM benchmark, the relevant figure of merit is the expected UCN production rate in the liquid-helium cells. The calculation therefore accounts for the full wavelength dependence of the UCN conversion process rather than simply counting neutrons within a narrow wavelength interval around 8.9~\AA.

The wavelength dependence of the UCN-production probability was obtained
from the calculated superfluid-helium production spectrum reported in
Ref.~\cite{KOROBKINA2002462}, which is based on the measured dynamic structure
factor. The published spectrum was converted into a relative UCN-production
weighting function as a function of incident-neutron wavelength and implemented
in the McStas analysis. An absolute normalization was subsequently applied
using the UCN-production coefficient adopted for the nEDM converter geometry
in Ref.~\cite{Ahmed_2019}.

For a slowly varying incident spectrum, the UCN production density can be written in terms of the spectral neutron flux at 8.9~\AA{} as~\cite{Ahmed_2019} 

\begin{equation}
R_{\rm UCN}
\simeq
2.2\times10^{-8}
\left.
\frac{d\phi}{d\lambda}
\right|_{8.9\,\AA}
\qquad
{\rm UCN\,cm^{-3}\,s^{-1}},
\end{equation}

where \(d\phi/d\lambda\) is expressed in
\(\mathrm{n\,s^{-1}\,cm^{-2}\,\AA^{-1}}\).
For the FINESSE nEDM geometry, the two converter cells have a total volume of approximately \(6000~\mathrm{cm^3}\).

Using the normalization adopted from Ref.~\cite{Ahmed_2019} together with the
wavelength dependence of the conversion process gives an estimated maximum
conversion probability of approximately $1.26\times10^{-6}$ UCN per incident
neutron near 8.9~\AA. The wavelength-dependent conversion function was implemented directly in the \textsc{McStas} monitors, allowing the UCN production rate to be calculated from the full neutron spectrum transported to the cells rather than from the 8.9~\AA{} flux alone.

With the choppers parked open, the baseline neutron-optics configuration gives a predicted production rate of approximately \(4.9\times10^3\) UCN/s summed over the two cells. Several combinations of bender curvature, background-scraper thickness, and downstream guide height were subsequently investigated. The resulting production rates are summarized in Table~\ref{tab:lowbackground_ucn}. This quantity represents the estimated in-situ UCN production rate and does not include subsequent UCN storage, wall losses, upscattering, extraction,
or detection efficiencies. The present calculation should therefore be regarded as an optics-level estimate based on the neutron spectrum delivered to the converter cells rather than as a prediction of the number of stored or detected UCN.

\begin{table}[H]
\centering
\caption{Comparison of Low-Background FINESSE guide configurations for the cryogenic nEDM benchmark at an ESS proton-beam power of 5~MW. The UCN production rates are calculated by applying the wavelength-dependent superfluid-helium UCN conversion probability to the neutron spectrum reaching the two converter cells.}
\label{tab:lowbackground_ucn}

\begin{tabular}{cccc}
\toprule
\textbf{Bend angle} &
\textbf{Scraper thickness} &
\textbf{Guide height} &
\textbf{UCN production rate [s$^{-1}$]} \\
\midrule
$3.48^\circ$ & 1.5~cm & 12~cm & 4900 \\
$3.48^\circ$ & 4.5~cm & 9~cm & 4580 \\
$4.48^\circ$ & 1.5~cm & 12~cm & 4430 \\
$4.48^\circ$ & 4.5~cm & 9~cm & 4210 \\
$2.65^\circ$ & 4.5~cm & 9~cm & 4590 \\
\bottomrule
\end{tabular}
\end{table}

The comparison shows that substantial changes in the guide geometry can be made without strongly degrading the predicted UCN production rate. Increasing the background-scraper thickness from 1.5 to 4.5~cm while reducing the downstream guide height from 12 to 9~cm decreases the calculated UCN yield from approximately \(4.9\times10^3\) to \(4.6\times10^3\)~s$^{-1}$, a reduction of only about 7\%. At the same time, the smaller guide cross section increases the neutron flux density and the thicker scraper provides additional shielding material in trajectories associated with the original source line of sight.

The low-curvature configuration provides a particularly interesting alternative. With a \(2.65^\circ\) bend, a 4.5~cm background scraper, and a 9~cm-high downstream guide, the predicted UCN yield remains approximately \(4.6\times10^3\)~s$^{-1}$. The weaker curvature improves transport of shorter-wavelength neutrons, making this configuration potentially attractive for experiments whose figure of merit extends over a broader cold-neutron wavelength range.

The wavelength distribution reaching the liquid-helium cells can be further controlled using a cascade of pulse-definition and frame-overlap choppers. An initial optimization based solely on maximizing the fraction of neutrons between 8.8 and 9.0~\AA{} was found to be unnecessarily restrictive because significant UCN conversion also occurs outside this narrow interval. The chopper optimization was therefore repeated using the full wavelength-dependent UCN conversion function.

For each sampled set of chopper phases and opening times, both the neutron transmission and the corresponding UCN production rate were calculated. This makes it possible to optimize directly the trade-off between UCN production and the total neutron flux reaching the cells, the latter providing a proxy for beam-related background. The studies show that most of the gain obtained by moving the second pulse-definition chopper downstream is achieved by about 42~m; increasing the distance further towards 60~m provides only a comparatively modest additional improvement when the full UCN conversion spectrum is taken into account.

Representative optimized configurations retain approximately 75--80\% of the maximum open-chopper UCN production while significantly reducing the total neutron population reaching the converter region. 

This demonstrates that the FINESSE timing system can be used not only for
wavelength definition but also to optimize the balance between predicted UCN
production and the transmitted neutron flux, the latter being used here as a
proxy for beam-related background.

\subsection{Low-background guides for other experiments}
\label{otherexp}

Although the cryogenic nEDM case provides a demanding benchmark for the optical design, the Low-Background FINESSE configuration is intended as a general-purpose beamline for precision neutron measurements. Different experiments place different requirements on the neutron beam and therefore require different figures of merit. Experiments using a spatially restricted neutron-decay volume benefit primarily from high neutron flux density, whereas capture experiments may depend on the integrated capture-weighted flux and cryogenic nEDM depends on the wavelength-weighted UCN production rate.

When realistic in-monolith extraction and E5 boundary conditions are imposed, the FINESSE low-background configuration reaches capture-flux densities of order \(10^{10}\)~n\,cm\(^{-2}\)\,s\(^{-1}\). Here the capture flux is defined as the neutron flux weighted by the ratio of the neutron capture cross section at a given energy to the capture cross section at thermal energy, 25~meV. The unpolarized values refer to the beam downstream of the second bender. For polarized experiments the appropriate figure of merit must instead be evaluated downstream of the V-cavity polarizer and any subsequent polarization-clean-up section.

The different guide configurations can therefore be compared using experiment-specific figures of merit. Representative examples from the present optimization are summarized in Table~\ref{tab:lowbackground_fom}. The table is not intended to define a single universally optimal FINESSE geometry, but rather to illustrate that the preferred guide configuration depends on the experimental observable.

\begin{table}[H]
\centering
\caption{Representative figures of merit for experiments considered for Low-Background FINESSE. The reference benchmarks correspond to existing or proposed experimental facilities, while the FINESSE values illustrate the dependence of performance on the low-background guide configuration. Different experiments require different figures of merit, and the values should therefore not be interpreted as defining a single universally optimal guide geometry. The nEDM values use the updated wavelength-dependent UCN-production calculation discussed in Sec.~\ref{sec:cryoedm_guide} and a beam power of 5 MW.}
\label{tab:lowbackground_fom}

\small
\setlength{\tabcolsep}{4pt}

\begin{tabularx}{\textwidth}{l p{2.6cm} c c c c c}
\toprule
\textbf{Experiment} &
\textbf{FOM} &
\textbf{Reference } &
\multicolumn{2}{c}{\textbf{Baseline}} &
\multicolumn{2}{c}{\textbf{High curvature}} \\
\cmidrule(lr){4-5}
\cmidrule(lr){6-7}

&
&
\textbf{benchmark}  & 
\textbf{1.5 cm} &
\textbf{4.5 cm} &
\textbf{1.5 cm} &
\textbf{4.5 cm} \\
\midrule

aSPECT &
Capture-flux density &
$2.0\times10^{10}$~\cite{Baessler2008aSPECT} &
$8.5\times10^{9}$ &
$9.0\times10^{9}$ &
$7.2\times10^{9}$ &
$7.7\times10^{9}$ \\

NPD$\gamma$ &
Polarized flux &
$1.2\times10^{11}$~\cite{MUSGRAVE201819} &
$2.7\times10^{11}$ &
$2.2\times10^{11}$ &
$2.18\times10^{11}$ &
$1.8\times10^{11}$ \\

nEDM &
UCN production rate &
$1.8\times10^{3}$~\cite{Ahmed_2019} &
4900 &
4580 &
4430 &
4210 \\

\bottomrule
\end{tabularx}
\end{table}

The comparison illustrates why no single geometrical optimization criterion is appropriate for the full low-background programme. For aSPECT-type measurements, reducing the guide height can increase the useful neutron flux density despite a reduction in total transmitted intensity. For NPD$\gamma$-type capture measurements, the relevant quantity is instead the polarized neutron flux delivered to the target over the useful wavelength interval. For nEDM, neither of these quantities alone is sufficient, because the neutron spectrum must additionally be weighted by the strongly wavelength-dependent probability for UCN production in superfluid helium.

Neutron beta decay provides a broad set of correlation observables that probe complementary aspects of the charged-current weak interaction. Measurements of the electron--antineutrino correlation \(a\), the beta asymmetry \(A\), the Fierz interference term \(b\), and spin-dependent or transverse-polarization observables provide sensitivity to the axial-vector coupling ratio \(\lambda=g_A/g_V\), CKM unitarity, and possible scalar, tensor, right-handed, or CP-violating interactions beyond the Standard Model. These observables require different experimental strategies and therefore different beamline figures of merit.

PERC~\cite{PERC}- and PERKEO~\cite{MARKISCH2009216}-type spectrometers are high-magnetic-field instruments optimized for high-statistics measurements of neutron-decay asymmetries and related correlation coefficients. They profit from the spatial localization of the neutron pulse, since decays can be selected only when the neutron pulse is fully contained within the well-defined response region of the spectrometer. For such measurements, the useful decay rate in a restricted fiducial volume is closely related to the neutron flux density rather than to the total integrated flux alone. Reduced-cross-section FINESSE guide configurations may therefore be advantageous for this class of measurement.

Nab, which is being operated at the Fundamental Neutron Physics Beamline at the Spallation Neutron Source, follows a complementary approach. It reconstructs the decay kinematics from the measured electron energy and proton time of flight in order to determine the electron--antineutrino correlation \(a\). The polarized follow-up configuration, pNab, would use the same apparatus with a polarized neutron beam to access the beta asymmetry and other spin-dependent correlations. Nab therefore provides an important reference example of a kinematic-reconstruction strategy, distinct from the high-field integrating approach used by PERC and PERKEO.

BRAND~\cite{BRAND} follows a further complementary strategy. It is designed to measure, in a single apparatus, a broad set of neutron beta-decay correlation coefficients, with particular emphasis on observables involving the transverse polarization of the emitted electron. Its detector concept is based on event-by-event reconstruction through electron tracking, electron energy measurement, proton detection in coincidence, and Mott polarimetry for the transverse electron spin. For BRAND, the relevant figure of merit is not simply the total neutron flux or the capture flux per unit area, but rather the rate of reconstructable neutron-decay events in the fiducial volume, together with the coincidence efficiency, polarization, analysing power, and achievable signal-to-background ratio.

The ultimate sensitivity of neutron beta-decay measurements at FINESSE will not be determined by beam intensity alone. Many of the dominant uncertainties are apparatus-limited, including detector calibration and linearity, backscattering, magnetic-field control, polarimetry, analysing powers, dead time, and instrumental asymmetries. The role of FINESSE is therefore not simply to provide a higher-intensity continuation of existing cold-neutron beamlines, but to offer a low-background, polarized, pulsed environment in which selected measurements can benefit from improved wavelength definition, pulse structure, phase-space control, and background rejection.

The same infrastructure can also support hadronic parity-violation measurements, spin-rotation experiments, beam-EDM concepts, and other experiments requiring wavelength information, pulse shaping, polarization capability, and stable low-background beam delivery. FINESSE therefore provides a flexible platform in which the guide configuration and beam conditioning can be matched to the experiment-specific figure of merit while retaining a common low-background transport concept.

\section{Exchangeable guide system}
\label{exchguide}

\begin{figure*}[t!]
\centering
\subcaptionbox{Bunker-region layout for the high-flux configuration. In this mode, the exchangeable guide section preserves the large phase-space acceptance required for flux-limited experiments.\label{fig:highflux_bunker}}{%
  \includegraphics[width=0.48\textwidth]{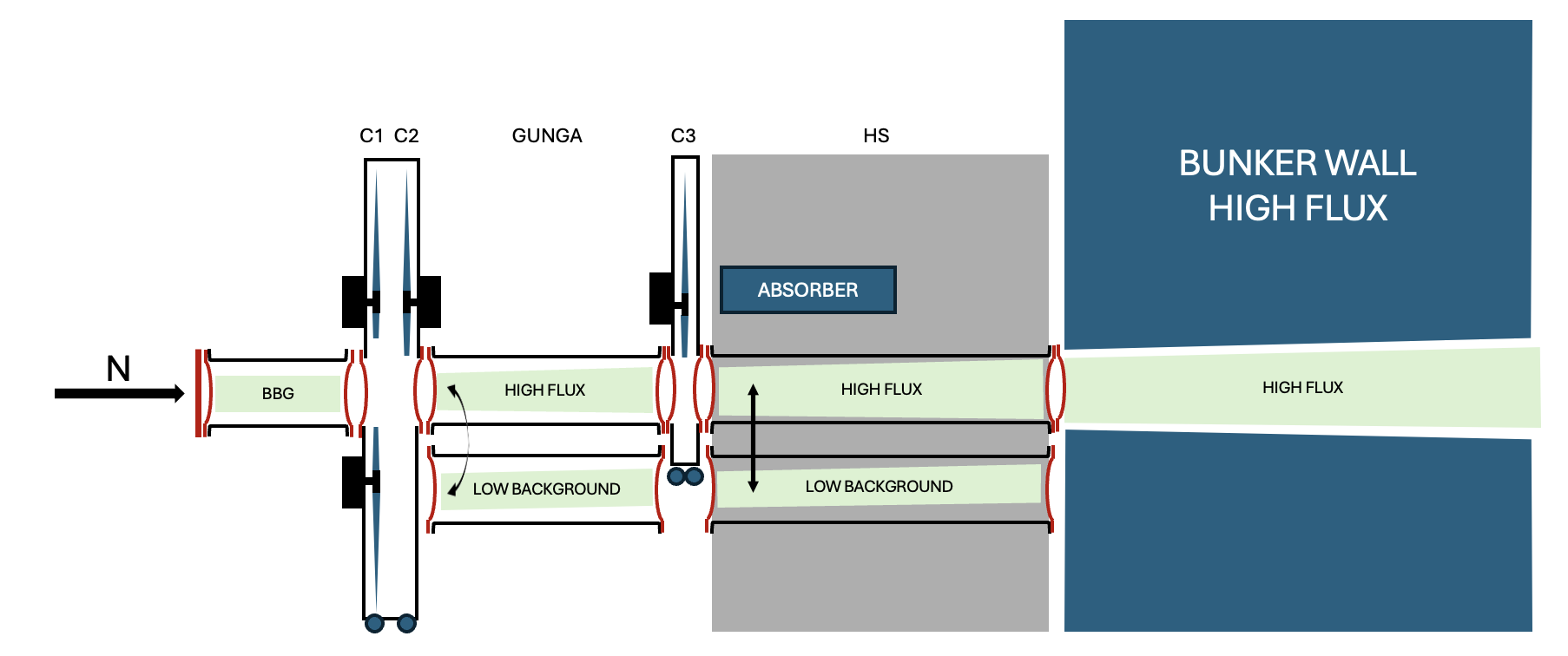}%
}
\hfill
\subcaptionbox{Bunker-region layout for the low-background configuration. In this mode, the exchangeable guide follows the curved path required to remove the direct line of sight to the moderator.\label{fig:lowbackground_bunker}}{%
  \includegraphics[width=0.48\textwidth]{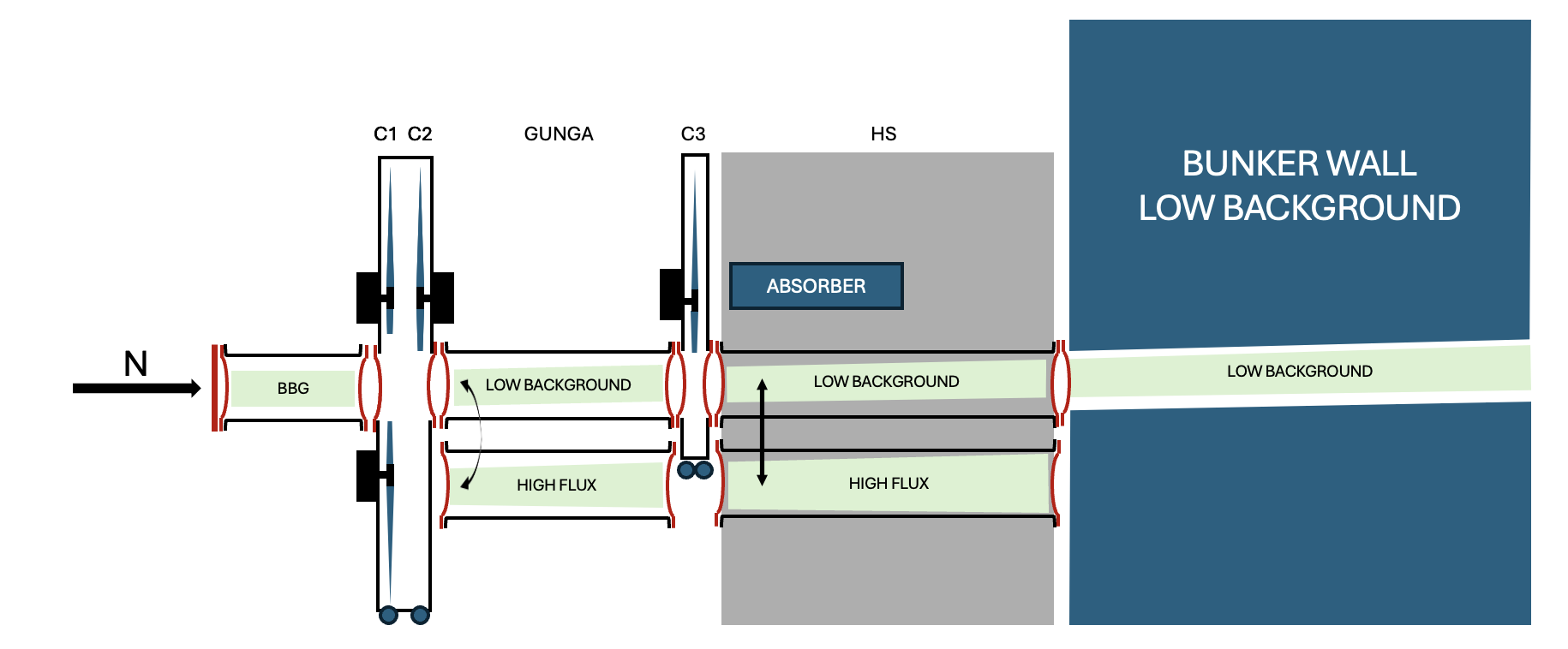}
}
\caption{Comparison of the bunker-region guide layouts for the two FINESSE operating modes. The high-flux configuration preserves maximum neutron transport, while the low-background configuration uses a curved guide path and shielding elements to suppress line-of-sight background. The movable exchangeable guide section labelled \textit{gunga} provides the transition between the two optical configurations inside the bunker. The region labelled HS corresponds to the heavy-shutter section, where a second exchangeable guide element is installed. At the bunker wall, the two configurations require different feedthrough openings, with only the opening corresponding to the selected configuration being present at a given time.}
\label{fig:guidebunker}
\end{figure*}
To operate in the two configurations described above, the FINESSE beamline must accommodate two alternative neutron guide paths: one for the high-flux configuration and one for the low-background configuration. These two bunker-region layouts are shown in Fig.~\ref{fig:guidebunker}. The selection between them is achieved through an exchangeable guide system located inside the ESS bunker, between the exit of the monolith and the bunker wall. Since this is among the least accessible parts of the beamline, the exchangeable system is designed to be robust, mechanically simple, and operated only during scheduled intervention periods.

In the current layout, the guide section inside the bunker starts approximately 5.5~m downstream of the source, just after the light shutter at the exit of the monolith. The light shutter is equipped with a beam bridge guide (BBG), which provides neutron transport through the shutter when it is in the open position. Downstream of the BBG, the first two choppers, a pulse-definition chopper (PDC1) and a frame-overlap chopper (FOC1), are located. The first exchangeable guide section starts downstream of these two choppers and extends up to the second frame-overlap chopper (FOC2), with a length of approximately 1.78~m. This guide section is foreseen to move horizontally between the two optical configurations. This choice is driven  by the need to avoid interference with the chopper drive motors. Although alternative arrangements, including relocation of the motors, remain possible, the horizontal-motion concept follows an established arrangement commonly used at ESS and is therefore adopted as the baseline solution. In the engineering layout shown in Fig.~\ref{fig:guidebunker}, this movable element is labelled \textit{gunga}, the Swedish word for ``swing''.

A second exchangeable guide section is located in the heavy-shutter region, downstream of FOC2 and just before the bunker wall. This section has an exchangeable guide length of approximately 2.32~m. The total exchangeable guide length is therefore limited to a few metres. This is a key advantage of the concept, since it allows the two optical configurations to share the same beam port, fixed chopper positions, and upstream extraction system, while limiting the exchange operation to short guide sections inside the bunker.

At the bunker wall, a dedicated insert or shielding block is required to accommodate the selected beam path. Two implementation options are under consideration. The first option is to use two dedicated bunker-wall blocks, one optimized for the high-flux configuration and one optimized for the low-background configuration. The second option is to develop a custom bunker-wall block with two openings, allowing both guide paths to be accommodated within a single shielding element. The final choice will depend on engineering integration, shielding performance, activation considerations, and the practical requirements for maintenance and exchange operations.

The baseline implementation therefore foresees two exchangeable guide sections inside the bunker: the horizontally movable \textit{gunga} section between the first chopper position and FOC2, and a second exchangeable section in the heavy-shutter region. This baseline concept already provides a concrete mechanical solution for switching between the two guide configurations. Alternative implementations remain possible and will be assessed during the final engineering design phase. These include translation stages with rails, elevator platforms, manual systems with kinematic supports and guide rods, or shutter-integrated guide assemblies. The final choice will depend on alignment reproducibility, shielding integration, vacuum interfaces, activation constraints, maintainability, and compatibility with ESS bunker access procedures.

Switching the operational mode involves exchanging only a few metres of neutron-guide elements inside the bunker and adapting the downstream end station to the measurements to be performed. Since the expected switching frequency is low, on the timescale of several months to years, neither rapid nor automated switching is required. The heavy-shutter section will require a drive system, while the movable \textit{gunga} section could be either motor-driven or manually actuated during scheduled intervention periods. The concept relies on established shutter and movable-guide technologies, with the main engineering requirements being reproducible alignment, shielding integration, vacuum compatibility, and reliable operation in the bunker environment.

The use of thin vacuum windows is also being considered. In particular, reducing the window thickness to approximately 0.5~mm would reduce neutron losses and improve transmission, especially for the wavelength ranges relevant to the low-background and nEDM configurations. The final window design will be optimized together with the vacuum layout, mechanical stability requirements, radiation environment, and neutron-transmission performance.

The chopper system is integrated into the common beamline infrastructure and provides the timing flexibility required by the different experimental modes. As currently foreseen, the choppers remain in fixed positions when switching between the high-flux and low-background guide configurations; only the guide elements are exchanged. This allows the chopper system to rely on established fixed-position engineering designs.

PDC1 and FOC1 are foreseen at approximately 6.7~m from the source, while FOC2 is located at approximately 8.5~m. Further downstream, a second pulse-definition chopper (PDC2) can be located at either 26~m or 42~m, together with a pulse-suppression chopper (PSC). For advanced pulse shaping, the PDC2--PSC system may be replaced by a pulse-multiplication chopper (PMC), a multislit chopper that divides the ESS long pulse into controlled sub-pulses.

The pulse-shaping chopper can be implemented as a double-disk system with phase-adjustable openings. The two disks may either share the same axis or be mounted on separate axes, since both solutions can provide equivalent opening control at the guide position. The final mechanical implementation will be defined during the engineering design phase.

The choppers can be parked open for high-flux experiments that do not require pulse shaping. In low-background operation, they provide wavelength selection, frame-overlap suppression, pulse cleaning, and time-of-flight information for background rejection and control of wavelength-dependent systematic effects.

The main parameters of the currently foreseen FINESSE chopper configuration are summarized in Table~\ref{tab:choppers}.

\begin{table}[H]
\centering
\caption{FINESSE chopper configuration. Distances are measured from the moderator. For the cryogenic nEDM optimization, the PDC2 position was additionally studied at 60 and 70~m.}
\begin{tabular}{lccc}
\toprule
\textbf{Chopper} & \textbf{Distance} & \textbf{Frequency} & \textbf{Opening} \\
\midrule
PDC1 & 6.7\,m & 14--70\,Hz & Variable (pair of co-rotating discs) \\
FOC1 & 6.7\,m & 14\,Hz & $66^\circ$ \\
FOC2 & 8.5\,m & 14\,Hz & $79^\circ$ \\
PDC2 & 26, 42, 60 or 70\,m & $n\times14$\,Hz (typically 70\,Hz) & Variable \\
PSC  & next to PDC2 & 14\,Hz & $60^\circ$ (default) \\
PMC  & replaces PDC2 & 14\,Hz & Multislit \\
\bottomrule
\end{tabular}
\label{tab:choppers}
\end{table}

Overall, the exchangeable guide concept allows FINESSE to implement two distinct optical configurations within a single ESS beam port, while relying on established engineering solutions and limiting the exchange operation to a short, well-defined section of the bunker.

\section{Detector concepts}
The broad FINESSE physics programme requires complementary detector systems matched to the different experimental signatures. Searches involving antineutrons require a dedicated annihilation detector capable of reconstructing the characteristic multi-particle final state produced by antineutron annihilation on a thin target. For neutron-based measurements, the detector requirements depend strongly on the observable and experimental configuration, with different applications requiring high-rate beam monitoring, transmission measurements, low-background neutron counting, or current-mode operation. The following sections summarize the detector concepts that have been developed for the main FINESSE benchmark measurements.

\subsection{Annihilation detector}
\label{anndetector}

The task of the annihilation detector is to identify the characteristic pionic signature arising from the annihilation of an antineutron in a thin carbon foil. The annihilation releases an energy of order
\(2m_n \simeq 1.88\)~GeV, producing an approximately isotropic multi-pion final state~\cite{ann1,ann2,ann3}. Since the experiment operates in the quasi-free low-magnetic-field region, track curvature cannot be used to estimate charged-particle momenta.

The principal detector requirements comprise precise reconstruction of a multi-track vertex on the annihilation foil, charged-particle identification from tracking and continuous ionisation energy loss (\(\frac{dE}{dx}\)), and calorimetric measurement of charged particles and photons, including precision reconstruction of \(\pi^0\rightarrow\gamma\gamma\) decays. Passive cosmic shielding and an active veto of cosmic-ray backgrounds are also necessary. Simulations of the NNBAR detector show that these observables can suppress backgrounds to a negligible level while retaining a signal efficiency of order \(70\)--\(75\%\)~\cite{Santoro:2024lvc}, compared with about \(50\%\) in the ILL experiment~\cite{Santoro:2024lvc,BaldoCeolin:1994nnbar}.

The detector size and transverse acceptance remain important elements of the overall \(n\rightarrow\bar n\) sensitivity, as discussed in Section~\ref{sec:neutron_oscillations}. The annihilation-detector design must therefore balance reconstruction performance, background rejection, and the large transverse acceptance required by the high-flux neutron beam.

Two detector implementations are being considered~\cite{Yiu:2022faw}. The \emph{minimal} option reuses the existing WASA CsI(Na) calorimeter, which surrounds a cylindrical Time Projection Chamber (TPC). A scintillator-fibre cosmic-ray veto supplements this system~\cite{Santoro:2023izd,wasa}. WASA provides nearly hermetic electromagnetic calorimetry~\cite{wasa} with a photon-energy resolution of
\[
\frac{\sigma_E}{E}\simeq \frac{0.05}{\sqrt{E[\mathrm{GeV}]}} ,
\]
giving precise \(\pi^0\) reconstruction. The TPC supplies three-dimensional tracking and \(\frac{dE}{dx}\)-based particle identification. Recent vertex-reconstruction studies obtain median transverse vertex errors of \(4\)--\(6\)~mm for the best-performing algorithms, compared with a 50\% containment radius of about 4.5~cm at ILL, corresponding to an approximately order-of-magnitude improvement~\cite{Burgman:2026fer}.

\begin{center}
  \includegraphics[width=0.72\linewidth]{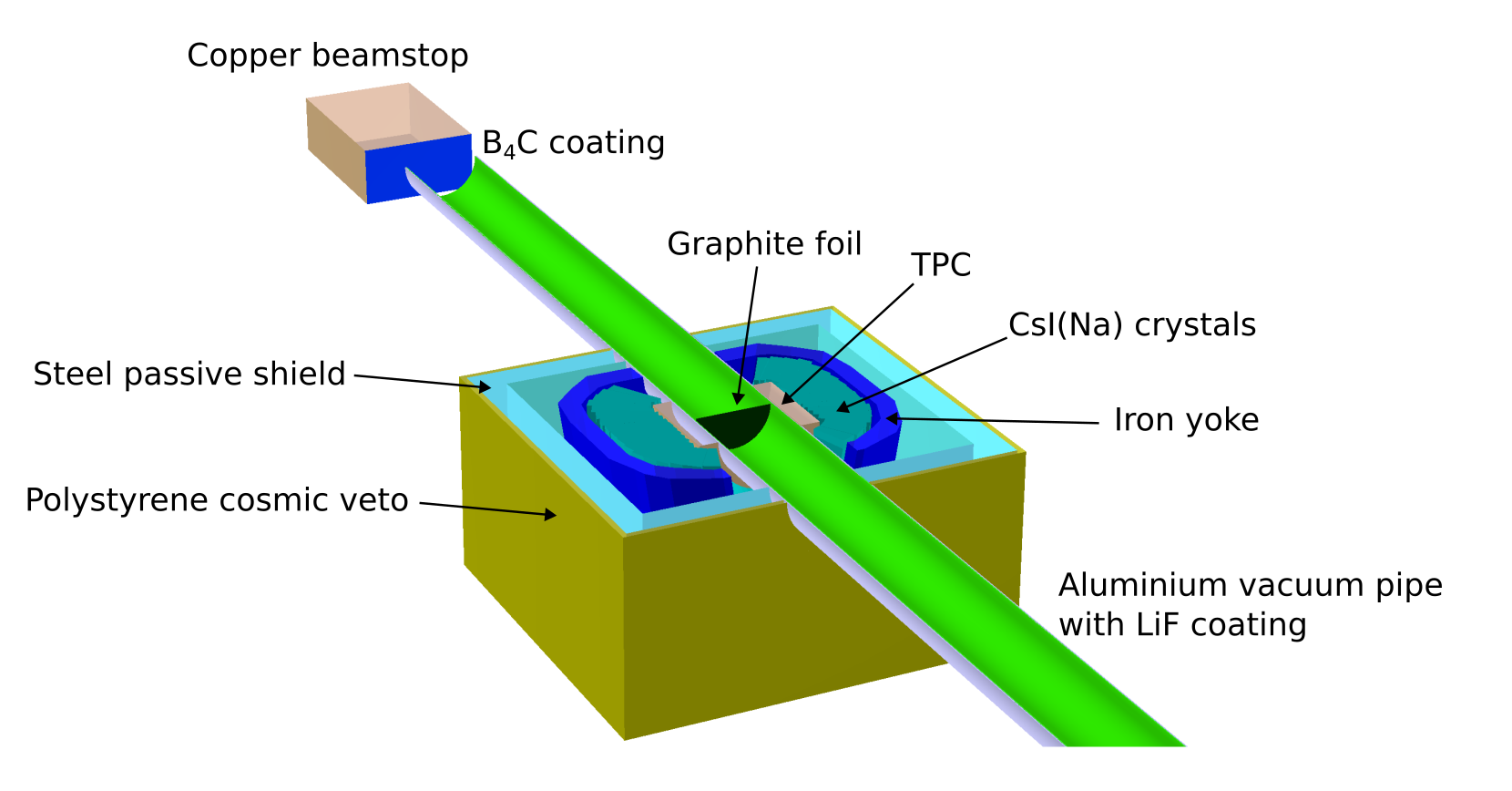}
  \\
  \refstepcounter{figure}\label{fig:hibeam_wasa_detector}
  \small\textbf{Figure \thefigure.} Geant4 model of the minimal annihilation detector, showing the graphite annihilation foil, cylindrical TPC, WASA CsI(Na) calorimeter, passive shielding, scintillator cosmic veto, and downstream beam stop.
\end{center}

A second, larger implementation follows the NNBAR detector concept~\cite{Santoro:2024lvc}. In this approach, the TPC is supplemented by a plastic-scintillator hadronic range detector (HRD), followed by a lead-glass electromagnetic calorimeter and an outer cosmic-veto system. The HRD allows low-energy protons and charged pions to range out, providing additional \(\frac{dE}{dx}\) information, range measurements, and coarse tracking information. An energy resolution of order \(20\)--\(30\%\) is expected for stopping charged particles in the annihilation final state.

The lead-glass calorimeter provides
\[
\frac{\sigma_E}{E}\simeq \frac{6\%}{\sqrt{E[\mathrm{GeV}]}}
\]
and approximately \(1\)~ns timing, enabling precision photon and \(\pi^0\) reconstruction~\cite{Santoro:2024lvc,Aphecetche:2003phx}. The performance is based on that demonstrated by the PHENIX lead-glass electromagnetic calorimeter~\cite{Aphecetche:2003phx}. Because this detector can be made larger than WASA, it permits a larger annihilation target and neutron-beam acceptance and therefore a higher \(n\rightarrow\bar n\) sensitivity.

Re-use of the PHENIX lead-glass supermodules has been studied. For a cylindrical barrel arrangement, a nearly uniform \(\pi^0\) acceptance of about 0.86 was obtained and, for realistic antineutron-annihilation events, the reconstructed \(\pi^0\) efficiency was about \(85\%\) with a mass resolution of 9.6~MeV.

Detector R\&D is ongoing for the various components~\cite{Dunne:2022calo,Rataj:2026suy}. At the Cyclotron Centre Bronowice (CCB), prototype scintillator and TPC systems have been exposed to low-energy fixed-target proton beams, complementing cosmic-ray measurements and providing tests with more highly ionising tracks~\cite{Rataj:2026suy}. These measurements are being used to benchmark the detector simulation, readout, and reconstruction, and to test the robustness of the technologies relevant for operation in the ESS environment.

\subsection{Neutron detectors}
\label{neutrondetector}

The FINESSE physics programme requires neutron detection over a broad range of
rates and background conditions. In particular, high-flux measurements such as
sterile-neutron disappearance searches, axion-like-particle searches, and
measurements of the neutron electric charge require stable monitoring of an
intense cold-neutron beam, whereas regeneration searches operate at much lower
rates and can use conventional low-background neutron detectors.

For the high-rate measurements, the FINESSE collaboration already has a
well-established baseline solution based on current-mode detection with a
\(^{3}\)He ionization chamber, following the concept developed for
HIBEAM~\cite{Santoro:2023izd}. The detector exploits the reaction
\[
n+{}^{3}\mathrm{He}\rightarrow{}^{3}\mathrm{H}+p+0.78~\mathrm{MeV},
\]
which provides a large ionization signal for slow-neutron capture while
retaining low sensitivity to gamma backgrounds. In current mode, the large
instantaneous neutron rates do not need to be resolved as individual events;
instead, the integrated ionization current provides a direct measure of the
incident neutron intensity. This makes the concept particularly well suited
to the disappearance measurement, for which detector linearity and long-term
stability are more important than event-by-event reconstruction.

A complementary \(^{4}\)He ionization chamber can be placed upstream of the
\(^{3}\)He chamber to measure the gamma-induced component of the signal.
Since the two chambers have nearly identical gamma response, subtraction of
the \(^{4}\)He signal from the \(^{3}\)He signal provides an effective method
for isolating the neutron contribution without relying on pulse-shape
discrimination. This approach has previously been demonstrated at a pulsed
spallation neutron source and provides a robust basis for the FINESSE
high-rate neutron detector~\cite{SZYMANSKI1994564}.

The ion chamber can be segmented both transversely and longitudinally to the
beam. Transverse segmentation provides information on the spatial distribution
of the beam and can reduce the impact of pulse-to-pulse phase-space
fluctuations, while longitudinal segmentation exploits the \(1/v\) dependence
of the \(^{3}\)He absorption cross section and can provide additional
information on the neutron spectrum~\cite{PhysRevC.100.015204}. For the sterile-neutron disappearance
measurement, the required stability of the total neutron signal is at the level
of \(10^{-7}\). Comparable \(^{3}\)He ionization-chamber technology has already
demonstrated precision at the \(10^{-8}\) level in neutron parity-violation
measurements~\cite{Gericke_2020}, providing confidence that the required
stability is technically achievable.

In parallel with this baseline, FINESSE is investigating the possibility of
operating in single-neutron-counting mode. Event-by-event detection would
provide additional information, including neutron time of flight, longitudinal
beam-profile measurements, and powerful internal consistency checks on
possible disappearance signals. It could therefore improve the control of
systematic effects and provide a complementary measurement strategy.

The current design involves \ce{^3He}-based gas detectors with a cylindrical
geometry for both the beam monitor and the sink detector. Similar technical
solutions are envisaged for both; therefore, the discussion below focuses on
the sink detector.

Due to the large footprint of the FINESSE neutron beam, the cylinder diameter
is \(1\,\mathrm{m}\). The detector operates in single-neutron-counting mode
with a padded readout, utilizing Gas Electron Multiplier (GEM) technology for
signal multiplication. Notably, the signal is carried by electrons rather than
ions, unlike in some previous proposals.

The electrode design is still to be finalized. Specifically, while the
standard approach for cylindrical gas detectors employs a single central wire
as the cathode, this is not feasible here: the high local field required to
maintain an adequate drift field throughout the detector volume would trigger
avalanche multiplication near the wire. Alternative designs are currently
under investigation, particularly multi-wire configurations.

Beyond the need for \ce{^3He}, the gas mixture remains to be defined. It must
ensure both adequate ion transport, to minimize space-charge effects, and
efficient electron transport, to achieve a fast response. Various options are
being studied, ranging from \ce{H2}, with its excellent ion mobility, to
\ce{CF4}, chosen for its electronic properties. Compatibility constraints must
also be met, particularly to prevent the formation of chemically aggressive
species such as \ce{HF} or organic radicals.

For the purpose of assessing scattering off the detector, the outer structural
shell is assumed to be made of S316 stainless steel. A \(2\,\mathrm{mm}\)
aluminum window is used for the beam entrance to ensure a low interaction
cross section. Finally, \(0.05\,\mathrm{mm}\)-thick Mylar sheets are employed
for longitudinal segmentation partitions.

The main technical challenges associated with the detector stem from high count
rates. These rates, in turn, result from the decision to operate in
single-neutron-counting mode. It is therefore appropriate to clarify the
advantages of this approach. Operating in the single-neutron-counting regime
enables longitudinal imaging. This offers two immediate advantages. The first
is a \emph{reduction in uncertainty}: since beam intensity fluctuations
produce identical fractional signal variations across the entire detector,
readings from different detector sections can be combined to reduce
uncertainties, in principle below the Poisson limit. Secondly,
\emph{consistency checks} can be performed by exploiting the beam intensity
and the spectral variation along the detector, the latter being known
\emph{a priori} from the \ce{^3He} cross sections. These checks can be applied
to both signal and noise.

While the aforementioned advantages, which rely solely on longitudinal imaging
capabilities, could also be achieved with a segmented detector operating in
current mode, other benefits depend crucially on single-neutron counting.
Indeed, this approach allows for the separation of different velocity
components: by recording the time of arrival (ToA) and longitudinal position
of each neutron, one can derive a time-of-flight (ToF) spectrum, thereby
counting neutrons at different velocities separately. This method enables the
use of the entire neutron spectrum to detect the disappearance phenomenon,
thereby improving statistics, while also allowing verification of the
signature \(1/v^2\) disappearance dependence, albeit with lower statistical
significance, thus helping to discriminate against spurious effects.

A representative neutron rate at the FINESSE beamline is
\(6.4\times10^{10}\,\mathrm{n\,s^{-1}}\) at an ESS linac power of 1~MW.
Assuming linear scaling with beam power, this corresponds to approximately
\(1.28\times10^{11}\,\mathrm{n\,s^{-1}}\) at the 2~MW benchmark used in this
paper. These values represent the average neutron rate over the full
measurement period; given the ESS pulse duration of a few milliseconds and a
repetition rate of 14~Hz, instantaneous rates on the order of
\(10^{12}\,\mathrm{s^{-1}}\) are expected. Maintaining a uniform
\ce{^3He} concentration throughout the detector would result in prohibitively
high count rates in the entrance region, for any realistic pad dimensions, or
would require an unreasonably long detector. Longitudinal segmentation is
therefore necessary.

Assuming a maximum pad count rate of \(10\,\mathrm{MHz}\), only slightly above
the current state of the art for GEM electronics, feasible designs can be
achieved. We consider two such designs: the first is \(1\,\mathrm{m}\) long,
featuring small pads along the entire length of the detector, while the second
is \(2\,\mathrm{m}\) long, with pad size decreasing along the detector's
length. Both designs utilize approximately \(10^5\) readout channels and a
substantial quantity of \ce{^3He}.

The large number of readout channels and the substantial quantity of
\ce{^3He} required represent important practical constraints. Alternative
hybrid configurations are therefore also under study. One possibility is to
use an initial section operating in current mode, with a relatively high
\ce{^3He} concentration and short longitudinal extent, to rapidly attenuate
the incoming neutron beam. Subsequent sections could then operate at lower
rates with larger readout pads and, potentially, in single-neutron-counting
mode. Such a configuration could preserve many of the advantages of
event-by-event detection while substantially reducing the requirements on
segmentation, readout rate, and \ce{^3He} inventory.

The uncertainty analysis conducted so far is rather rudimentary, as it was
intended more to assess the detector's feasibility than to evaluate its
performance in setting limits on the disappearance phenomenon. Sizing the
monitoring detector, which preferentially absorbs slower neutrons, as they are
also the most prone to the disappearance phenomenon, is ongoing.

Current-mode \(^{3}\)He detection therefore provides the established baseline
for the highest-flux FINESSE measurements, while single-neutron counting and
hybrid detector architectures provide additional possibilities for improving
spectral information, internal consistency checks, and systematic control.

\section{Short FINESSE: early physics and a path to the full beamline}
\label{shortfinesse}

The FINESSE programme can be implemented in stages, allowing an initial physics
programme to begin before completion of the full beamline. The first stage,
known as \textit{Short FINESSE}, requires only the upstream part of the
high-flux neutron guide and a compact experimental area located directly
outside the bunker. The overall beamline footprint is approximately 30~m, with
the neutron beam reaching the detector at about 25~m from the source. The
exchangeable-guide mechanics are included from the outset, preserving a direct
upgrade path to the full dual-mode FINESSE facility without major
re-engineering of the bunker region.

\begin{figure}[t!]
  \begin{center}
    \includegraphics[width=0.7\textwidth]{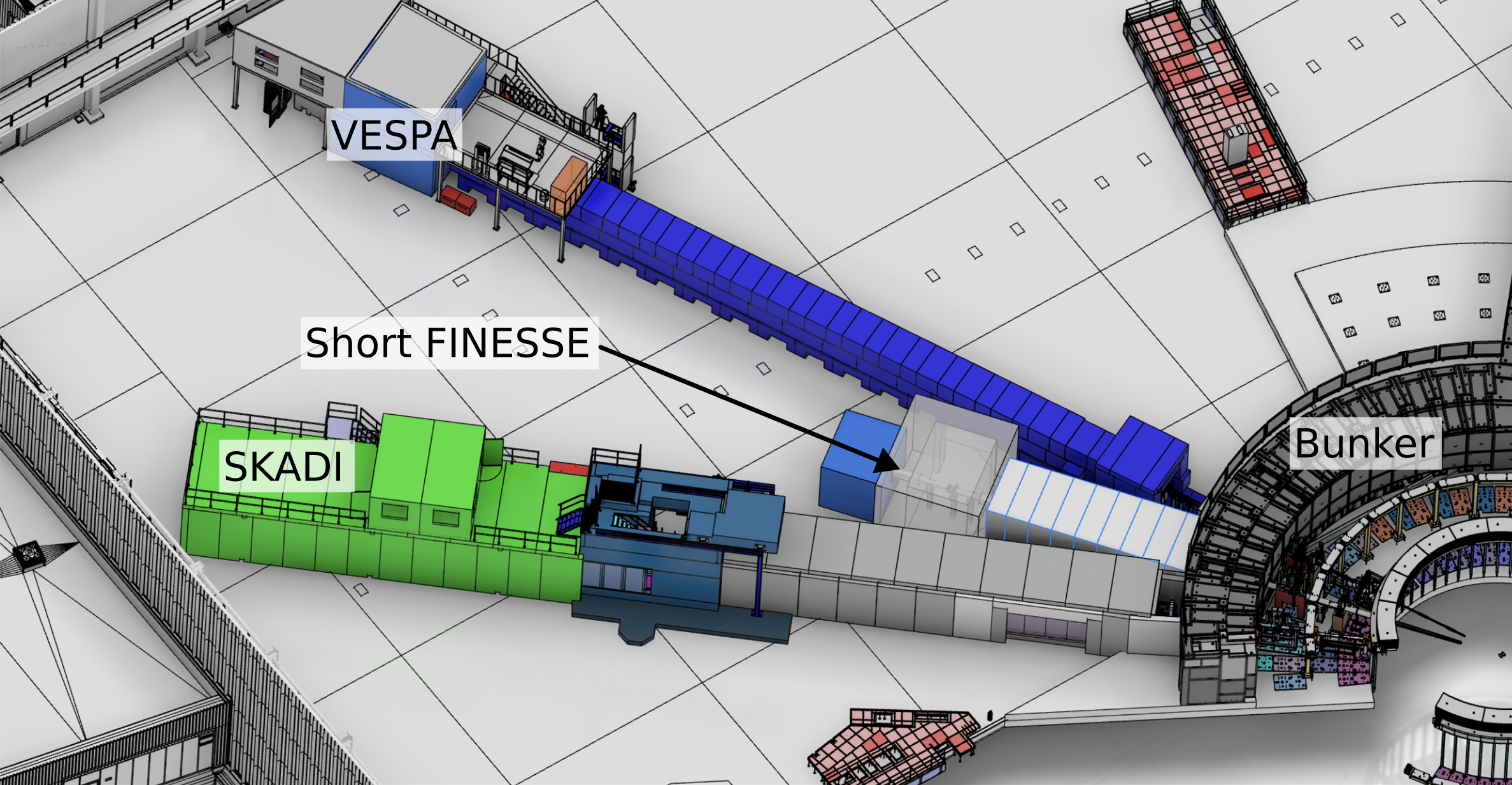}
    \caption{Overview of the Short FINESSE beamline in the ESS experimental hall.}
    \label{fig:shortbeamline}
  \end{center}
\end{figure}

\begin{figure}[H]
    \centering
    \includegraphics[width=0.7\textwidth]{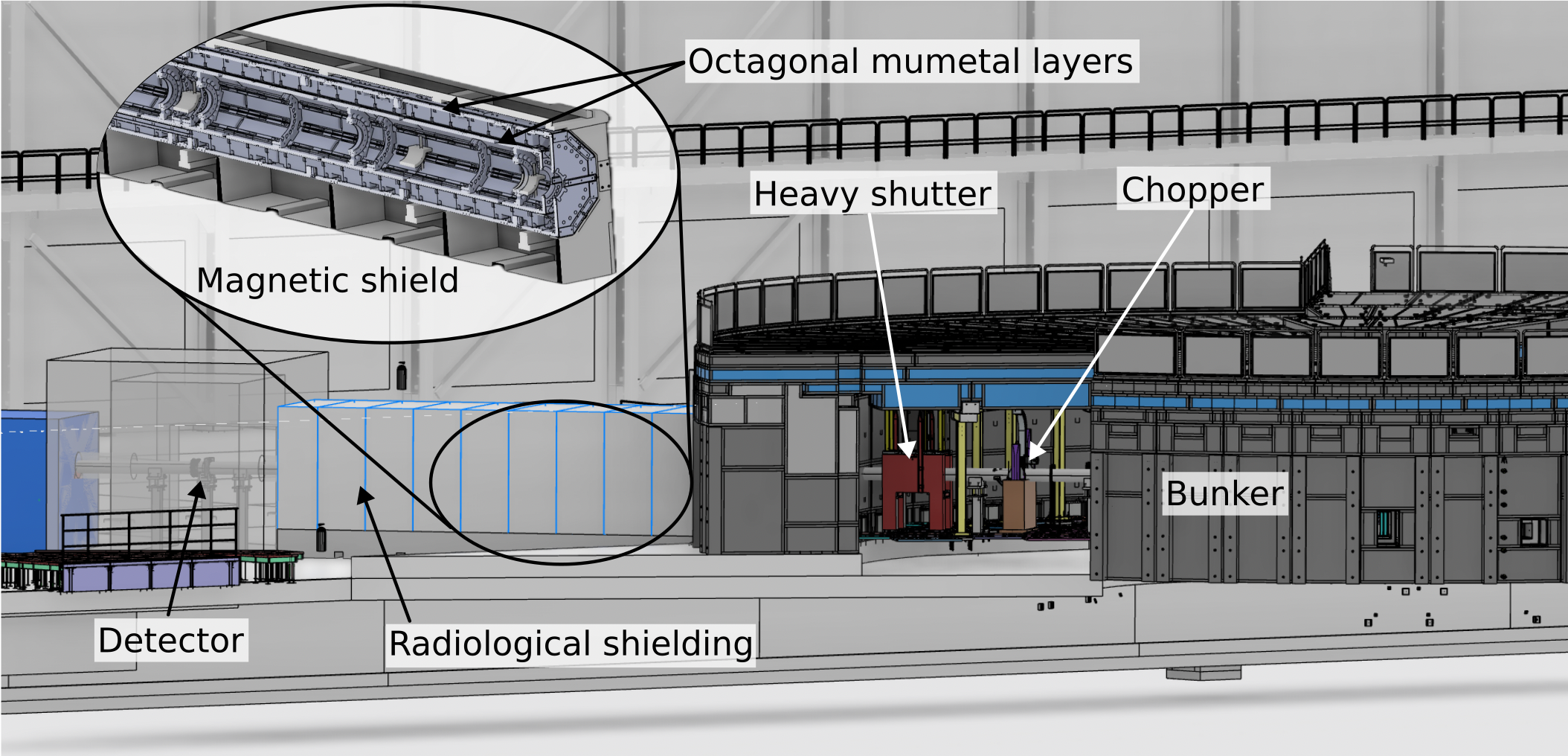}
    \caption{CAD model of ESS illustrating a Short FINESSE configuration for a
    search for axion-like particles. A 10~m-long, dual-layer octagonal
    mu-metal shield is positioned outside the bunker. The same short
    high-flux beamline can also support sterile-neutron searches, neutron
    electric-charge measurements, and R\&D for neutron--antineutron
    oscillation searches.}
    \label{fig:shortfinesse}
\end{figure}

This initial installation includes a 10-m-long magnetic shield located outside
the bunker, for which external funding has already been secured, together with
biological shielding and a compact experimental cave equipped with neutron
detectors. An overview of the Short FINESSE beamline in the experimental hall
is shown in Fig.~\ref{fig:shortbeamline}, while a representative experimental
configuration is shown in Fig.~\ref{fig:shortfinesse}.

The low-background guide, the complete downstream guide system, and the
full-length experimental area are not required for the initial Short FINESSE
physics programme.

The most mature early experiment foreseen for Short FINESSE is the Ramsey
neutron-beam search for ultralight axion-like dark matter described in
Section~\ref{ALP}~\cite{PhysRevLett.133.181001,RamseyProp}. The setup comprises
a \ce{^3He} polarizer, two RF coils, a uniform holding field generated by
cosine-theta coils, a \ce{^3He} spin analyzer, and a current-mode neutron
detector. Funding has also been secured for the RF-coil system, while
development of the \ce{^3He} polarization system is actively being pursued, as
discussed in Section~\ref{pol}. The principal remaining requirement is
therefore the beamline infrastructure. Since the ESS pulse structure can be
exploited for velocity compensation using time-dependent RF amplitude
modulation, the experiment does not require the complete chopper
system~\cite{RamseyProp}. The magnetic shielding and cosine-theta coils are
designed for reuse in subsequent neutron-oscillation experiments.

The same high-flux beam and compact experimental area can host the
sterile-neutron searches described in Section~\ref{sterile}. The magnetically
controlled region can be reused for disappearance or regeneration searches,
with absorbers, gas cells, and a dedicated low-background neutron detector
added as required. This would extend the experimental programme developed at
SNS and HFIR~\cite{sterile1,sterile2,sterile3}, taking advantage of the high
neutron intensity and controlled flight path available at ESS.

Short FINESSE can also support an early programme of free
neutron--antineutron oscillation studies. Its shorter propagation length limits
the ultimate discovery reach relative to the full High-Flux FINESSE
configuration, but it provides a realistic ESS beam environment for
commissioning the annihilation detector, characterizing beam-induced
backgrounds, and testing the magnetic shielding, reconstruction, and overall
experimental chain. As discussed in Section~\ref{sec:shortfinesse_optics},
neutron-optics simulations indicate that the uncollimated Short FINESSE
configuration reaches a beam figure of merit \(NT^2\) of the same order as that
of the historical ILL free-neutron search, making it a valuable environment for
\(n\rightarrow\bar n\) R\&D and validation of the experimental concept.

\subsection{Neutron-optics implementation for Short FINESSE}
\label{sec:shortfinesse_optics}

A central feature of the staged FINESSE concept is that Short FINESSE does not
require a separate neutron-guide design. Instead, it can be implemented using
the upstream part of the High-Flux FINESSE guide described in
Section~\ref{highfluxguide}. Neutrons are transported from the E5 extraction
system through the bunker to the entrance of the compact experimental region
at approximately 15~m from the moderator, followed by the approximately
10~m-long magnetically controlled region used for the first Short FINESSE
measurements.

This provides a direct path from the initial installation to the full
High-Flux FINESSE beamline. The guide installed for Short FINESSE forms part of
the final optical system and can remain in place when the beamline is
subsequently extended. Short FINESSE is therefore not based on a temporary or
independent neutron-optics solution, but on an early implementation of the
high-flux beamline itself.

A dedicated \textsc{McStas} study~\cite{9241726} was performed to evaluate the
performance of this shortened configuration for the initial physics programme,
in particular the Ramsey neutron-beam search for ultralight axion-like dark
matter described in Section~\ref{ALP}. The simulations used  a nominal linac power of 2~MW and considered the neutron
transport between the E5 extraction region and the entrance to the
magnetically controlled section.

For the Ramsey measurement, the relevant neutron-optics figure of merit is
\[
\mathrm{FOM}=NT^2,
\]
where \(N\) is the transmitted neutron intensity and \(T\) is the
free-precession time. The same quantity is also the conventional beam figure of
merit for free neutron--antineutron oscillation searches. For the uncollimated
Short FINESSE configuration, the simulations give
\[
NT^2=(5.1\pm0.03)\times10^8~\mathrm{n\,s}
\]
at 2~MW. This value is of the same order as the historical ILL free-neutron
search figure of merit,
\[
NT^2_{\mathrm{ILL}}\simeq1.5\times10^9~\mathrm{n\,s},
\]
and therefore supports the use of Short FINESSE for \(n\rightarrow\bar n\)
R\&D and commissioning under realistic ESS beam conditions.

\begin{figure}[H]
    \centering
    \includegraphics[width=0.5\textwidth]{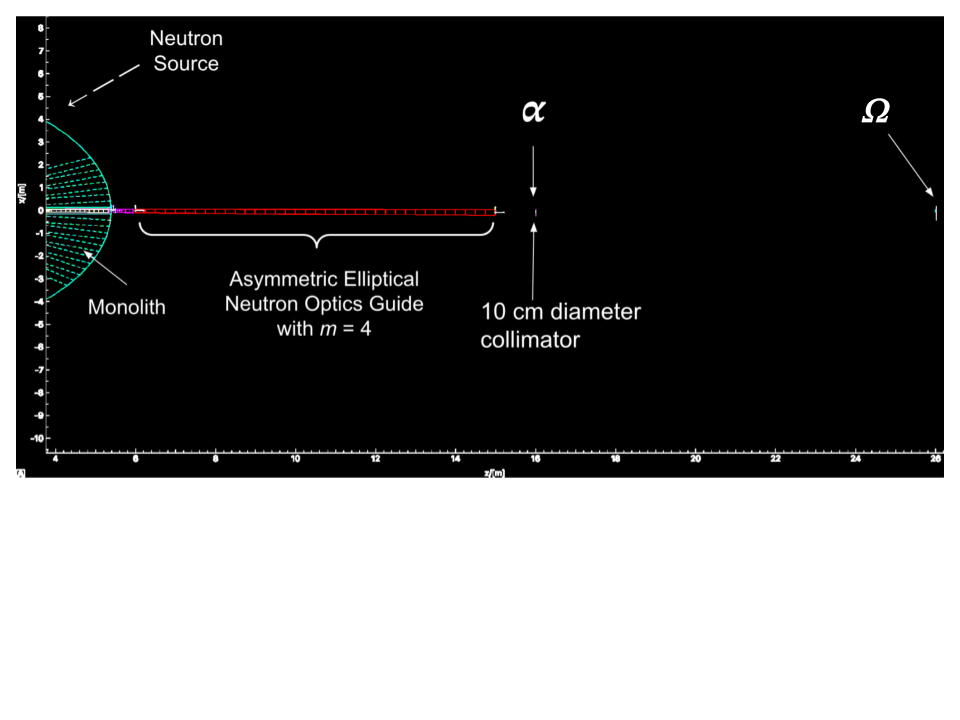}
    \caption{McStas model of the Short FINESSE neutron-optics configuration.
    The beamline uses the upstream part of the High-Flux FINESSE guide to
    transport neutrons from the E5 extraction system to the compact
    experimental region. For the Ramsey ALP configuration, the beam is matched
    to the approximately 10~cm acceptance of the polarization system before
    entering the approximately 10~m magnetically controlled interaction region.}
    \label{fig:guidedesignshortfinesse}
\end{figure}

For the Ramsey ALP configuration, the same high-flux optics can be matched to
the approximately 10~cm acceptance of the planned \ce{^3He} polarizer and
analyzer system through downstream phase-space selection and collimation near
the entrance to the magnetically controlled region. The resulting Short
FINESSE configuration is shown in Fig.~\ref{fig:guidedesignshortfinesse}.

The same optics installed for early operation can subsequently become part of
the full High-Flux FINESSE beamline, minimizing duplicated infrastructure and
providing a direct technical path from Short FINESSE to the complete
configuration.

\subsection{Polarisation capability for Short FINESSE}
\label{pol}

Polarized neutrons are required for a substantial fraction of the FINESSE physics programme, including neutron beta-decay measurements, hadronic parity-violation experiments, axion-like-particle searches, beam-EDM concepts, spin-rotation measurements, and other spin-dependent searches. The detailed polarization requirements depend on the specific experiment and on the FINESSE operating mode. In this section, we focus on the polarization system foreseen for Short FINESSE and, in particular, for the Ramsey axion-like-particle experiment.

Several methods are available for neutron-beam polarisation, including spin filters based on hyperpolarized \ce{^3He}, magnetic supermirrors, and Heusler-crystal polarizers. For the Short FINESSE Ramsey experiment described in Section~\ref{shortfinesse}, hyperpolarized \ce{^3He} is considered the primary option for both neutron polarisation and spin analysis.

A \ce{^3He} spin filter exploits the strong spin dependence of the neutron-absorption cross section in polarized \ce{^3He}. Efficient operation requires a high nuclear polarization of the gas, typically above 70\%. Hyperpolarization can be produced either by Spin-Exchange Optical Pumping (SEOP) or by Metastability-Exchange Optical Pumping (MEOP)~\cite{PhysRev.132.2561,RevModPhys.89.045004}. SEOP operates at relatively high gas pressure and is suitable for compact local systems, while MEOP provides highly polarized gas at low pressure which is subsequently compressed to the operating pressure.

The ESS plans to operate a central \ce{^3He} polarization facility based on MEOP, including a gas-compression and distribution system for supplying polarized \ce{^3He} to neutron instruments~\cite{refId0}. This central infrastructure is particularly well matched to the Short FINESSE requirements. The expected production rate is approximately 2~bar-litre per hour, with \ce{^3He} polarization approaching 80\%~\cite{TYREX2}. A compact SEOP system located close to the experiment could provide a complementary or backup source if required.

At Short FINESSE, polarized \ce{^3He} will be used in dedicated polarizer and analyzer cells located upstream and downstream of the Ramsey interaction region. The cells are foreseen to have an internal diameter of approximately 10~cm, matched to the neutron-beam cross section, and an active length of approximately 10~cm. Each cell will be surrounded by a highly uniform transverse holding field of order 1~mT, generated by a dedicated coil system, in order to preserve the \ce{^3He} polarization and define the neutron-spin quantization axis.

\begin{figure}[H]
    \centering
    \includegraphics[width=0.35\textwidth]{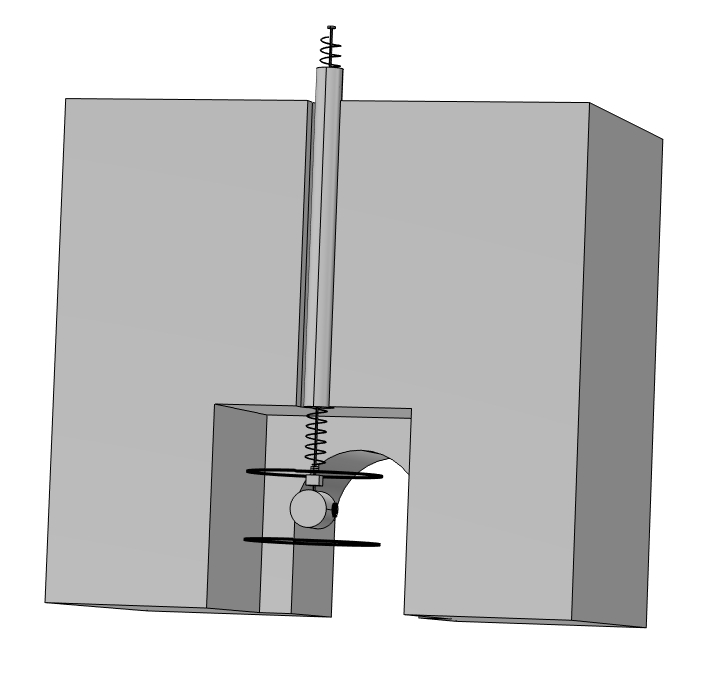}
    \caption{Conceptual arrangement of a \ce{^3He} spin-filter cell and its gas-handling connection inside the biological shielding.}
    \label{fig:cell}
\end{figure}

The polarizer and analyzer cells will be connected through narrow glass transfer tubes to a gas-handling system located outside the biological shielding, as illustrated in Fig.~\ref{fig:cell}. Polarized gas supplied by the ESS central facility can therefore be transferred to the cells without direct access to the beam enclosure. After approximately 24--48~hours, depending on the relaxation time of the individual cell, the gas can be recovered and returned to the ESS purification and polarization system.

The transfer lines will be surrounded by solenoidal holding-field coils to minimize depolarization during gas handling. The polarization of the \ce{^3He} in each cell will be monitored using nuclear magnetic resonance, with small RF coils tuned to the corresponding Larmor frequency. This arrangement provides a practical and reusable polarization system for the Short FINESSE axion experiment and can also support future spin-dependent measurements at the beamline.

\subsection{Neutron detector for Short FINESSE}
\label{sec:shortfinesse_detector}

The initial Short FINESSE programme requires a neutron detector capable of
operating at the high instantaneous flux expected in the E5 beam. As discussed
in Section~\ref{neutrondetector}, the baseline solution for such measurements is a
current-mode \ce{^3He} ionization chamber. An existing segmented \ce{^3He}
detector, originally developed for precision cold-neutron measurements
\cite{PENN2001332}, is available within the collaboration and provides a
natural starting point for Short FINESSE.

The detector geometry has been implemented in Geant4 and coupled to neutron
and photon spectra obtained from Short FINESSE beam-transport
simulations~\cite{9202314}. These studies confirm efficient absorption of the
cold-neutron component and have been used to investigate the influence of
beam-induced gamma radiation, fast neutrons, and scattering in the detector
structure. In particular, the simulations show that neutron scattering in the
structural materials can contribute significantly to the residual downstream
signal and should therefore be included in the final detector optimization.

\begin{figure}[H]
    \centering
    \begin{subfigure}[b]{0.45\textwidth}
        \centering
        \includegraphics[width=\textwidth]{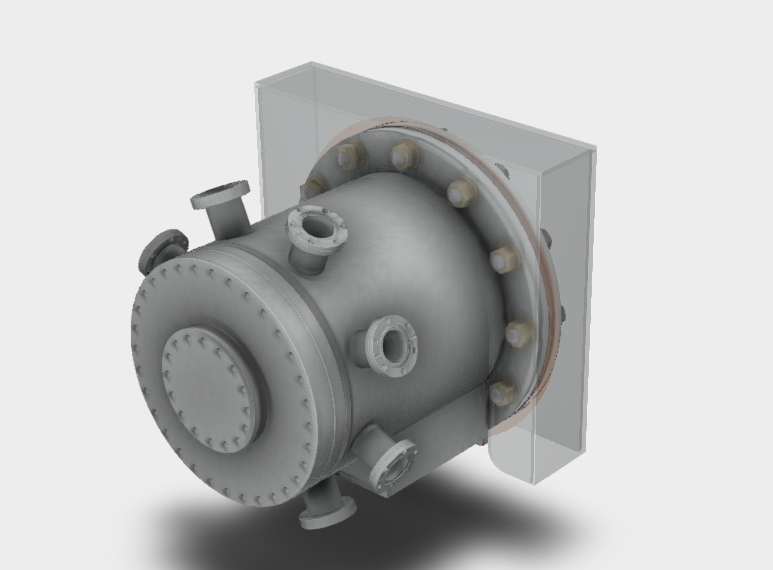}
        \caption{}
    \end{subfigure}
    \hfill
    \begin{subfigure}[b]{0.45\textwidth}
        \centering
        \includegraphics[width=\textwidth]{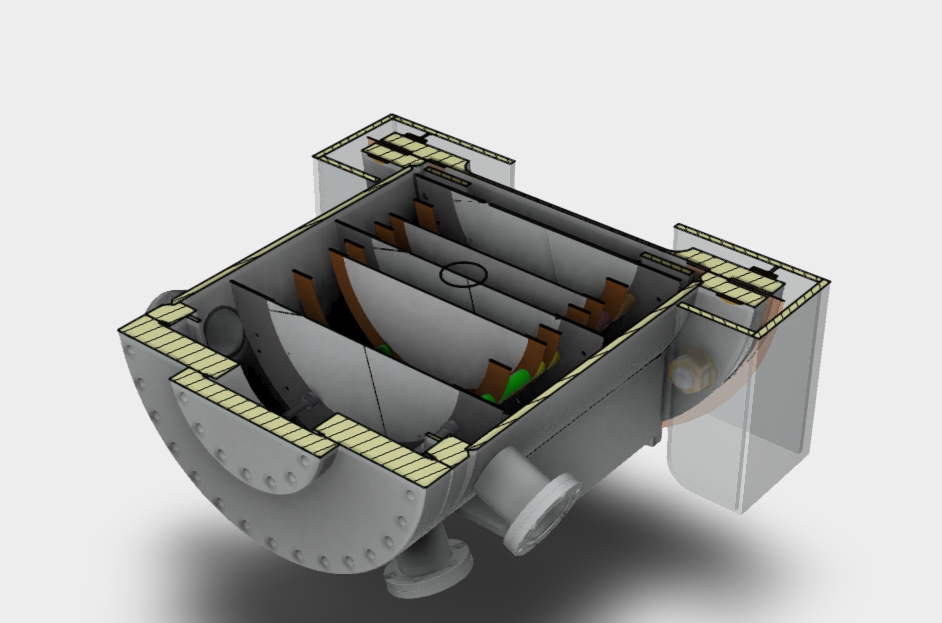}
        \caption{}
    \end{subfigure}
    \caption{Baseline \ce{^3He} current-mode neutron detector considered for
    Short FINESSE. \textbf{(a)} CAD model of the detector.
    \textbf{(b)} Cross-sectional view showing the internal collection plates
    and structural components.}
    \label{He3CAD}
\end{figure}

The final configuration will be optimized together with the Short FINESSE beam
profile, detector materials, shielding, and the measured E5 background
environment. Dedicated measurements during early beamline operation will be
important for validating the simulated fast-neutron and gamma backgrounds and
for defining the detector configuration used in the first physics measurements.

\section{Technical readiness and available hardware}

The beamline concept is supported by a substantial level of technical readiness. For the high-flux programme, detailed optics, shielding, magnetic-field, and detector studies have already been carried out. Prototype detector components and magnetic-shielding elements have been developed and tested, and a mature software environment exists for neutron transport, detector simulation, and event reconstruction.

Several key hardware components are already available or identified. These include the WASA crystal calorimeter for neutron--antineutron searches, magnetic shielding for a 10~m beam region, neutron detectors for sterile-neutron studies, and substantial cryogenic and vacuum infrastructure from the former SNS nEDM programme. These existing resources reduce both the implementation cost and the technical risk of the FINESSE programme.

\section{Conclusions}

FINESSE provides a technically realistic and scientifically broad concept for a
dedicated fundamental-physics beamline at ESS. Its dual-mode design combines a
high-flux configuration optimized for searches for rare neutron transformations
with a low-background configuration tailored to precision measurements, while
the staged Short FINESSE implementation provides a direct path toward early
physics and progressive realization of the full facility.

The studies presented here show that the same beamline can support a diverse
programme ranging from free neutron--antineutron oscillation searches and
sterile-neutron searches to ultralight dark-matter searches, measurements of
the neutron electric charge, neutron beta decay, parity-violation studies, and
a next-generation cryogenic neutron electric-dipole-moment experiment. The
combination of high neutron intensity, controlled magnetic environments,
low-background transport, flexible detector integration, and dedicated neutron
optics gives FINESSE the capability to address several of the most compelling
open questions in particle and fundamental physics with neutrons.

With its staged implementation strategy, reuse of infrastructure between the
different operating modes, and strong complementarity between precision and
rare-process measurements, FINESSE could establish ESS as a world-leading
facility for neutron-based particle physics and deliver a broad, long-term
programme at the intensity frontier.

\section{Acknowledgements }

D.~M. and V.~S. gratefully acknowledge support from the Council for Swedish
Research Infrastructure at the Swedish Research Council through the grant
``HIBEAM pre-studies.'' V.~S. gratefully acknowledges support from the Swedish
Research Council through the grants ``Development of an innovative neutron
detector for the European Spallation Source (ESS)'' and ``The First Particle
Physics Experiment at the ESS: Search for Axion-like Particles at the HIBEAM
Beamline.''  D.~M. and A.~B. are grateful for support from Olle Engkvists Stiftelse.
V.~S. also acknowledges support from Stiftelsen för Strategisk
Forskning through the grant ``Development of a magnetic control beamline for
fundamental physics and condensed matter science at the European Spallation
Source,'' as well as support from the Crafoord Foundation.

This work was also supported by the Italian Ministry of Foreign Affairs and
International Cooperation, grant number PGR SE24GR04. Detector
simulation computations were enabled by resources provided by LUNARC, the
Centre for Scientific and Technical Computing at Lund University. W.~M.~S.
acknowledges support from U.S. National Science Foundation grant PHY-2209481
and from the Indiana University Center for Spacetime Symmetries. L.J.B, J. R. and A.S. were supported by the U.S. Department of Energy (DOE), Office of Science, Office of Nuclear Physics (Contract No. DE-AC05-00OR22725).

\bibliographystyle{elsarticle-num} 
\bibliography{FINESSE_references_arxiv}

@article{Aphecetche:2003phx,
  author  = {Aphecetche, L. and others},
  title   = {PHENIX calorimeter},
  journal = {Nucl. Instrum. Meth. A},
  volume  = {499},
  number  = {2-3},
  pages   = {521--536},
  year    = {2003},
  doi     = {10.1016/S0168-9002(02)01954-X}
}

@article{PhysRev.132.2561,
  author  = {Colegrove, F. D. and Schearer, L. D. and Walters, G. K.},
  title   = {Polarization of {$^3$He} Gas by Optical Pumping},
  journal = {Phys. Rev.},
  volume  = {132},
  number  = {6},
  pages   = {2561--2572},
  year    = {1963},
  doi     = {10.1103/PhysRev.132.2561},
  url     = {https://doi.org/10.1103/PhysRev.132.2561}
}

@article{RevModPhys.89.045004,
  title = {Optically polarized $^{3}\mathrm{He}$},
  author = {Gentile, T. R. and Nacher, P. J. and Saam, B. and Walker, T. G.},
  journal = {Rev. Mod. Phys.},
  volume = {89},
  issue = {4},
  pages = {045004},
  numpages = {59},
  year = {2017},
  month = {Dec},
  publisher = {American Physical Society},
  doi = {10.1103/RevModPhys.89.045004},
  url = {https://link.aps.org/doi/10.1103/RevModPhys.89.045004}
}

@article{ refId0,
	author = {{Lee, Wai Tung} and {Hagman, Joel} and {Martin Rodriguez, Damian} and {Stellhorn, Annika} and {Backs, Alex} and {Arnold, Thomas} and {Blackburn, Elizabeth} and {Deen, Pascale} and {Durniak, Celine} and {Feygenson, Mikhail} and {Holmes, Alexander T.} and {Houston, Judith} and {Jaksch, Sebastian} and {Kirstein, Oliver} and {Mannix, Dan} and {Månsson, Martin} and {Morgano, Manuel} and {Nilsen, Gøran} and {Noferini, Daria} and {Nylander, Tommy} and {Orlov, Dmytro} and {Santoro, Valentina} and {Schmidt, Søren} and {Schulz, Michael} and {Schweika, Werner} and {Strobl, Markus} and {Tartaglione, Aureliano} and {Toft-Petersen, Rasmus} and {Villacorta, Félix J.} and {Willendrup, Peter} and {Wolff, Maximillian} and {Woracek, Robin}},
	title = {Polarisation Development at the European Spallation Source},
	DOI= "10.1051/epjconf/202328603004",
	url= "https://doi.org/10.1051/epjconf/202328603004",
	journal = {EPJ Web Conf.},
	year = 2023,
	volume = 286,
	pages = "03004",
}

@article{HADDOCK2018227,
title = {A search for possible long range spin dependent interactions of the neutron from exotic vector boson exchange},
journal = {Physics Letters B},
volume = {783},
pages = {227-233},
year = {2018},
issn = {0370-2693},
doi = {10.1016/j.physletb.2018.06.066},
url = {https://www.sciencedirect.com/science/article/pii/S0370269318305227},
author = {C. Haddock and J. Amadio and E. Anderson and L. Barrón-Palos and B. Crawford and C. Crawford and D. Esposito and W. Fox and I. Francis and J. Fry and H. Gardiner and A. Holley and K. Korsak and J. Lieffers and S. Magers and M. Maldonado-Velázquez and D. Mayorov and J.S. Nico and T. Okudaira and C. Paudel and S. Santra and M. Sarsour and H.M. Shimizu and W.M. Snow and A. Sprow and K. Steffen and H.E. Swanson and F. Tovesson and J. Vanderwerp and P.A. Yergeau}
}

@misc{geerits2024measuringangularmomentumneutron,
      title={Measuring the Angular Momentum of a Neutron Using Earth's Rotation}, 
      author={Niels Geerits and Stephan Sponar and Kyle E. Steffen and William M. Snow and Steven R. Parnell and Giacomo Mauri and Gregory N. Smith and Robert M. Dalgliesh and Victor de Haan},
      year={2024},
      eprint={2407.09307},
      archivePrefix={arXiv},
      primaryClass={quant-ph},
      url={https://arxiv.org/abs/2407.09307}, 
}

@misc{mulkey2025newupperboundsexotic,
      title={New Upper Bounds on Exotic Neutron Spin-Electron Spin Interactions via Neutron Spin Rotation Measurements in a Compensated Ferrimagnet}, 
      author={T. Mulkey and K. N. Lopez and C. D. Hughes and B. Hill and M. Van Meter and H. Wijeratne and J. C. Long and M. Sarsour and W. M. Snow and K. Li and R. Parajuli and S. Samiei and D. V. Baxter and M. Luxnat and Y. Zhang and C. Jiang and E. Stringfellow and J. Torres and R. Hobbs},
      year={2025},
      eprint={2508.14243},
      archivePrefix={arXiv},
      primaryClass={nucl-ex},
      url={https://arxiv.org/abs/2508.14243}, 
}

@article{Ahmed_2019,
doi = {10.1088/1748-0221/14/11/P11017},
url = {https://doi.org/10.1088/1748-0221/14/11/P11017},
year = {2019},
month = {nov},
volume = {14},
number = {11},
pages = {P11017},
author = {Ahmed, M.W. and Alarcon, R. and Aleksandrova, A. and Baeßler, S. and Barron-Palos, L. and Bartoszek, L.M. and Beck, D.H. and Behzadipour, M. and Berkutov, I. and Bessuille, J. and Blatnik, M. and Broering, M. and Broussard, L.J. and Busch, M. and Carr, R. and Cianciolo, V. and Clayton, S.M. and Cooper, M.D. and Crawford, C. and Currie, S.A. and Daurer, C. and Dipert, R. and Dow, K. and Dutta, D. and Efremenko, Y. and Erickson, C.B. and Filippone, B.W. and Fomin, N. and Gao, H. and Golub, R. and Gould, C.R. and Greene, G. and Haase, D.G. and Hasell, D. and Hawari, A.I. and Hayden, M.E. and Holley, A. and Holt, R.J. and Huffman, P.R. and Ihloff, E. and Imam, S.K. and Ito, T.M. and Karcz, M. and Kelsey, J. and Kendellen, D.P. and Kim, Y.J. and Korobkina, E. and Korsch, W. and Lamoreaux, S.K. and Leggett, E. and Leung, K.K.H. and Lipman, A. and Liu, C.Y. and Long, J. and MacDonald, S.W.T. and Makela, M. and Matlashov, A. and Maxwell, J.D. and Mendenhall, M. and Meyer, H.O. and Milner, R.G. and Mueller, P.E. and Nouri, N. and O'Shaughnessy, C.M. and Osthelder, C. and Peng, J.C. and Penttila, S.I. and Phan, N.S. and Plaster, B. and Ramsey, J.C. and Rao, T.M. and Redwine, R.P. and Reid, A. and Saftah, A. and Seidel, G.M. and Silvera, I. and Slutsky, S. and Smith, E. and Snow, W.M. and Sondheim, W. and Sosothikul, S. and Stanislaus, T.D.S. and Sun, X. and Swank, C.M. and Tang, Z. and Dinani, R. Tavakoli and Tsentalovich, E. and Vidal, C. and Wei, W. and White, C.R. and Williamson, S.E. and Yang, L. and Yao, W. and Young, A.R.},
title = {A new cryogenic apparatus to search for the neutron electric dipole moment},
journal = {Journal of Instrumentation}
}

@article{KOROBKINA2002462,
title = {Production of UCN by downscattering in superfluid He4},
journal = {Physics Letters A},
volume = {301},
number = {5},
pages = {462-469},
year = {2002},
issn = {0375-9601},
doi = {10.1016/S0375-9601(02)01052-6},
url = {https://www.sciencedirect.com/science/article/pii/S0375960102010526},
author = {E. Korobkina and R. Golub and B.W. Wehring and A.R. Young}
}

@article{PENN2001332,
title = {A low-noise 3He ionization chamber for measuring the energy spectrum of a cold neutron beam},
journal = {Nuclear Instruments and Methods in Physics Research Section A: Accelerators, Spectrometers, Detectors and Associated Equipment},
volume = {457},
number = {1},
pages = {332-337},
year = {2001},
issn = {0168-9002},
doi = {10.1016/S0168-9002(00)00748-8},
url = {https://www.sciencedirect.com/science/article/pii/S0168900200007488},
author = {S.D. Penn and E.G. Adelberger and B.R. Heckel and D.M. Markoff and H.E. Swanson}
}

@misc{9202314,
  author       = {{Henriksson Nordqvist, Hugo}},
  language     = {{eng}},
  note         = {{Student Paper}},
  title        = {{Studies of the HIBEAM Neutron Detector for Axion-Like Particle Searches at the European Spallation Source}},
  year         = {{2025}},
}

@misc{9241726,
  author       = {{Kuhl, Marianne Pauline}},
  language     = {{eng}},
  note         = {{Student Paper}},
  title        = {{Design of an optics system for the short-HIBEAM beamline at the European Spallation Source}},
  year         = {{2026}},
}

@misc{TYREX2,
  author       = {{Institut Laue-Langevin}},
  title        = {{TYREX2 characteristics}},
  howpublished = {ILL instrument webpage},
  year         = {2026},
  note         = {Accessed 2026-09-04},
  url          = {https://www.ill.eu/en/for-ill-users/instruments/instruments-list/tyrex-2/characteristics/}
}

@article{Piegsa2012,
  author  = {Piegsa, Florian M. and Pignol, Guillaume},
  title   = {Limits on the Axial Coupling Constant of New Light Bosons},
  journal = {Physical Review Letters},
  volume  = {108},
  pages   = {181801},
  year    = {2012},
  doi     = {10.1103/PhysRevLett.108.181801}
}

@article{Altarev2009,
  author  = {Altarev, I. and others},
  title   = {Test of Lorentz Invariance with Spin Precession of Ultracold Neutrons},
  journal = {Physical Review Letters},
  volume  = {103},
  pages   = {081602},
  year    = {2009},
  doi     = {10.1103/PhysRevLett.103.081602}
}

@article{Parnell2020,
  author  = {Parnell, S. R. and van Well, A. A. and Plomp, J. and Dalgliesh, R. M. and Steinke, N.-J. and Cooper, J. F. K. and Geerits, N. and Steffen, K. E. and Snow, W. M. and de Haan, V. O.},
  title   = {Search for Exotic Spin-Dependent Couplings of the Neutron with Matter Using Spin-Echo Based Neutron Interferometry},
  journal = {Physical Review D},
  volume  = {101},
  pages   = {122002},
  year    = {2020},
  doi     = {10.1103/PhysRevD.101.122002}
}

@article{Chu2013,
  author  = {Chu, P.-H. and Dennis, A. and Fu, C. B. and Gao, H. and Khatiwada, R. and Laskaris, G. and Li, K. and Smith, E. and Snow, W. M. and others},
  title   = {Laboratory Search for Spin-Dependent Short-Range Force from Axionlike Particles Using Optically Polarized $^{3}$He Gas},
  journal = {Physical Review D},
  volume  = {87},
  pages   = {011105},
  year    = {2013},
  doi     = {10.1103/PhysRevD.87.011105}
}

@article{Abel2017,
  author  = {Abel, C. and others},
  title   = {Search for Axionlike Dark Matter through Nuclear Spin Precession in Electric and Magnetic Fields},
  journal = {Physical Review X},
  volume  = {7},
  pages   = {041034},
  year    = {2017},
  doi     = {10.1103/PhysRevX.7.041034}
}

@article{Schulthess2022,
  author  = {Schulthess, Ivo and Chanel, Estelle and Fratangelo, Anastasio and Gottstein, Alexander and Gsponer, Andreas and Hodge, Zachary and Pistillo, Ciro and Ries, Dieter and Soldner, Torsten and Thorne, Jacob and Piegsa, Florian M.},
  title   = {New Limit on Axionlike Dark Matter Using Cold Neutrons},
  journal = {Physical Review Letters},
  volume  = {129},
  pages   = {191801},
  year    = {2022},
  doi     = {10.1103/PhysRevLett.129.191801}
}

@article{Cong2025,
  author  = {Cong, Lei and Ji, Wei and Fadeev, Pavel and Ficek, Filip and Jiang, Min and Flambaum, Victor V. and Guan, Haosen and Jackson Kimball, Derek F. and Kozlov, Mikhail G. and Stadnik, Yevgeny V. and Budker, Dmitry},
  title   = {Spin-Dependent Exotic Interactions},
  journal = {Reviews of Modern Physics},
  volume  = {97},
  pages   = {025005},
  year    = {2025},
  doi     = {10.1103/RevModPhys.97.025005}
}

@article{Wang2026,
  author  = {Wang, Yuanhong and Huang, Ying and Kang, Xiang and Chang, Dangui and Xu, Jiaxuan and Chen, Yifan and Pustelny, Szymon and Zheng, Wenqiang and Jiang, Min and Peng, Xinhua and Du, Jiangfeng},
  title   = {Constraints on Axion Dark Matter by Distributed Intercity Quantum Sensors},
  journal = {Nature},
  volume  = {650},
  pages   = {314--319},
  year    = {2026},
  doi     = {10.1038/s41586-025-10034-w}
}

@article{Burgman:2026fer,
    author = "Burgman, Alexander and others",
    title = "{Vertex reconstruction for a search for neutron-antineutron conversions with HIBEAM}",
    eprint = "2608.28314",
    archivePrefix = "arXiv",
    primaryClass = "hep-ex",
    month = "8",
    year = "2026"
}

@article{BRAND,
	author = {{Bodek, K.} and {De Keukeleere, L.} and {Kolodziej, M.} and {Kozela, A.} and {Kuzniak, M.} and {Lojek, K.} and {Perkowski, M.} and {Przybilski, H.} and {Pysz, K.} and {Rozpedzik, D.} and {Severijns, N.} and {Soldner, T.} and {Young, A.R.} and {Zejma, J.}},
	title = {BRAND – Search for BSM physics at TeV scale by exploring transverse polarization of electrons emitted in neutron decay},
	DOI= "10.1051/epjconf/201921904001",
	url= "https://doi.org/10.1051/epjconf/201921904001",
	journal = {EPJ Web Conf.},
	year = 2019,
	volume = 219,
	pages = "04001",
}

@article{MARKISCH2009216,
title = {The new neutron decay spectrometer Perkeo III},
journal = {Nuclear Instruments and Methods in Physics Research Section A: Accelerators, Spectrometers, Detectors and Associated Equipment},
volume = {611},
number = {2},
pages = {216-218},
year = {2009},
note = {Particle Physics with Slow Neutrons},
issn = {0168-9002},
doi = {10.1016/j.nima.2009.07.066},
url = {https://www.sciencedirect.com/science/article/pii/S0168900209015290},
author = {B. Märkisch and H. Abele and D. Dubbers and F. Friedl and A. Kaplan and H. Mest and M. Schumann and T. Soldner and D. Wilkin}
}

@article{PERC,
title = {A clean, bright, and versatile source of neutron decay products},
journal = {Nuclear Instruments and Methods in Physics Research Section A: Accelerators, Spectrometers, Detectors and Associated Equipment},
volume = {596},
number = {2},
pages = {238-247},
year = {2008},
issn = {0168-9002},
doi = {10.1016/j.nima.2008.07.157},
url = {https://www.sciencedirect.com/science/article/pii/S0168900208011285},
author = {D. Dubbers and H. Abele and S. Baeßler and B. Märkisch and M. Schumann and T. Soldner and O. Zimmer}
}

@article{Soldner_2019,
   title={ANNI – A pulsed cold neutron beam facility for particle physics at the ESS},
   volume={219},
   ISSN={2100-014X},
   url={http://dx.doi.org/10.1051/epjconf/201921910003},
   DOI={10.1051/epjconf/201921910003},
   journal={EPJ Web of Conferences},
   publisher={EDP Sciences},
   author={Soldner, Torsten and Abele, Hartmut and Konrad, Gertrud and Märkisch, Bastian and Piegsa, Florian M. and Schmidt, Ulrich and Theroine, Camille and Sánchez, Pablo Torres},
   editor={Jenke, T. and Degenkolb, S. and Geltenbort, P. and Jentschel, M. and Nesvizhevsky, V.V. and Rebreyend, D. and Roccia, S. and Soldner, T. and Stutz, A. and Zimmer, O.},
   year={2019},
   pages={10003} }

@article{Santoro:2024lvc,
    author = "Santoro, V. and others",
    title = "{HighNESS conceptual design report: Volume II. The NNBAR experiment.}",
    doi = "10.3233/jnr-230951",
    journal = "J. Neutron Res.",
    volume = "25",
    number = "3-4",
    pages = "315--406",
    year = "2024"
}

@article{Mohapatra:2024mirrorDM,
  author  = {Mohapatra, Rabindra N.},
  title   = {Dark Matter and Mirror World},
  journal = {Entropy},
  volume  = {26},
  number  = {4},
  pages   = {282},
  year    = {2024},
  doi     = {10.3390/e26040282}
}

@article{Kamyshkov:2022nmn,
  author        = {Kamyshkov, Yuri and Ternullo, James and Varriano, Louis and Berezhiani, Zurab},
  title         = {Neutron-Mirror Neutron Oscillations in Absorbing Matter},
  journal       = {Symmetry},
  volume        = {14},
  number        = {2},
  pages         = {230},
  year          = {2022},
  doi           = {10.3390/sym14020230},
  eprint        = {2111.01791},
  archivePrefix = {arXiv}
}

@article{Canetti:2012zc,
  author        = {Canetti, Laurent and Drewes, Marco and Shaposhnikov, Mikhail},
  title         = {Matter and Antimatter in the Universe},
  journal       = {New J. Phys.},
  volume        = {14},
  pages         = {095012},
  year          = {2012},
  doi           = {10.1088/1367-2630/14/9/095012},
  eprint        = {1204.4186},
  archivePrefix = {arXiv}
}

@article{Mohapatra:2022nnbarBaryo,
  author  = {Mohapatra, Rabindra N.},
  title   = {Baryogenesis with Observable Neutron--Antineutron Oscillation},
  journal = {Acta Phys. Polon. B Proc. Suppl.},
  volume  = {15},
  pages   = {2-A12},
  year    = {2022},
  doi     = {10.5506/APhysPolBSupp.15.2-A12}
}

@article{Babu:2013yca,
  author  = {Babu, K. S. and Bhupal Dev, P. S. and Fortes, Elaine C. F. S. and Mohapatra, R. N.},
  title   = {Post-Sphaleron Baryogenesis and an Upper Limit on the Neutron-Antineutron Oscillation Time},
  journal = {Phys. Rev. D},
  volume  = {87},
  number  = {11},
  pages   = {115019},
  year    = {2013},
  doi     = {10.1103/PhysRevD.87.115019}
}

@article{Mmcapture,
    author = "Meirose, B. and Milstead, D.",
    title = "{Probing Sub-MeV Dark Matter with Neutron-Capture $\gamma$ Spectroscopy}",
    eprint = "2603.26472",
    archivePrefix = "arXiv",
    primaryClass = "nucl-ex",
    month = "3",
    year = "2026"
}

@article{Massarczyk2022M1Axion,
  author       = {Massarczyk, Ralph and Chu, P.-H. and Elliott, S. R.},
  title        = {Axion Emission from Nuclear Magnetic Dipole Transitions},
  journal      = {Phys. Rev. D},
  volume       = {105},
  number       = {1},
  pages        = {015031},
  year         = {2022},
  doi          = {10.1103/PhysRevD.105.015031},
  eprint       = {2112.08285},
  archivePrefix= {arXiv},
  primaryClass = {hep-ph}
}

@book{ESPP2026BriefingBook,
  author        = {de Blas, Jorge and others},
  title         = {Physics Briefing Book: Input for the 2026 update of the European Strategy for Particle Physics},
  series        = {CERN Yellow Reports: Monographs},
  volume        = {8},
  reportNumber  = {CERN-2025-008, CERN-ESU-2025-001},
  year          = {2025},
  eprint        = {2511.03883},
  archivePrefix = {arXiv},
  primaryClass  = {hep-ex},
  doi           = {10.17181/CERN.35CH.2O2P},
  url           = {https://cds.cern.ch/record/2944678}
}

@techreport{ESPP2026Recommendations,
  author        = {{The European Strategy Group}},
  title         = {The European Strategy for Particle Physics: 2026 Update --- Recommendations by the European Strategy Group},
  reportNumber  = {CERN-ESU-2025-002},
  address       = {Monte Verit\`a/Ascona, Switzerland},
  year          = {2025},
  doi           = {10.17181/CERN.423R.S20Z},
  url           = {https://cds.cern.ch/record/2950671}
}

@misc{NuPECCLRP2024,
  author        = {{NuPECC}},
  title         = {NuPECC Long Range Plan 2024 for European Nuclear Physics},
  year          = {2025},
  eprint        = {2503.15575},
  archivePrefix = {arXiv},
  primaryClass  = {nucl-ex},
  doi           = {10.48550/arXiv.2503.15575},
  url           = {https://arxiv.org/abs/2503.15575}
}

@article{Barbieri:1981yr,
    author = "Barbieri, Riccardo and Mohapatra, Rabindra N.",
    title = "{Spontaneous Breaking of Global Baryon Number Symmetry}",
    reportNumber = "CERN-TH-3140",
    doi = "10.1007/BF01574002",
    journal = "Z. Phys. C",
    volume = "11",
    pages = "175",
    year = "1981"
}

@article{Nesvizhevsky:2018tft,
    author = "Nesvizhevsky, V. V. and Gudkov, V. and Protasov, K. V. and Snow, W. M. and Voronin, A. Yu",
    title = "{Experimental Approach to Search for Free Neutron-Antineutron Oscillations Based on Coherent Neutron and Antineutron Mirror Reflection}",
    eprint = "1810.04988",
    archivePrefix = "arXiv",
    primaryClass = "hep-ex",
    doi = "10.1103/PhysRevLett.122.221802",
    journal = "Phys. Rev. Lett.",
    volume = "122",
    number = "22",
    pages = "221802",
    year = "2019"
}

@article{Gudkov:2021wvn,
    author = "Gudkov, Vladimir and others",
    title = "{A Possible Neutron-Antineutron Oscillation Experiment at PF1B at the Institut Laue Langevin}",
    doi = "10.3390/sym13122314",
    journal = "Symmetry",
    volume = "13",
    number = "12",
    pages = "2314",
    year = "2021"
}

@article{Byron2023CRES,
    author = "Byron, W. and others",
    collaboration = "He6-CRES",
    title = "{First Observation of Cyclotron Radiation from MeV-Scale e{\ensuremath{\pm}} following Nuclear {\ensuremath{\beta}} Decay}",
    eprint = "2209.02870",
    archivePrefix = "arXiv",
    primaryClass = "nucl-ex",
    doi = "10.1103/PhysRevLett.131.082502",
    journal = "Phys. Rev. Lett.",
    volume = "131",
    number = "8",
    pages = "082502",
    year = "2023"
}

@article{Bodek:2021BrandReview,
  author       = {Bodek, Kazimierz and Kozela, Adam},
  title        = {Measurement of the transverse polarization of electrons emitted in neutron decay -- nTRV experiment},
  journal      = {SciPost Phys. Proc.},
  volume       = {5},
  pages        = {015},
  year         = {2021},
  doi          = {10.21468/SciPostPhysProc.5.015}
}

@article{Gericke:2011qoo,
  author       = {Gericke, M. T. and others},
  title        = {Measurement of parity-violating $\gamma$-ray asymmetry in the capture of polarized cold neutrons on protons},
  journal      = {Phys. Rev. C},
  volume       = {83},
  pages        = {015505},
  year         = {2011},
  doi          = {10.1103/PhysRevC.83.015505}
}

@article{BaldoCeolin:1994nnbar,
    author = "Baldo-Ceolin, M. and others",
    title = "{A New experimental limit on neutron - anti-neutron oscillations}",
    reportNumber = "DFPD-94-EP-13",
    doi = "10.1007/BF01580321",
    journal = "Z. Phys. C",
    volume = "63",
    pages = "409--416",
    year = "1994"
}

@article{Snow2012PVRotation,
  author  = {Snow, W. M. and others},
  title   = {Parity violating neutron spin rotation in He-4 and H},
  journal = {Nuovo Cimento C},
  year    = {2012},
  volume  = {35},
  number  = {4},
  pages   = {57--62},
  doi     = {10.1393/ncc/i2012-11291-8}
}

@article{Meirose:2025fzq,
    author = "Meirose, B. and Nieuwenhuis, R. and Pasechnik, R. and Gisbert, H. and Vale Silva, L. and Milstead, D.",
    title = "{Searching for long-lived particles in free neutron experiments}",
    eprint = "2506.08701",
    archivePrefix = "arXiv",
    primaryClass = "hep-ex",
    doi = "10.1088/1361-6471/ade3f2",
    journal = "J. Phys. G",
    volume = "52",
    number = "7",
    pages = "075001",
    year = "2025"
}

@article{Siemensen:2018charge,
  title = {Search for an electric charge of the neutron},
  author = {Siemensen, C. and D\"using, C. and Geltenbort, P. and Giebel, C. and Reich, T. and Plonka, C.},
  journal = {Phys. Rev. D},
  volume = {97},
  issue = {5},
  pages = {052004},
  numpages = {9},
  year = {2018},
  month = {Mar},
  publisher = {American Physical Society},
  doi = {10.1103/PhysRevD.97.052004},
  url = {https://link.aps.org/doi/10.1103/PhysRevD.97.052004}
}

@article{Falkowski2021Beta,
  author       = {Adam Falkowski and Mart{\'i}n Gonz{\'a}lez-Alonso and Oscar Naviliat-Cuncic},
  title        = {Comprehensive analysis of beta decays within and beyond the Standard Model},
  journal      = {JHEP},
  year         = {2021},
  volume       = {04},
  pages        = {126},
  doi          = {10.1007/JHEP04(2021)126},
  eprint       = {2010.13797},
  archivePrefix= {arXiv},
  primaryClass = {hep-ph}
}

@article{Cirigliano2024CLEW,
  author       = {Vincenzo Cirigliano and Wouter Dekens and Jordy de Vries and Emanuele Mereghetti and Tom Tong},
  title        = {Anomalies in global SMEFT analyses: a case study of first-row CKM unitarity},
  journal      = {JHEP},
  year         = {2024},
  volume       = {03},
  pages        = {033},
  doi          = {10.1007/JHEP03(2024)033},
  eprint       = {2311.00021},
  archivePrefix= {arXiv},
  primaryClass = {hep-ph}
}

@article{Hayen2024Review,
  author       = {Leendert Hayen},
  title        = {Opportunities and Open Questions in Modern Beta Decay},
  journal      = {Annual Review of Nuclear and Particle Science},
  year         = {2024},
  volume       = {74},
  pages        = {497--528},
  doi          = {10.1146/annurev-nucl-121423-100730},
  eprint       = {2403.08485},
  archivePrefix= {arXiv},
  primaryClass = {nucl-th}
}

@article{ATLAS2025ccDY,
  author       = {{ATLAS Collaboration}},
  title        = {Measurement of double-differential charged-current Drell--Yan cross-sections at high transverse masses in $pp$ collisions at $\sqrt{s}=13$ TeV with the ATLAS detector},
  journal      = {JHEP},
  year         = {2025},
  volume       = {07},
  pages        = {026},
  doi          = {10.1007/JHEP07(2025)026},
  eprint       = {2502.21088},
  archivePrefix= {arXiv},
  primaryClass = {hep-ex},
  reportNumber = {CERN-EP-2025-032}
}

@article{Baessler2024Nab,
  author       = {Stefan Baessler and others},
  title        = {Study of neutron beta decay with the Nab experiment},
  journal      = {EPJ Web of Conferences},
  year         = {2024},
  volume       = {303},
  pages        = {05001},
  doi          = {10.1051/epjconf/202430305001},
  note         = {MENU 2023 proceedings}
}

@article{Broussard2025NabOps,
  author       = {L. J. Broussard and others},
  title        = {From Commissioning to Precision Data-Taking: Resolving Operational Challenges in the Nab Detector Systems},
  year         = {2025},
  eprint       = {2511.16678},
  archivePrefix= {arXiv},
  primaryClass = {physics.ins-det},
  doi          = {10.48550/arXiv.2511.16678}
}

@article{mirror1,
  title = {The Early Mirror Universe: Inflation, Baryogenesis, Nucleosynthesis and Dark Matter},
  shorttitle = {The Early Mirror Universe},
  author = {Berezhiani, Zurab and Comelli, Denis and Villante, Francesco L.},
  year = 2001,
  month = mar,
  journal = {Physics Letters B},
  volume = {503},
  number = {3},
  pages = {362--375},
  issn = {0370-2693},
  doi = {10.1016/S0370-2693(01)00217-9},
  urldate = {2026-09-07}
}

@article{mirror2,
  title = {Mirror World and Its Cosmological Consequences},
  author = {Berezhiani, Zurab},
  year = 2004,
  month = sep,
  journal = {International Journal of Modern Physics A},
  volume = {19},
  number = {23},
  pages = {3775--3806},
  publisher = {World Scientific Publishing Co.},
  issn = {0217-751X},
  doi = {10.1142/S0217751X04020075},
  urldate = {2026-09-07}
}

@article{Fornal:2023review,
  author       = {Fornal, Bartosz},
  title        = {Neutron Dark Decay},
  journal      = {Universe},
  volume       = {9},
  number       = {10},
  pages        = {449},
  year         = {2023},
  doi          = {10.3390/universe9100449},
  eprint       = {2306.11349},
  archivePrefix= {arXiv},
  primaryClass = {hep-ph}
}

@misc{McCrea:2020n3He,
  author       = {McCrea, M. and others},
  title        = {The n$^3$He Experiment: Parity Violation in Polarized Neutron Capture on $^3$He},
  year         = {2020},
  eprint       = {2004.10889},
  archivePrefix= {arXiv},
  primaryClass = {nucl-ex}
}

@article{Berezhiani:2015afa,
    author = "Berezhiani, Zurab",
    title = "{Neutron{\textendash}antineutron oscillation and baryonic majoron: low scale spontaneous baryon violation}",
    eprint = "1507.05478",
    archivePrefix = "arXiv",
    primaryClass = "hep-ph",
    doi = "10.1140/epjc/s10052-016-4564-0",
    journal = "Eur. Phys. J. C",
    volume = "76",
    number = "12",
    pages = "705",
    year = "2016"
}

@article{PERCProgress2012,
  author  = {Konrad, G. and others},
  title   = {Neutron Decay with {PERC}: A Progress Report},
  journal = {Journal of Physics: Conference Series},
  volume  = {340},
  pages   = {012048},
  year    = {2012},
  doi     = {10.1088/1742-6596/340/1/012048},
  url     = {https://doi.org/10.1088/1742-6596/340/1/012048}
}

@misc{MaerkischCKM2023,
  author       = {M{\"a}rkisch, B.},
  title        = {Neutron beta decay with pulsed cold neutron beams: {PERKEO III} and {PERC} (CKM 2023 talk)},
  howpublished = {Indico: 12th International Workshop on the CKM Unitarity Triangle (CKM 2023)},
  year         = {2023},
  url          = {https://indico.cern.ch/event/1184945/contributions/5540848/attachments/2719237/4723864/Maerkisch%20CKM%202023%20-1.pdf}
}

@article{Roick2019,
  author       = {Roick, C. and Saul, H. and Abele, H. and M{\"a}rkisch, B.},
  title        = {Undetected Electron Backscattering in {PERKEO III}},
  year         = {2019},
  eprint       = {1905.10189},
  archivePrefix= {arXiv},
  primaryClass = {nucl-ex},
  url          = {https://arxiv.org/abs/1905.10189}
}

@misc{MaerkischPSI2022,
  author       = {M{\"a}rkisch, B.},
  title        = {Neutron beta decay with pulsed cold neutron beams (PSI 2022 talk)},
  howpublished = {Indico: Physics of Fundamental Symmetries and Interactions (PSI 2022)},
  year         = {2022},
  url          = {https://indico.psi.ch/event/11742/contributions/38334/attachments/23314/41273/Maerkisch%20-%20PSI2022.pdf}
}

@article{Saul2020_b,
  author       = {Saul, H. and Roick, C. and Abele, H. and others},
  title        = {Limit on the Fierz Interference Term $b$ from a Measurement of the Beta Asymmetry in Neutron Decay},
  journal      = {Phys. Rev. Lett.},
  volume       = {125},
  pages        = {112501},
  year         = {2020},
  doi          = {10.1103/PhysRevLett.125.112501},
  eprint       = {1911.01766},
  archivePrefix= {arXiv},
  primaryClass = {nucl-ex},
  url          = {https://arxiv.org/abs/1911.01766}
}

@article{Abe:2020ywm,
    author = "Abe, K. and others",
    collaboration = "Super-Kamiokande",
    title = "{Neutron-antineutron oscillation search using a 0.37 megaton$\cdot$year exposure of Super-Kamiokande}",
    eprint = "2012.02607",
    archivePrefix = "arXiv",
    primaryClass = "hep-ex",
    doi = "10.1103/PhysRevD.103.012008",
    journal = "Phys. Rev. D",
    volume = "103",
    pages = "012008",
    year = "2021"
}

@article{Yiu:2022faw,
    author = "Yiu, Sze-Chun and others",
    title = "{Status of the Design of an Annihilation Detector to Observe Neutron-Antineutron Conversions at the European Spallation Source}",
    doi = "10.3390/sym14010076",
    journal = "Symmetry",
    volume = "14",
    number = "1",
    pages = "76",
    year = "2022"
}

@article{Barrow:2025rhm,
    author = "Barrow, Joshua L. and Fierlinger, Peter and Kamyshkov, Yuri and Meirose, Bernhard and Milstead, David and Mohapatra, Rabindra N. and Santoro, Valentina",
    title = "{Avoiding Blindness in Baryon Number Violating Processes: Free-Beam and Intranuclear Paths to Neutron-Antineutron Transitions}",
    eprint = "2511.01829",
    archivePrefix = "arXiv",
    primaryClass = "hep-ph",
    month = "11",
    year = "2025"
}

@article{Piegsa2013_BeamEDM,
  author       = {Piegsa, Florian M.},
  title        = {New concept for a neutron electric dipole moment search using a pulsed beam},
  journal      = {Physical Review C},
  volume       = {88},
  pages        = {045502},
  year         = {2013},
  doi          = {10.1103/PhysRevC.88.045502},
  eprint       = {1309.1959},
  archivePrefix= {arXiv},
  primaryClass = {nucl-ex}
}

@article{Chanel2019_PulsedBeamEDM,
  author       = {Chanel, E. and others},
  title        = {The pulsed neutron beam EDM experiment},
  journal      = {EPJ Web of Conferences},
  volume       = {219},
  pages        = {02004},
  year         = {2019},
  doi          = {10.1051/epjconf/201921902004},
  eprint       = {1812.03987},
  archivePrefix= {arXiv},
  primaryClass = {physics.ins-det}
}

@article{Abel2020_PSI_nEDM,
  author       = {Abel, C. and Ayres, N. J. and Ban, G. and others},
  title        = {Measurement of the Permanent Electric Dipole Moment of the Neutron},
  journal      = {Physical Review Letters},
  volume       = {124},
  pages        = {081803},
  year         = {2020},
  doi          = {10.1103/PhysRevLett.124.081803},
  eprint       = {2001.11966},
  archivePrefix= {arXiv},
  primaryClass = {hep-ex}
}

@article{Chupp2019_EDMReview,
  author       = {Chupp, T. E. and Fierlinger, P. and Ramsey-Musolf, M. J. and Singh, J. T.},
  title        = {Electric dipole moments of atoms, molecules, nuclei, and particles},
  journal      = {Reviews of Modern Physics},
  volume       = {91},
  pages        = {015001},
  year         = {2019},
  doi          = {10.1103/RevModPhys.91.015001},
  eprint       = {1710.02504},
  archivePrefix= {arXiv},
  primaryClass = {physics.atom-ph}
}

@article{Santoro:2023izd,
    author = "Santoro, V. and others",
    title = "{The HIBEAM instrument at the European spallation source}",
    eprint = "2311.08326",
    archivePrefix = "arXiv",
    primaryClass = "physics.ins-det",
    doi = "10.1088/1361-6471/adc8c2",
    journal = "J. Phys. G",
    volume = "52",
    number = "4",
    pages = "040501",
    year = "2025"
}

@article{PhysRevLett.133.181001,
  title = {Proposal for a Ramsey Neutron-Beam Experiment to Search for Ultralight Axion Dark Matter at the European Spallation Source},
  author = {Fierlinger, P. and Holl, M. and Milstead, D. and Santoro, V. and Snow, W. M. and Stadnik, Y. V.},
  journal = {Phys. Rev. Lett.},
  volume = {133},
  issue = {18},
  pages = {181001},
  numpages = {7},
  year = {2024},
  month = {Oct},
  publisher = {American Physical Society},
  doi = {10.1103/PhysRevLett.133.181001},
  url = {https://link.aps.org/doi/10.1103/PhysRevLett.133.181001}
}

@article{Abele:2022iml,
    author = "Abele, H. and others",
    title = "{Particle Physics at the European Spallation Source}",
    eprint = "2211.10396",
    archivePrefix = "arXiv",
    primaryClass = "physics.ins-det",
    doi = "10.1016/j.physrep.2023.06.001",
    journal = "Phys. Rept.",
    volume = "1023",
    pages = "1--84",
    year = "2023"
}

@article{zurab2,
    author = "Berezhiani, Zurab",
    title = "{More about neutron - mirror neutron oscillation}",
    eprint = "0804.2088",
    archivePrefix = "arXiv",
    primaryClass = "hep-ph",
    doi = "10.1140/epjc/s10052-009-1165-1",
    journal = "Eur. Phys. J. C",
    volume = "64",
    pages = "421--431",
    year = "2009"
}

@article{Bertone:2004pz,
    author = "Bertone, Gianfranco and Hooper, Dan and Silk, Joseph",
    title = "{Particle dark matter: Evidence, candidates and constraints}",
    eprint = "hep-ph/0404175",
    archivePrefix = "arXiv",
    reportNumber = "FERMILAB-PUB-04-047-A",
    doi = "10.1016/j.physrep.2004.08.031",
    journal = "Phys. Rept.",
    volume = "405",
    pages = "279--390",
    year = "2005"
}

@article{nTMM,
    author = "Berezhiani, Zurab and Biondi, Riccardo and Kamyshkov, Yuri and Varriano, Louis",
    title = "{On the Neutron Transition Magnetic Moment}",
    eprint = "1812.11141",
    archivePrefix = "arXiv",
    primaryClass = "nucl-th",
    doi = "10.3390/physics1020021",
    journal = "MDPI Physics",
    volume = "1",
    number = "2",
    pages = "271--289",
    year = "2019"
}

@article{disappearance,
    author = "Berezhiani, Zurab and Frost, Matthew and Kamyshkov, Yuri and Rybolt, Ben and Varriano, Louis",
    title = "{Neutron Disappearance and Regeneration from Mirror State}",
    eprint = "1703.06735",
    archivePrefix = "arXiv",
    primaryClass = "hep-ex",
    doi = "10.1103/PhysRevD.96.035039",
    journal = "Phys. Rev. D",
    volume = "96",
    number = "3",
    pages = "035039",
    year = "2017"
}

@article{ann1,
    author = "Golubeva, E. S. and Kondratyuk, L. A.",
    editor = "Koch, H. and Kunze, M. and Peters, K.",
    title = "{Annihilation of low energy antineutrons on nuclei}",
    doi = "10.1016/S0920-5632(97)00260-0",
    journal = "Nucl. Phys. B Proc. Suppl.",
    volume = "56",
    pages = "103--107",
    year = "1997"
}

@article{ann2,
    author = "Golubeva, Elena S. and Barrow, Joshua L. and Ladd, Charles G.",
    title = "{Model of $\bar n$ annihilation in experimental searches for $\bar n$ transformations}",
    eprint = "1804.10270",
    archivePrefix = "arXiv",
    primaryClass = "hep-ex",
    doi = "10.1103/PhysRevD.99.035002",
    journal = "Phys. Rev. D",
    volume = "99",
    number = "3",
    pages = "035002",
    year = "2019"
}

@article{ann3,
    author = "Barrow, Joshua L. and Golubeva, Elena S. and Paryev, Eduard and Richard, Jean-Marc",
    title = "{Progress and simulations for intranuclear neutron-antineutron transformations in ${}^{40}_{18} Ar$}",
    eprint = "1906.02833",
    archivePrefix = "arXiv",
    primaryClass = "hep-ex",
    reportNumber = "FERMILAB-PUB-19-298-V",
    doi = "10.1103/PhysRevD.101.036008",
    journal = "Phys. Rev. D",
    volume = "101",
    number = "3",
    pages = "036008",
    year = "2020"
}

@article{Davis:2016uyk,
    author = "Davis, E. David and Young, Albert R.",
    title = "{Neutron-antineutron oscillations beyond the quasifree limit}",
    eprint = "1611.04205",
    archivePrefix = "arXiv",
    primaryClass = "nucl-ex",
    doi = "10.1103/PhysRevD.95.036004",
    journal = "Phys. Rev. D",
    volume = "95",
    number = "3",
    pages = "036004",
    year = "2017"
}

@article{wasa,
    author = "Bargholtz, Chr. and others",
    collaboration = "CELSIUS/WASA",
    title = "{The WASA Detector Facility at CELSIUS}",
    eprint = "0803.2657",
    archivePrefix = "arXiv",
    primaryClass = "nucl-ex",
    doi = "10.1016/j.nima.2008.06.011",
    journal = "Nucl. Instrum. Meth. A",
    volume = "594",
    pages = "339--350",
    year = "2008"
}

@article{Bertelsen2017GuideBot,
  author  = {Bertelsen, Mads},
  title   = {The automatic neutron guide optimizer guide\_bot},
  journal = {Nuclear Instruments and Methods in Physics Research Section A: Accelerators, Spectrometers, Detectors and Associated Equipment},
  volume  = {867},
  pages   = {195--203},
  year    = {2017},
  doi     = {10.1016/j.nima.2017.06.012}
}

@article{mcstas,
  author  = {Willendrup, Peter Kj{\ae}r and Lefmann, Kim},
  title   = {{McStas} (i): Introduction, use, and basic principles for ray-tracing simulations},
  journal = {Journal of Neutron Research},
  volume  = {22},
  number  = {1},
  pages   = {1--16},
  year    = {2020},
  doi     = {10.3233/JNR-190108},
  url     = {https://doi.org/10.3233/JNR-190108}
}

@article{Rataj:2026suy,
    author = "Rataj, Blahoslav and Santoro, Valentina and Oskarsson, Anders and Holl, Matthias and Hehl, Verena and Silvermyr, David",
    title = "{Development and Characterization of a Time Projection Chamber Prototype for Neutron Oscillation Searches at the European Spallation Source}",
    eprint = "2605.28558",
    archivePrefix = "arXiv",
    primaryClass = "physics.ins-det",
    month = "5",
    year = "2026"
}

@article{ill,
    author = "Baldo-Ceolin, M. and others",
    title = "{A new experimental limit on neutron-anti neutron transitions}",
    reportNumber = "PD-1989-97",
    doi = "10.1016/0370-2693(90)90601-2",
    journal = "Phys. Lett. B",
    volume = "236",
    pages = "95--101",
    year = "1990"
}

@article{sterile1,
  title = {Experimental {{Search}} for {{Neutron}} to {{Mirror Neutron Oscillations}} as an {{Explanation}} of the {{Neutron Lifetime Anomaly}}},
  author = {Broussard, L. J. and Barrow, J. L. and {DeBeer-Schmitt}, L. and Dennis, T. and Fitzsimmons, M. R. and Frost, M. J. and Gilbert, C. E. and Gonzalez, F. M. and Heilbronn, L. and Iverson, E. B. and Johnston, A. and Kamyshkov, Y. and Kline, M. and Lewiz, P. and Matteson, C. and Ternullo, J. and Varriano, L. and Vavra, S.},
  year = 2022,
  month = may,
  journal = {Physical Review Letters},
  volume = {128},
  number = {21},
  pages = {212503},
  publisher = {American Physical Society},
  doi = {10.1103/PhysRevLett.128.212503},
  urldate = {2026-09-07}
}

@article{sterile2,
  author = {Gonzalez, Francisco M. and Rock, Cary and Broussard, Leah J. and {DeBeer-Schmitt}, Lisa and Frost, Matthew J. and Heilbronn, Lawrence and Iverson, Erik B. and Kamyshkov, Yuri and Kline, Michael and Milstead, David and Powers, Devyn and Rogers, James and Santoro, Valentina and Vavra, Shaun},
  title = {Improved Limits on {$n\rightarrow n^{\prime}$} Transformation from the {Spallation Neutron Source}},
  journal = {Physical Review D},
  volume = {110},
  number = {7},
  pages = {072022},
  year = {2024},
  doi = {10.1103/PhysRevD.110.072022},
  url = {https://doi.org/10.1103/PhysRevD.110.072022}
}

@article{RamseyProp,
  title = {A Simulation Framework for {{Ramsey}} Interferometry},
  author = {Persson, Linus B. and Fierlinger, Peter and Holl, Matthias and Santoro, Valentina},
  year = 2026,
  month = aug,
  journal = {Journal of Instrumentation},
  volume = {21},
  number = {08},
  pages = {P08004},
  publisher = {IOP Publishing},
  issn = {1748-0221},
  doi = {10.1088/1748-0221/21/08/P08004},
  urldate = {2026-09-07},
  langid = {english}
}

@misc{sterile3,
  title = {New Limits on Neutron-to-Mirror-Neutron Transformation from the {HFIR} Cold Neutron Beam},
  author = {Rogers, James M. and Broussard, Leah J. and Crawford, Christopher B. and {DeBeer-Schmitt}, Lisa and Frost, Matthew J. and Gonzalez, Francisco M. and Haviland, Carolyn O. and Heilbronn, Lawrence and Iverson, Erik B. and Kamyshkov, Yuri and Khan, Mubasshir and Mullins, Andrew and Milstead, David and Persson, Linus B. and Rock, Cary and Santoro, Valentina and Saunders, Alexander and Vavra, Shaun and Whittington, Nathan D.},
  year = 2026,
  month = aug,
  number = {arXiv:2608.21120},
  eprint = {2608.21120},
  primaryclass = {hep-ex},
  publisher = {arXiv},
  doi = {10.48550/arXiv.2608.21120},
  urldate = {2026-09-07},
  archiveprefix = {arXiv}
}

@article{PhysRevC.100.015204,
  title = {Experimental upper bound and theoretical expectations for parity-violating neutron spin rotation in $^{4}\mathrm{He}$},
  author = {Swanson, H. E. and Heckel, B. R. and Bass, C. D. and Bass, T. D. and Dawkins, J. M. and Horton, J. C. and Luo, D. and Snow, W. M. and Walbridge, S. B. and Crawford, B. E. and Gan, K. and Micherdzinska, A. M. and Huffer, C. and Markoff, D. M. and Mumm, H. P. and Nico, J. S. and Sarsour, M. and Sharapov, E. I. and Zhumabekova, V.},
  journal = {Phys. Rev. C},
  volume = {100},
  issue = {1},
  pages = {015204},
  numpages = {14},
  year = {2019},
  month = {Jul},
  publisher = {American Physical Society},
  doi = {10.1103/PhysRevC.100.015204},
  url = {https://link.aps.org/doi/10.1103/PhysRevC.100.015204}
}

@article{HardyTowner2020,
  author       = {Hardy, J. C. and Towner, I. S.},
  title        = {Superallowed $0^{+}\to 0^{+}$ nuclear $\beta$ decays 2020: Critical survey, with implications for $V_{ud}$ and CKM unitarity},
  journal      = {Phys. Rev. C},
  volume       = {102},
  pages        = {045501},
  year         = {2020},
  doi          = {10.1103/PhysRevC.102.045501}
}

@inproceedings{Holl2022_BeamlineSimulation,
  author       = {Holl, M. and others},
  title        = {Beamline Simulation for the NNBAR Experiment at the European Spallation Source},
  booktitle    = {Proceedings of the International Conference on Radiation Shielding and Topical Meeting on Radiation Protection and Shielding (ICRS/RPSD 2022)},
  pages        = {431--434},
  year         = {2022},
  doi          = {10.13182/ICRSRPSD22-39421},
  eprint       = {2209.08997},
  archivePrefix= {arXiv},
  primaryClass = {physics.ins-det}
}

@inproceedings{Dunne:2022calo,
  author       = {Dunne, K. and Meirose, B. and Milstead, D. and Oskarsson, A. and Santoro, V. and Silverstein, S. and Yiu, S.-C.},
  title        = {The HIBEAM/NNBAR Calorimeter Prototype},
  booktitle    = {Proceedings of the 2021 International Conference on Technology and Instrumentation in Particle Physics (TIPP 2021)},
  volume       = {2374},
  number       = {1},
  pages        = {012014},
  year         = {2022},
  doi          = {10.1088/1742-6596/2374/1/012014},
  eprint       = {2107.02147},
  archivePrefix= {arXiv},
  primaryClass = {physics.ins-det}
}

@article{MUSGRAVE201819,
title = {Measurement of the absolute neutron beam polarization from a supermirror polarizer and the absolute efficiency of a neutron spin rotator for the NPDGamma experiment using a polarized  3He neutron spin-filter},
journal = {Nuclear Instruments and Methods in Physics Research Section A: Accelerators, Spectrometers, Detectors and Associated Equipment},
volume = {895},
pages = {19-28},
year = {2018},
issn = {0168-9002},
doi = {10.1016/j.nima.2018.03.055},
url = {https://www.sciencedirect.com/science/article/pii/S0168900218304078},
author = {M.M. Musgrave and S. Baeßler and S. Balascuta and L. Barrón-Palos and D. Blyth and J.D. Bowman and T.E. Chupp and V. Cianciolo and C. Crawford and K. Craycraft and N. Fomin and J. Fry and M. Gericke and R.C. Gillis and K. Grammer and G.L. Greene and J. Hamblen and C. Hayes and P. Huffman and C. Jiang and S. Kucuker and M. McCrea and P.E. Mueller and S.I. Penttilä and W.M. Snow and E. Tang and Z. Tang and X. Tong and W.S. Wilburn}
}

@article{Gericke_2020,
  title     = {First Precision Measurement of the Parity Violating Asymmetry in Cold Neutron Capture on {$^{3}$He}},
  volume    = {125},
  number    = {13},
  journal   = {Physical Review Letters},
  publisher = {American Physical Society},
  author    = {Gericke, M. T. and Bae{\ss}ler, S. and Barr{\'o}n-Palos, L. and Birge, N. and Bowman, J. D. and Calarco, J. and Cianciolo, V. and Coppola, C. E. and Crawford, C. B. and Fomin, N. and Garishvili, I. and Greene, G. L. and Hale, G. M. and Hamblen, J. and Hayes, C. and Iverson, E. and Kabir, M. L. and McCrea, M. and Plemons, E. and Ram{\'i}rez-Morales, A. and Mueller, P. E. and Novikov, I. and Penttila, S. and Scott, E. M. and Watts, J. and Wickersham, C.},
  year      = {2020},
  month     = sep,
  doi       = {10.1103/PhysRevLett.125.131803},
  url       = {https://doi.org/10.1103/PhysRevLett.125.131803}
}

@article{SZYMANSKI1994564,
title = {Ion chamber system for neutron flux measurements},
journal = {Nuclear Instruments and Methods in Physics Research Section A: Accelerators, Spectrometers, Detectors and Associated Equipment},
volume = {340},
number = {3},
pages = {564-571},
year = {1994},
issn = {0168-9002},
doi = {10.1016/0168-9002(94)90139-2},
url = {https://www.sciencedirect.com/science/article/pii/0168900294901392},
author = {J.J. Szymanski and J.D. Bowman and P.P.J. Delheij and C.M. Frankle and J. Knudson and S. Penttilä and S.J. Seestrom and S.H. Yoo and V.W. Yuan and X. Zhu}
}

@article{Baessler2008aSPECT,
  author  = {Bae{\ss}ler, S. and Ayala Guardia, F. and Borg, M. and Gl{\"u}ck, F. and Heil, W. and Konrad, G. and Konorov, I. and Mu{\~n}oz Horta, R. and Petzoldt, G. and Rich, D. and Simson, M. and Sobolev, Yu. and Wirth, H.-F. and Zimmer, O.},
  title   = {First measurements with the neutron decay spectrometer aSPECT},
  journal = {The European Physical Journal A},
  volume  = {38},
  pages   = {17--26},
  year    = {2008},
  doi     = {10.1140/epja/i2008-10660-0}
}

@misc{ONeill:SINE2026,
  author       = {O'Neill, G.},
  title        = {Scientific Investigations of Nuclei at ESS (SINE)},
  howpublished = {Proposal submitted to the ESS Instrument Roadmap},
  year         = {2026},
  note         = {SINE proposal}
}

@misc{FINESSEProposal2026,
  author       = {{FINESSE Collaboration}},
  title        = {{FINESSE: A Versatile Beamline for Fundamental Interactions with Neutrons at the European Spallation SourcE}},
  howpublished = {ESS Instrument Roadmap proposal},
  year         = {2026},
  note         = {Submitted to the European Spallation Source Instrument Roadmap call}
}

@article{GavilanMartin2025SpinInterferometer,
  author  = {Gavilan-Martin, Daniel and {\L}ukasiewicz, Grzegorz and Padniuk, Mikhail and Klinger, Emmanuel and Smolis, Magdalena and Figueroa, Nataniel L. and Jackson Kimball, Derek F. and Sushkov, Alexander O. and Pustelny, Szymon and Budker, Dmitry and Wickenbrock, Arne},
  title   = {Searching for dark matter with a spin-based interferometer},
  journal = {Nature Communications},
  volume  = {16},
  pages   = {4953},
  year    = {2025},
  doi     = {10.1038/s41467-025-60178-6}
}

@article{Zhang2025AxionReanalysis,
  author  = {Zhang, K. Y. and Wu, L. Y. and Yan, H.},
  title   = {New Limits on Ultralight Axionlike Dark Matter from Reanalyzed Data},
  journal = {Physical Review Letters},
  volume  = {135},
  pages   = {131001},
  year    = {2025},
  doi     = {10.1103/l8kd-yhlg}
}

@article{Abel2021,
  title = {A Search for Neutron to Mirror-Neutron Oscillations Using the {{nEDM}} Apparatus at {{PSI}}},
  author = {Abel, C. and Ayres, N. J. and Ban, G. and Bison, G. and Bodek, K. and Bondar, V. and Chanel, E. and Chiu, P. -J. and Crawford, C. and Daum, M. and Dinani, R. T. and Emmenegger, S. and Flaux, P. and {Ferraris-Bouchez}, L. and Griffith, W. C. and Gruji{\'c}, Z. D. and Hild, N. and Kirch, K. and Koch, H. -C. and Koss, P. A. and Kozela, A. and Krempel, J. and Lauss, B. and Lefort, T. and Leredde, A. and Mohanmurthy, P. and {Naviliat-Cuncic}, O. and Pais, D. and Piegsa, F. M. and Pignol, G. and Rawlik, M. and Rebreyend, D. and Rien{\"a}cker, I. and Ries, D. and Roccia, S. and Rozpedzik, D. and {Schmidt-Wellenburg}, P. and Severijns, N. and Thorne, J. and Weis, A. and Wursten, E. and Zejma, J. and Zsigmond, G.},
  year = 2021,
  month = jan,
  journal = {Physics Letters B},
  volume = {812},
  pages = {135993},
  issn = {0370-2693},
  doi = {10.1016/j.physletb.2020.135993},
  urldate = {2026-10-08}
}

@article{Altarev2009b,
  title = {Neutron to Mirror-Neutron Oscillations in the Presence of Mirror Magnetic Fields},
  author = {Altarev, I. and Baker, C. A. and Ban, G. and Bodek, K. and Daum, M. and Fierlinger, P. and Geltenbort, P. and Green, K. and {van der Grinten}, M. G. D. and Gutsmiedl, E. and Harris, P. G. and Henneck, R. and Horras, M. and Iaydjiev, P. and Ivanov, S. and Khomutov, N. and Kirch, K. and Kistryn, S. and Knecht, A. and Knowles, P. and Kozela, A. and Kuchler, F. and Ku{\'z}niak, M. and Lauer, T. and Lauss, B. and Lefort, T. and Mtchedlishvili, A. and {Naviliat-Cuncic}, O. and Paul, S. and Pazgalev, A. and Pendlebury, J. M. and Petzoldt, G. and Pierre, E. and {Plonka-Spehr}, C. and Qu{\'e}m{\'e}ner, G. and Rebreyend, D. and Roccia, S. and Rogel, G. and Severijns, N. and Shiers, D. and Sobolev, {\relax Yu}. and Stoepler, R. and Weis, A. and Zejma, J. and Zenner, J. and Zsigmond, G.},
  year = 2009,
  month = aug,
  journal = {Physical Review D},
  volume = {80},
  number = {3},
  pages = {032003},
  publisher = {American Physical Society},
  doi = {10.1103/PhysRevD.80.032003},
  urldate = {2026-10-08}
}

@misc{Ayres2026,
  title = {Laboratory Constraints on {{peV}}-Scale Mass Splitting between Ordinary and Sterile Neutron States},
  author = {Ayres, N. J. and Berezhiani, Z. and Bison, G. and Bodek, K. and Bondar, V. and Chiu, P.-J. and Daum, M. and Doorenbos, C. B. and Emmenegger, S. and Kirch, K. and Kletzl, V. and Krempel, J. and Lauss, B. and Pais, D. and Rien{\"a}cker, I. and Ries, D. and Rozp{\k e}dzik, D. and {Schmidt-Wellenburg}, P. and Tanaka, K. S. and Zejma, J. and Ziehl, N. and Zsigmond, G.},
  year = 2026,
  month = sep,
  number = {arXiv:2608.12173},
  eprint = {2608.12173},
  primaryclass = {hep-ex},
  publisher = {arXiv},
  doi = {10.48550/arXiv.2608.12173},
  urldate = {2026-10-08},
  archiveprefix = {arXiv}
}

@article{Ban2007,
  title = {Direct {{Experimental Limit}} on {{Neutron}}--{{Mirror-Neutron Oscillations}}},
  author = {Ban, G. and Bodek, K. and Daum, M. and Henneck, R. and Heule, S. and Kasprzak, M. and Khomutov, N. and Kirch, K. and Kistryn, S. and Knecht, A. and Knowles, P. and Ku{\'z}niak, M. and Lefort, T. and Mtchedlishvili, A. and {Naviliat-Cuncic}, O. and Plonka, C. and Qu{\'e}m{\'e}ner, G. and Rebetez, M. and Rebreyend, D. and Roccia, S. and Rogel, G. and Tur, M. and Weis, A. and Zejma, J. and Zsigmond, G.},
  year = 2007,
  month = oct,
  journal = {Physical Review Letters},
  volume = {99},
  number = {16},
  pages = {161603},
  publisher = {American Physical Society},
  doi = {10.1103/PhysRevLett.99.161603},
  urldate = {2026-10-08}
}

@article{Ban2023,
  title = {Search for {{Neutron}}-to-{{Hidden-Neutron Oscillations}} in an {{Ultracold Neutron Beam}}},
  author = {Ban, G. and Chen, J. and Lefort, T. and {Naviliat-Cuncic}, O. and {Saenz-Arevalo}, W. and Chiu, P.-J. and Cl{\'e}ment, B. and Larue, P. and Pignol, G. and Roccia, S. and Guigue, M. and Jenke, T. and Perriolat, B. and {Schmidt-Wellenburg}, P.},
  year = 2023,
  month = nov,
  journal = {Physical Review Letters},
  volume = {131},
  number = {19},
  pages = {191801},
  publisher = {American Physical Society},
  doi = {10.1103/PhysRevLett.131.191801},
  urldate = {2026-10-08}
}

@article{Berezhiani2018,
  title = {New Experimental Limits on Neutron -- Mirror Neutron Oscillations in the Presence of Mirror Magnetic Field},
  author = {Berezhiani, Z. and Biondi, R. and Geltenbort, P. and Krasnoshchekova, I. A. and Varlamov, V. E. and Vassiljev, A. V. and Zherebtsov, O. M.},
  year = 2018,
  month = sep,
  journal = {The European Physical Journal C},
  volume = {78},
  number = {9},
  pages = {717},
  issn = {1434-6052},
  doi = {10.1140/epjc/s10052-018-6189-y},
  urldate = {2026-10-08},
  langid = {english}
}

@article{Serebrov2009,
  title = {Search for Neutron--Mirror Neutron Oscillations in a Laboratory Experiment with Ultracold Neutrons},
  author = {Serebrov, A. P. and Aleksandrov, E. B. and Dovator, N. A. and Dmitriev, S. P. and Fomin, A. K. and Geltenbort, P. and Kharitonov, A. G. and Krasnoschekova, I. A. and Lasakov, M. S. and Murashkin, A. N. and Shmelev, G. E. and Varlamov, V. E. and Vassiljev, A. V. and Zherebtsov, O. M. and Zimmer, O.},
  year = 2009,
  month = dec,
  journal = {Nuclear Instruments and Methods in Physics Research Section A: Accelerators, Spectrometers, Detectors and Associated Equipment},
  series = {Particle {{Physics}} with {{Slow Neutrons}}},
  volume = {611},
  number = {2},
  pages = {137--140},
  issn = {0168-9002},
  doi = {10.1016/j.nima.2009.07.041},
  urldate = {2026-10-08}
}

\end{document}